\documentclass[a4paper, 10pt, oneside,superscriptaddress,showkeys]{revtex4-1}

\usepackage[latin1]{inputenc}
\usepackage{amssymb}
\usepackage{amsmath}
\usepackage{amsthm}
\usepackage{bbold}
\usepackage{braket}
\usepackage{graphicx}
\usepackage{simplewick}
\usepackage[toc,page]{appendix}
\usepackage{enumerate}
\usepackage{framed}

\usepackage[usenames,dvipsnames]{color}
\usepackage{hyperref}
\hypersetup{
  colorlinks=true,
 linkcolor=blue,
 citecolor=red}

\newcommand{\be}{\begin{equation}}
\newcommand{\ee}{\end{equation}}

\begin{document}

\title{Cluster expansion for ground states of local Hamiltonians}

\author{Alvise Bastianello} 
\affiliation{SISSA, via Bonomea 265, 34136 Trieste, Italy}
\affiliation{INFN, Sezione di Trieste, Italy}

\author{Spyros Sotiriadis}
\affiliation{SISSA, via Bonomea 265, 34136 Trieste, Italy}
\affiliation{INFN, Sezione di Trieste, Italy}
\affiliation{Institut de Math\'{e}matiques de Marseille, (I2M) Aix Marseille Universit\'{e}, CNRS, 
Centrale Marseille, UMR 7373, 39, rue F. Joliot Curie, 13453, Marseille, France }
\affiliation{University of Roma Tre, Department of Mathematics and Physics, 
L.go S. L. Murialdo 1, 00146 Roma, Italy}
\thanks {Corresponding author at: SISSA, via Bonomea 265, 34136 Trieste, Italy.
E-mail address: abastia@sissa.it (A. Bastianello).}

\date{\today}

\begin{abstract}
A central problem in many-body quantum physics is the determination of the ground state of a thermodynamically large physical system. We construct a cluster expansion for ground states of local Hamiltonians, which naturally incorporates physical requirements inherited by locality as conditions on its cluster amplitudes. 
Applying a diagrammatic technique we derive the relation of these amplitudes to thermodynamic quantities and local observables. 
Moreover we derive a set of functional equations that determine the cluster amplitudes for a general Hamiltonian, verify the consistency with perturbation theory and discuss non-perturbative approaches. 
Lastly we verify the persistence of locality features of the cluster expansion under unitary evolution with a local Hamiltonian and provide applications to out-of-equilibrium problems: a simplified proof of equilibration to the GGE and a cumulant expansion for the statistics of work, for an interacting-to-free quantum quench.

\end{abstract}

\maketitle

\newpage

\tableofcontents

\newpage

\section{Introduction}
\label{introduction}
One of the main goals of quantum many body physics is to determine the ground state of physically interesting models. Whenever the ground state wavefunction is known, all other ground state properties and observables of interest can be derived. In practise in order to extract physically important information about the ground state, it is valuable to know it in a form that reflects its general physical properties and allows a systematic expansion of observables from more to less relevant information. For example, Renormalisation Group methods are supposed to capture the universal ground state properties of physical systems in the thermodynamic limit. 
One of the most fundamental physical principles that characterise physical systems is \emph{locality} and this is clearly reflected on the properties of their ground state. Indeed physical Hamiltonians are expected to involve only local couplings or interactions between near different points and this imposes constrains on the form of their ground state. In particular, ground states of local Hamiltonians possess an extensive number of particles, energy and other thermodynamic quantities; relative fluctuations of such extensive quantities about their average values decay rapidly with the system size; local observables, like multi-point correlations, in general have a well-defined limit for large system sizes. 
These properties are deeply linked to the extensivity of a suitably defined `free energy' from which all other thermodynamic quantities and local observables can be calculated.

The extensive part of this quantity can be expressed conveniently as a \emph{cluster expansion} which, roughly speaking, is an expansion in terms of `joint cumulants' or connected multi-point correlation functions in the state. Cluster expansions have been used extensively in statistical physics, both in classical and quantum systems, typically for the study of the partition function of thermal states \cite{mayer,old1,old2}. They have also played an important role in the constructive approach to quantum field theory \cite{constructiveQFT,malyshev} and systematic applications have been developed in various fields, from nuclear physics \cite{nuclear} to quantum optics \cite{qoptics}. Although most studies focused mainly on the partition function, it is also possible to define such expansions directly for the quantum states themselves, which is the approach adopted in the present work. 

This approach has the advantage of expressing the state in an exact form, yet and more importantly organising its information content in order of decreasing physical relevance, while clearly demonstrating its locality properties at any order. It also provides the possibility to study the thermodynamic properties of overlaps between different states. Such overlaps are useful in quantum information theory where they are known as \emph{fidelity} between two states, which has been shown to manifest criticality by exhibiting non-analytic behaviour when one of the two states approaches a critical point \cite{fid1,fid2,fid3,fid4,fid5,fid6}. It is also useful in out-of-equilibrium problems where a quantum system undergoes a rapid change in some parameter, a process known as \emph{quantum quench} \cite{cc06}. {Quantum quenches have attracted significant interest partially due to experimental implementation in cold atom systems \cite{exp1,exp2,exp3,exp4,exp5,exp6,exp7,exp8} and due to the possibility of studying thermalisation and more generally equilibration in closed quantum systems \cite{thr1a,thr1b,thr1c,thr2,thr3,thr4,thr5,thr5a,thr5b,thr6} (for a review \cite{revq2}). In this context, overlaps between the initial state and post-quench excited states are relevant in the study of} out-of-equilibrium quantum thermodynamics as they capture information about quantum fluctuation relations \cite{QF} and in particular the statistics of work done by such quenches \cite{Silva,SotiriadisGambassiSilva}. 
As we will see, the cluster expansion provides an exact spectral expansion of the state, allowing the systematic derivation of thermodynamic quantities, like the quench work statistics, as well as their quantum fluctuations.
We suppose the existence of fields $\psi^{\dagger}(\vec{x})$ and $\psi(\vec{x})$ in a $d$ dimensional space that satisfy canonical bosonic commutation rules
\begin{equation}
[\psi(\vec{x}),\psi(\vec{y})]=0\hspace{3pc}[\psi^{\dagger}(\vec{x}),\psi(\vec{y})]=\delta(\vec{x}-\vec{y})\label{intro1}
\end{equation}

We suppose the existence of a unique translational invariant vacuum $\ket{0}$ annihilated by $\psi(\vec{x})$ and assume that the Hilbert space can be described acting with $\psi^{\dagger}$ on the vacuum. In this perspective, the field $\psi^{\dagger}(\vec{x})$ has the meaning of creating a particle in position $\vec{x}$.
A general cluster expansion of a quantum state $|\Psi\rangle$ is of the form
\be
|\Psi\rangle = \mathcal{N} \exp\left[  \sum_{n=1}^\infty \frac{1}{n!}\int \prod_{i=1}^n d^{d}x_i \; W^{c}_n(\vec{x}_1,\vec{x}_2,...,\vec{x}_n) \prod_{i=1}^n \psi^\dagger(\vec{x}_i)  \right] |0\rangle\label{intro0}
\ee

The functions $W^{c}_n(\vec{x}_1,\vec{x}_2,...,\vec{x}_n)$ will be called \emph{cluster amplitudes} and are related to the correlations between different points. The simplest examples of states written in this form are the coherent and squeezed coherent states, in which the expansion terminates after one or two terms respectively. In general any state that is Gaussian in terms of the field operators $\psi^\dagger(\vec{x})$ can be written in the above form with only the first two cluster terms
\be
|\Psi\rangle = \mathcal{N} \exp\left[  \int d^{d}x\; W^{c}_1(\vec{x}) \psi^\dagger(\vec{x}) +\frac{1}{2} \int d^{d}x \hspace{1pt}d^{d}y \; W^{c}_2(\vec{x},\vec{y}) \psi^\dagger(\vec{x}) \psi^\dagger(\vec{y}) \right] |0\rangle\label{intro2}
\ee

Indeed only states of this form satisfy Wick's theorem, i.e. their multi-point correlation functions can be decomposed into combinations of two-point and one point correlation functions in all possible ways, or equivalently, connected correlation functions between more than two points vanish. Higher order terms in the above expansion give rise to non-zero multi-point connected correlation functions. In this respect the cluster expansion is clearly analogous to the cumulant expansion of probability theory. 

{Even though (\ref{intro0}) is written in the continuum limit, the general construction can be applied to lattice systems as well, once we substitute the spatial integrals with sums over the lattice sites. 
The above expansion can be equivalently written in terms of the Fourier transforms of the fields $\psi^{\dagger}(\vec{x})$ and $\psi(\vec{x})$, respectively $a^{\dagger}_{\vec{k}}$ and $a_{\vec{k}}$, that create and annihilate particles with definite momentum $\vec{k}$. In determining the ground state of an interacting Hamiltonian, the $a_{\vec{k}}^{\dagger}$ and $a_{\vec{k}}$ operators can be chosen to diagonalize the free part of the Hamiltonian, so that (\ref{intro0}) is an expansion of the interacting ground state in terms of the eigenmodes of the free Hamiltonian, but other choices are in principle allowed and sometimes more suitable. Also note that, although we will focus on bosonic systems, our method could also be extended to systems of fermions or other particle excitations obeying different statistics which will be subject of future studies. The general construction can be applied to condensed matter systems and relativistic quantum field theories; for explicit calculations we mainly focus on the latter context.}

{As we will see below, the locality of the underlying Hamiltonian imposes general thermodynamic conditions on their ground states that do not rely on the details of the particular system. We call states that satisfy these properties \emph{local quantum states}. 

The cluster amplitudes for the ground state of a specific model can be determined by solving a set of functional equations that are equivalent to the time-independent Schr\"odinger equation and similarly time evolution can be studied by a set of time-dependent equations. Both sets of equations are consistent with the locality properties at all orders and in particular the evolution under a local Hamiltonian preserves these properties.}
Apart from the potential of exact description of ground states, the cluster expansion is also valuable in designing approximation schemes, to a large extend independently of the parameter or quality that is considered small in a particular approximation. As we will see, the cluster expansion reproduces efficiently results from standard perturbation theory. More importantly, it is also suitable for non-perturbative methods, like self-consistent approximation or variational methods, but it could be also used to study out of equilibrium problems. Lastly, it can be useful in designing efficient numerical methods \cite{dmrg0,dmrg1,dmrg2,dmrg3}. In this context the cluster expansion provides a convenient way to express any ground state that automatically satisfies locality requirements. In general any approximation based on suitable truncation of the cluster expansion captures correctly the qualitative features related to locality and gives increasingly better quantitative results as the truncation order increases.
In the next section we present the structure of the paper and a detailed summary of the main results.

\section{Summary of results}
\label{summary} 

{The paper is organized in three parts. The first part presents the construction of the cluster expansion in a general abstract way and explains its main properties based on the physical requirements related to locality. The second part shows how the cluster expansion for ground states of general physical models can be derived by means of truncation schemes and recursive approximations, but also perturbative methods with which its consistency is demonstrated. The third part discuss the time-dependent formulation and applies results of the previous parts to out-of-equilibrium problems, especially quantum quenches.} 
Below we summarize the main results of each section:

\begin{enumerate}
\item \textbf{Construction and properties of the cluster expansion (Sec. \ref{clusterexp}-\ref{grtool} and their subsections)}
\begin{enumerate}
\item We identify the main consequences of locality in physical systems to the structure of their ground state. One of them is the cluster decomposition principle, i.e. the principle that connected multi-point correlation functions decay with the distance between the points. This property is valid for infinite sized systems. Considering instead large finite systems with local or short-range interactions, we expect that their macroscopic properties are described by a (suitably defined) extensive logarithmic partition function $\log \mathcal{Z}$. More generally for such systems we expect that, in the thermodynamic limit, thermodynamic quantities exhibit vanishing fluctuations and correlations functions of local observables are well-defined. We use these general physical requirements to identify a class of \emph{local quantum states} in which ground states of local Hamiltonians belong.
\item We define the \emph{partition function} $\mathcal{Z}[\Psi]$ of a quantum state $|\Psi\rangle$ through its norm, as compared to the vacuum of the bosonic fields over which we expand it and which is supposed to have non zero overlap with $\ket{\Psi}$. 
\item We construct the \emph{cluster expansion} for a pure state, whose information are completely encoded in suitable functions called \emph{cluster amplitudes}. 
\item We construct suitable graphical tools, similar to Feynman diagrams, to compute $\mathcal{Z}[\Psi]$ and the correlation functions in terms of the cluster amplitudes $W_n^c$.
\item We obtain a sufficient condition in order for the cluster amplitudes $W_n^c$ to describe a local quantum state at any order in the graphical expansion, specifically they must decay sufficiently fast with the distance.
\item Through the relation between $W_n^c$ and the correlators of $\psi$ and $\psi^{\dagger}$ we show that the cluster decomposition property is automatically satisfied at any order of the graphical expansion when the cluster amplitudes $W_n^c$ satisfy the above condition.
\end{enumerate}

\item \textbf{Derivation of cluster expansion for ground states (Sec. \ref{perturb} and its subsections)}
\begin{enumerate}
\item We derive the cluster amplitudes $W_n^c$ for the ground state of a general class of physical models by means of a perturbative approach. This provides a check of the general statements of the previous section and gives a natural truncation of the cluster expansion at each perturbation order.
\item We derive general functional equations for the cluster amplitudes of ground states. These constitute the time-independent Schr\"odinger equation for the cluster expansion of the ground state wavefunction.
\item We demonstrate the application of the above to the case of a free quantum field theory, where we easily derive the known `squeezed vacuum' form of the ground state, as well as to the interacting $\phi^4$ theory. We verify the consistency with standard perturbation theory. 
\item We comment on possible approximation schemes to describe ground states through the new tool of the cluster expansion. In particular the exact information encoded in the cluster expansion of local quantum states opens up the possibility of variational approaches, which could give insight beyond standard perturbation theory.
\end{enumerate}

\item \textbf{Applications to out of equilibrium thermodynamics (Sec. \ref{dynamic} and its subsections)}
\begin{enumerate}
\item We verify that, when a local quantum state evolves according to a local Hamiltonian, it remains a local quantum state during the evolution, that is to say the general properties of the cluster amplitudes of this class of states are not spoiled along the time evolution.
\item The evolution of quantum systems are studied in the Schr\"odinger picture through the time-dependent Schr\"odinger equation expressed in terms of the cluster amplitudes. This could give insight in new approximation schemes to describe systems out of equilibrium.
\item In the context of quantum quenches, it has been earlier shown that initial states that are ground states of free or interacting models and evolve under massive free Hamiltonians, tend for large times to some generalized sort of equilibrium, described by the so-called Generalized Gibbs Ensemble (GGE), which is expected to be valid more generally for the evolution under an integrable Hamiltonian. This can be shown much easier by using the general information about local quantum states.
\item A direct consequence of the cluster expansion properties is also that in the case of quenches where the initial state consists solely of pairs of particles of opposite momenta (which is a very common case) the initial state is in fact a squeezed state. This simplifies dramatically the study of its time evolution.
\item Starting from the knowledge of the cluster expansion, we compute the statistics of the work done during a quantum quench from an interacting to a free quantum field theory.
\end{enumerate}
\end{enumerate}

\section{General construction of cluster expansion}
\label{clusterexp}

This section is mainly divided in two parts: after having introduced the cluster expansion for general states, in Section \ref{thprop} we describe the thermodynamic properties that a ground state of a local or short range Hamiltonian is supposed to possess. Through these properties, in Section \ref{LQS} we will introduce the class of \emph{local quantum states} and anticipate the main restrictions imposed by locality requirements on their cluster expansion, leaving the technical details to Section \ref{grtool}. In order to construct the cluster expansion of a general state, we consider a system of local bosonic fields $\psi(\vec{x})$ defined in a finite box of length $L$ with periodic boundary conditions and we will later take the limit $L\rightarrow\infty$.
We define creation and annihilation operators of the bosonic fields in momentum space through the Fourier transform of the fields in coordinate space $\psi(\vec{x})$:
\begin{equation}
a_{\vec{k}}=\frac{1}{L^{d/2}}\int d^{d}x \hspace{2pt}e^{i\vec{k}\vec{x}}\psi(\vec{x})\label{construct12}
\end{equation}
The momenta ${\vec{k}}$ are quantized and the bosonic operators $a_{\vec{k}}$ satisfy canonical commutation rules
\begin{equation}
[a_{\vec{k}},a_{\vec{q}}]=0 \, ;\hspace{2pc}[a_{\vec{k}},a^{\dagger}_{\vec{q}}]=\delta_{\vec{k},\vec{q}}\, ;\hspace{3pc} \text{with }\vec{k}=\frac{2\pi}{L}(n_{1},..,n_{d})
\end{equation} 
where $\delta_{\vec{k},\vec{q}}$ is the Kronecker delta over all momentum components and $n_{i}$ are integers.
Any state $\ket{\Psi}$ can be constructed by acting with the $\psi^{\dagger}(\vec{x})$ operators on the vacuum of the theory, therefore we can expand it as
\begin{equation}
\ket{\Psi}=W_{0}\ket{0}+\int d^{d}x \hspace{2pt}W_{1}(\vec{x})\psi^{\dagger}(\vec{x})\ket{0}+\int d^{d}x\hspace{1pt}d^{d}y \hspace{2pt}W_{2}(\vec{x},\vec{y})\psi^{\dagger}(\vec{x})\psi^{\dagger}(\vec{y})\ket{0}+...\label{coordexp}
\end{equation}

The integrations from now on are meant to be in the finite box of length $L$.
In the following, we will suppose $\braket{0|\Psi}\ne 0$, so we can normalize the state with respect the reference state $\ket{0}$, that is to say $\braket{0|\Psi}=W_{0}=1$. It can be simply shown that any state such that $\braket{0|\Psi}\ne0$ can be written in the exponential form:
\begin{equation}
\ket{\Psi}=\exp\left[\sum_{n=1}^{\infty}\frac{1}{n!}\int d^{dn}x\hspace{3pt}W^{c}_{n}(\vec{x}_{1}..\vec{x}_{n})\psi^{\dagger}(\vec{x}_{1})..\psi^{\dagger}(\vec{x}_{n})\right]\ket{0}\label{constr17}
\end{equation}
where the $W_{n}^{c}$ is the \emph{connected} part of the $W_{n}$ function, defined in a way analogous to the cumulant expansion. The first few terms are:
\begin{equation}
\nonumber W^{c}_{1}(\vec{x})=W_{1}(\vec{x})\hspace{4pc}W^{c}_{2}(\vec{x},\vec{y})=W_{2}(\vec{x},\vec{y})-W_{1}(\vec{x})W_{2}(\vec{y})
\end{equation}
\begin{equation}
W^{c}_{3}(\vec{x},\vec{y},\vec{z})=W_{3}(\vec{x},\vec{y},\vec{z})-\left[W_{2}(\vec{x},\vec{y})W_{1}(\vec{z})+\text{permutations}\right] + 2 W_{1}(\vec{x})W_{1}(\vec{y})W_{1}(\vec{z})\label{cluster8}
\end{equation}

The consistency check of (\ref{coordexp}) and (\ref{constr17}) is a straightforward calculation. 
From now on, we will focus on states invariant under coordinate translations, thus each $W_{n}^{c}$ must be translationally invariant. It is useful to express the above expansion also in momentum space, which is simply done by a Fourier transform of the $W_{n}^{c}$ functions and of the $\psi^{\dagger}$ fields:
\begin{equation}
\ket{\Psi}=\exp\left[\sum_{n=1}^{\infty}\frac{1}{ L^{d(n-2)/2}}\frac{1}{n!}\sum_{\vec{k}_{1}+..+\vec{k}_{n}=0}\mathcal{K}_{n}(\vec{k}_{1},..,\vec{k}_{n})a^{\dagger}_{\vec{k}_{1}}..a^{\dagger}_{\vec{k}_{n}}\right]\ket{0}\label{expk}
\end{equation}
where $\mathcal{K}_{n}$ is linked to the Fourier transform of $W_{n}^{c}$ in the infinite volume limit:
\begin{equation}
\int d^{nd}x\hspace{3pt}e^{-i\sum_{j=1}^{n}\vec{k}_{j}\vec{x}_{j}}W_{n}^{c}(\vec{x}_{1},..,\vec{x}_{n})=(2\pi)^{d}\delta\left(\sum_{j=1}^{n}\vec{k}_{j}\right)\mathcal{K}_{n}(\vec{k}_{1},..,\vec{k}_{n})\label{wfourier}
\end{equation}

In the limit of infinite system size $L \to \infty$ we can write:
\be
|\Psi\rangle = \exp\left[  \sum_{n=1}^\infty \frac{(2\pi)^{d}}{n!}\int \prod_{i=1}^n \frac{d^{d}k_i}{(2\pi)^{d}} \; \delta\left(\sum_{i=1}^{n}\vec{k}_{i}\right)\mathcal{K}_n(\vec{k}_1,\vec{k}_2,...,\vec{k}_n) \prod_{i=1}^n a^\dagger(\vec{k}_i)  \right] |0\rangle\label{momentaexp}
\ee
where now the commutator of $a(\vec{k})$ is a Dirac delta function instead of a Kronecker delta, i.e. $[a(\vec{k}),a^{\dagger}(\vec{q})]=(2\pi)^{d}\delta^{d}(\vec{k}-\vec{q})$. 
Notice that in the above expression we have explicitly separated the Dirac delta function which is due to the translational invariance. Also notice that any explicit dependence on $L$ has disappeared. 
The cluster expansion can be written for all those states with non zero overlap with the bare vacuum, but ground states of local Hamiltonians are not arbitrary states and satisfy precise thermodynamic requirements: the next section is devoted to describe these properties and to the introduction of the \emph{local quantum states}.

\subsection{Thermodynamic properties of ground states and local quantum states}
\label{thprop}

As we already pointed out in the introduction, ground states of local Hamiltonians are not random states of the Hilbert space, but satisfy certain conditions following from locality. Below we enlist some common manifestations of locality.

\

\paragraph{Extensive thermodynamic quantities:} 

\

Ground states of extended systems are characterised by well-defined extensive thermodynamic quantities. Extensivity of the energy, for example, means that the expectation value of the system's Hamiltonian increases linearly with the system's volume $V=L^d$ in the limit $L\to\infty$. The same behaviour is also expected for other thermodynamic quantities that correspond to spatial integrals of local density operators over the whole system 
\be
Q = \int_V d^dx \;  \mathcal{Q}(\vec{x}) 
\label{ext}
\ee
where $\mathcal{Q}(x)$ is considered as `local' if it only involves operators acting on a neighbourhood of the point $\vec{x}$. In a lattice model this would mean operators acting on the site $x$ and those within a finite radius around it. In the continuum limit, this is equivalent to considering operators that may include spatial derivatives of any finite order. 
\ \\

\paragraph{Cluster decomposition and vanishing relative fluctuations:} 

\

Another crucial property of such states is that relative fluctuations of thermodynamic quantities vanish in the thermodynamic limit  $L\to\infty$. This property is the basis of an important ingredient of standard statistical physics, the equivalence of statistical ensembles. It is also related to locality, more precisely to the cluster decomposition principle, which is another fundamental requirement that ground states of local Hamiltonians are supposed to satisfy. It states that connected correlation functions between observables localised at large spatial separations must vanish e.g. for two point correlations
\be
\langle \mathcal{O}(\vec{x}) \mathcal{O}(\vec{x}+\vec{R}) \rangle \xrightarrow{|\vec{R}|\to\infty} \langle \mathcal{O}(\vec{x})\rangle \langle \mathcal{O}(\vec{x}+\vec{R}) \rangle\label{thprop4}
\ee

This principle originates from quantum field theory in infinite space and expresses the requirement that quantum measurements at distant spatial points must be independent from each other. In the context of statistical physics, where we are interested in the behaviour of systems of finite size $L$ in the thermodynamic limit, it refers to distances $R$ that tend to infinity not faster than the system size $L$ i.e.
\be
\langle \mathcal{O}(\vec{x}) \mathcal{O}(\vec{x}+ \vec{R}) \rangle_L - \langle \mathcal{O}(\vec{x})\rangle_L \; \langle \mathcal{O}(\vec{x}+\vec{R}) \rangle_L \xrightarrow[|\vec{R}|\ll L]{|\vec{R}|,L\to\infty }  0
\label{CD2}
\ee
where the subscript `$L$' means that the expectation value is calculated in a system of finite size $L$. 

The vanishing of relative fluctuations of extensive quantities in the thermodynamic limit is a direct consequence of the above relation. Indeed due to (\ref{CD2}) the variance of $Q$ as defined by (\ref{ext}) scales slower than $L^2$ and therefore the relative variance decays in the thermodynamic limit
\be
\frac{\langle Q^2 \rangle - \langle Q \rangle^2}{\langle Q \rangle^2} 
= \frac{\int_0^L dx \int_0^L dy \; \big ( \langle \mathcal{Q}(\vec{x}) \mathcal{Q}(\vec{y}) \rangle - \langle \mathcal{Q}(\vec{x}) \rangle \langle \mathcal{Q}(\vec{y}) \rangle \big )}{\big ( \int_0^L dx \; \langle \mathcal{Q}(\vec{x}) \rangle \big )^2 } \xrightarrow{L\to\infty }  0
\ee

\

\paragraph{Extensivity of `free energy':}

\

In thermal ensembles, thermodynamic quantities are in general derived as logarithmic derivatives of the partition function $\mathcal{Z}_{th}$ with respect to intensive parameters \cite{Amit}. 
For example, for a thermal state of temperature $\beta^{-1}$ the partition function is $\mathcal{Z}_{th}=\text{Tr} \, e^{-\beta H}$ where $H$ is the system's Hamiltonian, and the extensive energy is $\langle H\rangle = - \partial(\log\mathcal{Z}_{th})/\partial \beta$. 
It is useful to introduce the concept of partition function also for a generic pure state $|\Psi\rangle$. We will suppose that the state $\ket{\Psi}$ has non zero overlap with the vacuum $\ket{0}$ defined as the reference state in Section \ref{introduction}. Then we define its partition function as the norm
\be
\mathcal{Z}[\Psi] = \langle \Psi|\Psi\rangle\label{deffreeenergy}
\ee
when the normalization is fixed in reference to the vacuum of the theory
\begin{equation}
\braket{0|\Psi}=1\label{dar17}
\end{equation}
It should be mentioned that the above normalization presupposes a suitable UV regularization. Indeed in most cases in relativistic QFT the above overlap, when both states are normalized to one, is UV divergent in perturbation theory.
We will see this fact also in our framework and see that, assuming the normalization $\braket{0|\Psi}=1$ with an explicit UV cut-off, the norm $\braket{\Psi|\Psi}$ presents UV divergences when the cut-off is taken to infinity.
The aim of the present work is to study general thermodynamical properties of pure states that are linked to the infrared behavior of the theory and are not related to their UV behavior. UV properties are, in some sense, less general than the infrared ones: UV divergences and the proper renormalization scheme depend both on the dimensionality and on the specific model, but different systems share the same general thermodynamic properties.
Moreover, if the field theory is the continuous description of a lattice system, its lattice space provides a natural cut off: actually, in the presence of a lattice, our construction can be still applied provided we replace the integrals in the coordinate space with the proper summations.
Of course, a proper treatment of the UV singularities is necessary to compute physical expectation values in actual models and this problem deserves further investigation, but it is beyond the scope of the present work that aims to a description of general thermodynamic properties.
Because of the above, in the following we will assume that all the expressions are UV finite by means of a suitable cut off.
As a general remark, note that a proper renormalization scheme must guarantee finite physical quantities (i.e. correlation functions), but the overlaps in the renormalized theory could be still UV ill-defined when we take the cutoff to infinity, since they are not physical observables.

In the following, we define $\log\mathcal{Z}[\Psi]$ as the `free energy' of the state.
This definition for the free energy has some similarities with the usual thermal one and in particular it is expected to be an extensive quantity.
We can justify this statement when $\ket{\Psi}$ is a non degenerate ground state of some local Hamiltonian $H$ and the vacuum $\ket{0}$ is the ground state of another local Hamiltonian. For example, if $H$ is made by a free part plus an interaction, $\ket{0}$ could be the ground state of the free part of the Hamiltonian, but this is not the only possibility.
In this case, the free energy $\log\mathcal{Z}[\Psi]$ becomes a well known object in the study of the Casimir effect \cite{criticalcasimirgambassi} known as surface free energy. 
Consider a new quantity $\mathcal{Z}_{0}$ so defined:
\begin{equation}
\mathcal{Z}_{0}\equiv \log\bra{0}e^{-\beta H}\ket{0}\label{corr18}
\end{equation}

In the same way that a thermal partition function of a quantum system can be seen as the partition function of a classical system in the $d+1$ dimensional cylinder of circumference $\beta$, in the the present case $\mathcal{Z}_{0}$ can be seen as that of a slab geometry.
The width of this slab is given by $\beta$ and the boundary conditions on the slab are fixed by $\ket{0}$, that can be interpreted as a boundary state \cite{SotiriadisGambassiSilva}. 
The free energy $\log\mathcal{Z}_{0}$ is the sum of several terms \cite{criticalcasimirgambassi}. The leading term for large $\beta$ is extensive in the whole volume of the classical system, therefore proportional to the volume of the $d$ dimensional quantum system and to the length $\beta$ of the extra dimension: this is the bulk contribution to the free energy. A second term due to the boundaries is also present and it is usually called surface free energy: this quantity scales with the size of the boundaries, but it is not affected by their distance. Passing from the quantum to the classical system, the size of the boundary is simply the volume of the $d$ dimensional quantum system: the surface free energy is independent from $\beta$, but it is extensive in the volume of the quantum system. Subleading terms in $\beta\to\infty$ are generically present, but we are not interested in them.
We are going to see that the free energy $\log\mathcal{Z}[\Psi]$ is nothing else than the surface free energy in the classical system, therefore it must be extensive in the volume of the quantum system. In order to do so, we let $\beta\to\infty$ in (\ref{corr18}): in the zero temperature limit, the thermal ensemble becomes a projector on the ground state.
\begin{equation}
e^{-\beta H}\hspace{1pc}\xrightarrow {\beta\to\infty}\hspace{1pc}\frac{e^{-\beta E_{G}}}{\braket{\Psi|\Psi}}\ket{\Psi}\bra{\Psi}+..\label{thproj}
\end{equation}

Above, $E_{G}$ is the energy of the ground state $\ket{\Psi}$. If the theory presents a gap between the ground state and the first excited state, further corrections in (\ref{thproj}) are exponentially suppressed. We explicitly insert the normalization factor $\braket{\Psi|\Psi}$ because we are using the non standard normalization $\braket{0|\Psi}=1$.
Thanks to (\ref{thproj}) we can immediately evaluate the zero temperature limit of $\mathcal{Z}_{0}$:
\begin{equation}
\log\mathcal{Z}_{0}\xrightarrow{\beta\to\infty}\log\left(e^{-\beta E_{G}}\frac{1}{\braket{\Psi|\Psi}}\right)=-\beta E_{G}-\log\braket{\Psi|\Psi}
\label{thprop10}
\end{equation}

Above, further corrections are exponentially damped when $\beta\to\infty$.
Looking at (\ref{thprop10}) we can exactly recognize the contributions described before. The term $\beta E_{G}$ is proportional both to the length of the classical extra dimension and to the volume of the $d$ dimensional quantum system, since the ground state energy is an extensive quantity. Therefore, $-\beta E_{G}$ is the bulk free energy of the classical system on the slab. The next term is independent from $\beta$ and therefore it must be recognized as the surface free energy, extensive in the volume of the quantum system: apart from a minus sign, it is exactly the free energy of the state defined through its partition function (\ref{deffreeenergy}).

\

\paragraph{Well-defined correlation functions in the thermodynamic limit:}

\

Another property commonly required in quantum field theory is that correlation functions have a well-defined limit when the volume becomes infinite. In fact cluster expansion methods have been developed in the past primarily in order to check the validity of this requirement \cite{constructiveQFT}. In this context the system size $L$ plays the role of an infrared regularization and in the end all physically meaningful quantities must be independent from it as $L\rightarrow\infty$. 
In many physical contexts, this property may be considered as a consequence of cluster decomposition, in the following sense.  Cluster decomposition means that regions of the system very far apart are completely uncorrelated to each other. 
A small increase of the system size means physically that we modify the system only at its boundaries and therefore a measurement performed deep in the bulk should not be affected by this small change. As a result, bulk correlations should be independent from the system size when this is already very large. 
However we will recognize this property as a separate one in order to avoid restrictions coming from such more special considerations.

\subsection{Local quantum states}
\label{LQS}

Motivated by the general arguments of the previous section, we define as \emph{local quantum states} the states that share these thermodynamic properties. As we saw the various properties are not all independent but instead they are related to each other. We therefore identify three general properties which are more fundamental, in the sense that other properties, like the extensivity of thermodynamic quantities, vanishing of relative fluctuations and ensemble equivalence in the thermodynamic limit, can be deduced from them. Also we can consider these three properties as rather independent from each other, in the sense that one does not necessarily imply the other, except in more special settings.

\begin{center}
\framebox[15cm]{
\begin{minipage}{14cm}
\hspace{1pt} \ \\
\begin{center}
\textbf{Local quantum states and their properties:}
\end{center}
\ \\
We denote a state $\ket{\Psi}$ as a \emph{local quantum state} if the following properties are satisfied:
\begin{itemize}
\item \textbf{Property 1:} The correlators of the fundamental local fields $\psi$, $\psi^{\dagger}$ have a well-defined $L\rightarrow\infty$ limit, that is, in this limit they are independent from the system size. 
\item \textbf{Property 2:} The cluster decomposition property for the $\psi$, $\psi^{\dagger}$ fields holds.
\item\textbf{Property 3:} The `free energy' $\log\mathcal{Z}[\Psi]$ is an extensive quantity.
\end{itemize}
\hspace{1pt} \ \\
\end{minipage}}
\end{center}
\ \\

Property 2 is the cluster decomposition property expressed in terms of the fields $\psi$: if $\psi$ is a physical field itself, then this requirement is obvious. We will see at the beginning of Section \ref{perturb} how to properly define $\psi$ in the context of common relativistic field theories.

For the sake of clarity, we anticipate the connection of Properties 1, 2, 3 with the form of cluster amplitudes of local quantum states. In Section \ref{grtool} we introduce a diagrammatic technique for the expansion of the free energy and correlation functions in terms of cluster amplitudes and then we find a sufficient condition to ensure that any graph of this expansion is consistent with Properties 1, 2 and 3. We require the cluster amplitudes to be independent of the system size and to decay fast enough such that:
\begin{equation}
\int d^{d(n-j)}y\hspace{5pt} \left|W_{n}^{c}(\vec{x}_{1},..,\vec{x}_{j},\vec{y}_{1},..,\vec{y}_{n-j})\right|<M_{n} <\infty\hspace{2pc}\forall j\ge1,\hspace{5pt}\forall \vec{x}_{i}, \hspace{3pt}i\in\{1,...,j\}\label{suffcond}
\end{equation}
The above condition must be interpreted as follows: any UV singularities in $W_{n}^{c}$ must be regularized by means of an explicit UV cut off, in such a way that $W_{n}^{c}$ does not have any such singularity. In this case the bound $M_{n}$ depends explicitly on the cut off, as it can be seen in the special case $j=n$ of (\ref{suffcond}):
\begin{equation}
|W_{n}^{c}(\vec{x}_{1},...,\vec{x}_{n})|<M_{n},\hspace{2pc}\forall \vec{x}_{i},\; i\in\{1,...,n\}\label{suffcondn}
\end{equation}

Once $W_{n}^{c}$ becomes bounded, condition (\ref{suffcond}) becomes a constraint on the decay of the cluster amplitude at large distances: it must decay fast enough to have that the function we obtain integrating out some coordinates in $|W_{n}^{c}|$ is still bounded. Of course, if $W_{n}^{c}$ has some singularities, after we remove the UV cut off (\ref{suffcondn}) is no longer valid. In this case the diagrams of Section \ref{grtool} are not guaranteed to be finite any longer, but eventual divergences are not due to the thermodynamic limit $L\to\infty$, rather to UV singularities of the cluster amplitudes.

Condition (\ref{suffcond}) has some non trivial consequences in Fourier space:
\begin{equation}
\mathcal{K}_{n}\left(-\sum_{j=2}^{n}\vec{k}_{j},\vec{k}_{2},...,\vec{k}_{n}\right)=\int d^{d(n-1)}x\hspace{2pt}e^{-i\sum_{j=2}^{n}\vec{k}_{j}\vec{x}_{j}}W_{n}^{c}(\vec{0},\vec{x}_{2},..,\vec{x}_{n})
\end{equation} 
The Dirac delta in the definition (\ref{wfourier}) of $\mathcal{K}_{n}$ has been eliminated by fixing one of the coordinates of $W_{n}^{c}$ and integrating on the others. Using (\ref{suffcond}) with $j=1$ we have:
\begin{equation}
\left|\mathcal{K}_{n}\left(-\sum_{j=2}^{n}\vec{k}_{j},\vec{k}_{2},...,\vec{k}_{n}\right)\right|\le\int d^{d(n-1)}x\hspace{2pt}\left|W_{n}^{c}(\vec{0},\vec{x}_{2},..,\vec{x}_{n})\right|<\infty\label{new24}
\end{equation}

Note that (\ref{suffcond}) is a sufficient condition that ensures Property 1, 2, 3 at any order of the expansion, but the latter could not converge: in principle there could be local quantum states whose cluster amplitudes do sot satisfy (\ref{suffcond}).  Nevertheless, in Section \ref{scatmatsec} we provide an explicit computation of cluster amplitudes of ground states in perturbation theory. At any order in perturbation theory, $\mathcal{K}_{n}$ are non singular functions of their momenta. Moreover, in Section \ref{sectionexacteq} non singular functional equations for the cluster amplitudes of ground states are derived: these evidences strongly suggest that (\ref{suffcond}) could be even a necessary condition to have Property 1, 2 and 3.
We note that in the case of states with a fixed number of particles $N$ the properties 1, 2, 3 are violated at first sight. For example, the cluster property is spoiled: if we measure a many-body observable and find that all particles are in a subset of space then obviously further measurements would find no particle anywhere else, independently of the distance, i.e. the measurements are strongly dependent. In such systems with fixed number of particles, the correct definition of the thermodynamic limit is that both the number of particles $N$ and the system size $L$ tend to infinity so that the density $N/L$ remains constant.
With this definition, any observable that measures a finite number of particles would satisfy the cluster property and our analysis is again applicable. 
In Appendix \ref{finitep} we outline how to apply our formalism in systems with a finite number of particles through a familiar example, the prototypical model of the interacting Bose Gas in $d$ dimensions.

\section{Diagrammatic technique}
\label{grtool}

In this section we present a diagrammatic technique, analogous to the Feynman diagrams, which is our main tool for the study of the cluster expansion. Only a basic exposition is presented here, leaving further considerations to Appendix \ref{grtoolperturb}.
In Section \ref{sectech} we use this formalism to study the effects of locality requirements on the cluster expansion of local quantum states and derive the results we outlined in the previous section.

The fundamental quantities we would like to compute are the free energy and the correlators of the fields $\psi$, $\psi^{\dagger}$. These quantities, apart from the particular case of gaussian states that have only $W_{1,2}^{c}\ne 0$ (see Appendix \ref{gaussian} for the explicit calculations), cannot be computed in a closed form and we will have a sum of complicated terms that involve products and integrations which can be expressed conveniently as graphs.
We first construct the graphical rules to evaluate the norm $\braket{\Psi|\Psi}$; next we generalize them to include also correlators.

\begin{figure}
\begin{center}
\begin{minipage}[c]{0.4\textwidth}
\begin{center}
\includegraphics[scale=0.22]{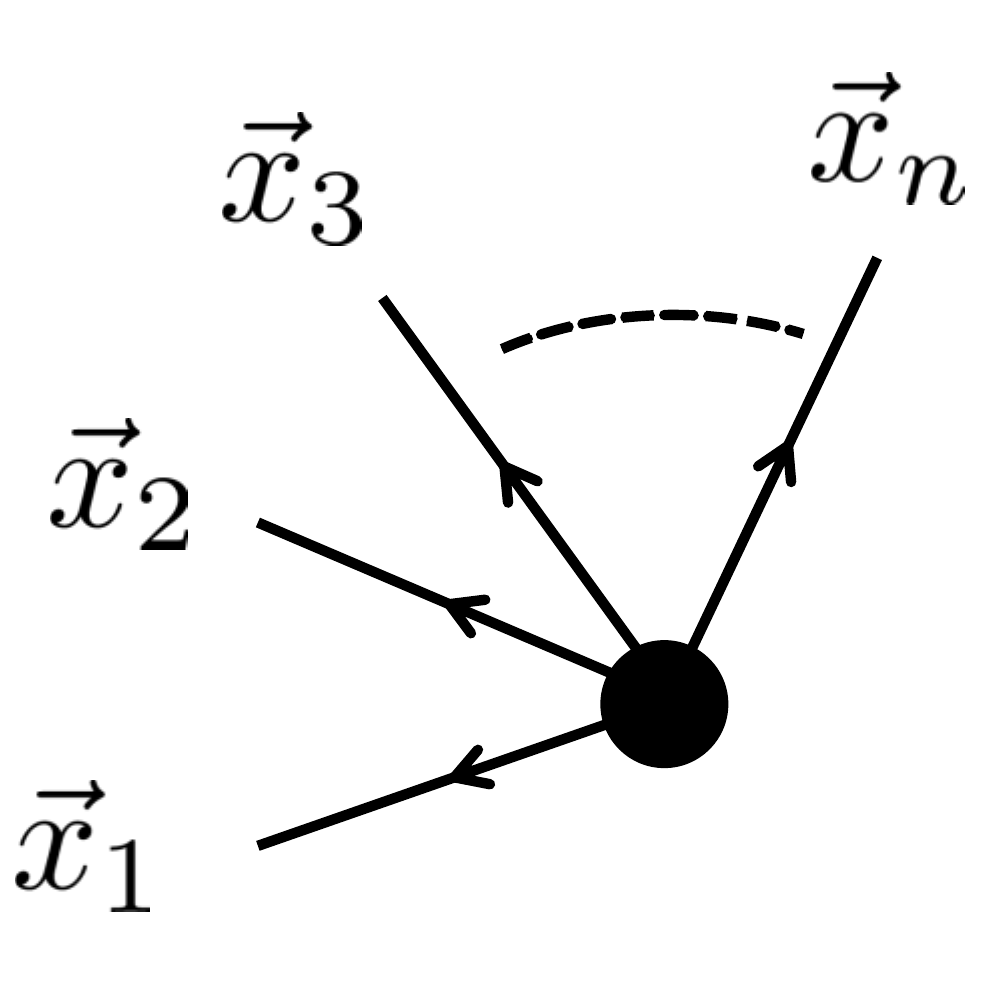}
\end{center}
\caption{ $W_{n}^{c}(\vec{x_{1}},..\vec{x}_{n})$}\label{figgrtool1}
\end{minipage}
\begin{minipage}[c]{0.1\textwidth}
\hspace{1pc}
\end{minipage}
\begin{minipage}[c]{0.4\textwidth}
\begin{center}
\includegraphics[scale=0.22]{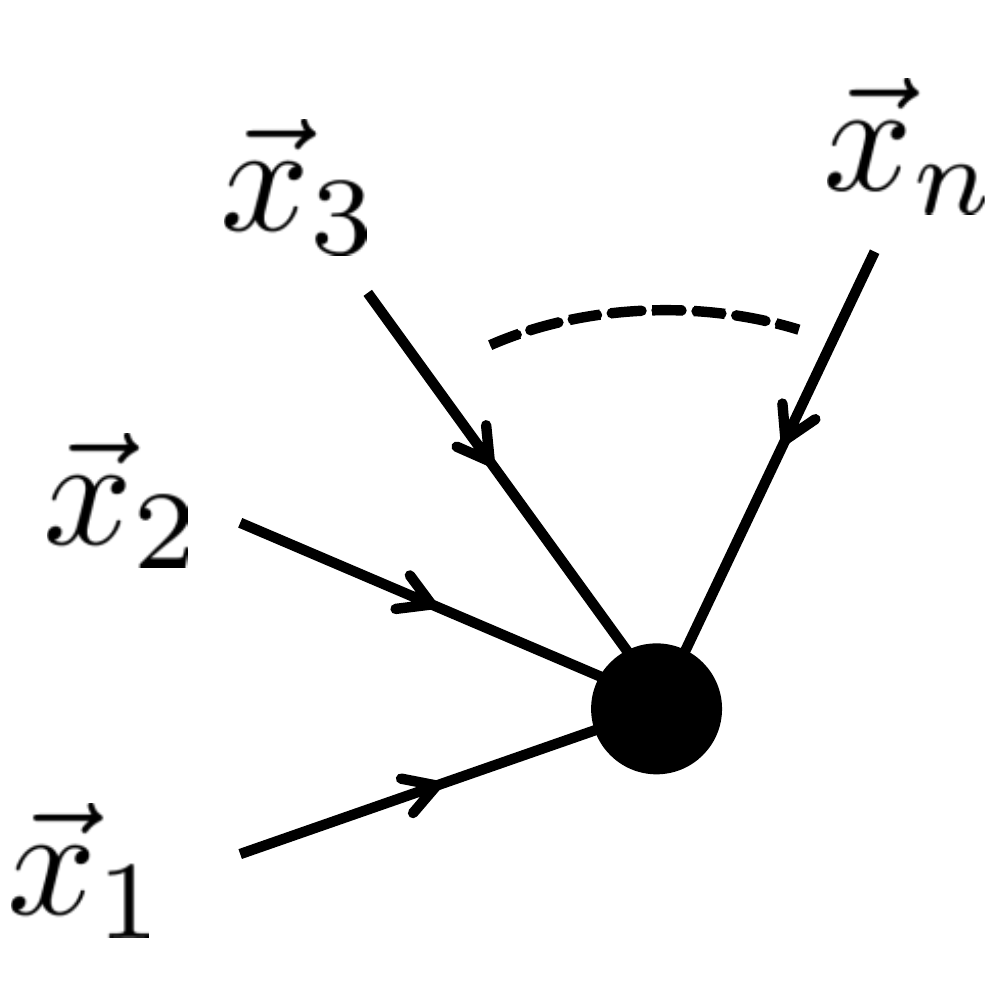}
\end{center}
\caption{$[W_{n}^{c}(\vec{x}_{1},..,\vec{x}_{n})]^{*}$}\label{figgrtool2}
\end{minipage}
\end{center}
\end{figure}

As can be guessed, the `interaction' vertices in our diagrams will represent the $W_{n}^{c}$ functions. When we compute $\braket{\Psi|\Psi}$ we have $W_{n}^{c}$ coming from the ket and $[W_{n}^{c}]^{*}$ coming from the bra, therefore we need two different kinds of vertices. To distinguish them, instead of using different vertex notation, we attach an arrow to the lines and the convention is that a vertex with $n$ outward pointing arrows is associated to $W^{c}_{n}$, while a vertex with $n$ inward pointing arrows is associated to $[W_{n}^{c}]^{*}$ (Figure \ref{figgrtool1}-\ref{figgrtool2}). To the edges of each leg is associated a coordinate.

\begin{figure}
\begin{center}
\begin{minipage}[c]{0.4\textwidth}
\begin{center}
\includegraphics[scale=0.3]{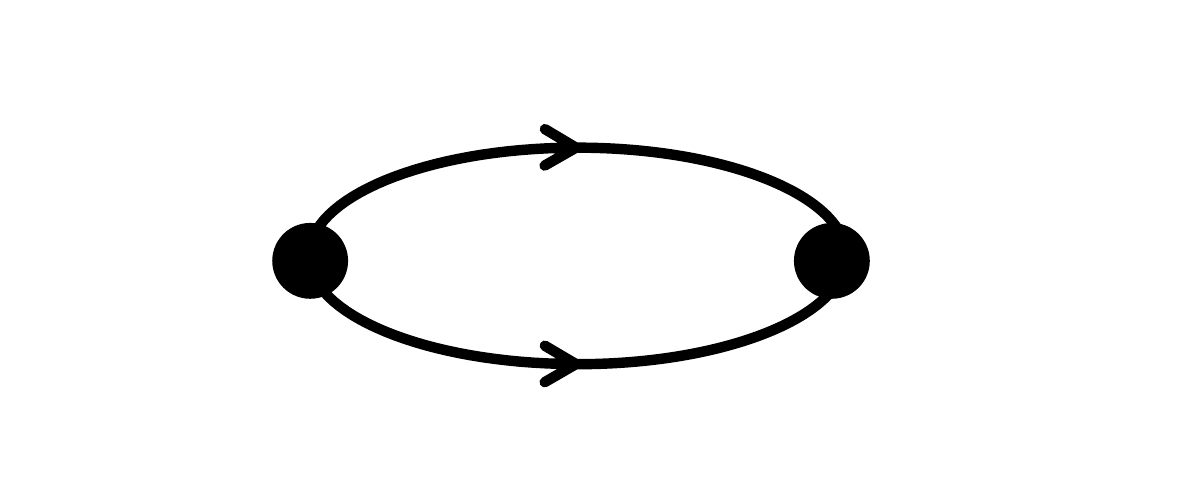}
\end{center}
\end{minipage}
\begin{minipage}[c]{0.4\textwidth}
\begin{center}
\includegraphics[scale=0.3]{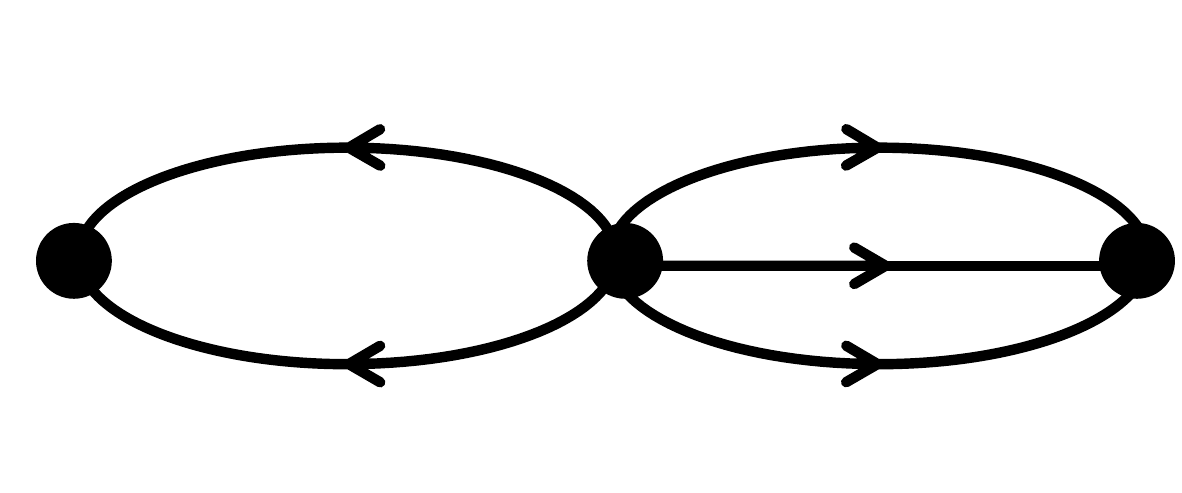}
\end{center}
\end{minipage}
\end{center}
\caption{These two graphs contributes to the free energy, their respective values, strating from the left, are $\frac{1}{2}\int d^{d}x_{1}d^{d}x_{2} |W_{2}^{c}(\vec{x}_{1},\vec{x}_{2})|^{2}$ and $\frac{1}{2}\frac{1}{3!}\int d^{5d}x W_{5}^{c}(\vec{x}_{1},\vec{x}_{2},\vec{x}_{3},\vec{x}_{4},\vec{x}_{5})[W_{2}^{c}(\vec{x}_{1},\vec{x}_{2})]^{*}[W_{3}^{c}(\vec{x}_{3},\vec{x}_{4},\vec{x}_{5})]^{*}$}\label{figgrtool3}
\end{figure}

The contraction of the fields $\psi$ and $\psi^{\dagger}$ in the bra and ket means that we are inserting delta functions on the coordinates; we can indicate this by joining two legs of the vertices. Notice that we can contract only fields that come from the ket with fields that come from the bra, which is the reason why we use arrows: each vertex has either only ingoing arrows or only outgoing arrows; mixed configurations are not allowed. This recipe ensures that we are contracting the right fields and each vertex representing a term of the ket will be joined with a vertex representing a term of the bra, as it should. 
When we want to compute $\mathcal{Z}[\Psi]$, all the $W_{n}^{c}$ and $[W_{n}^{c}]^{*}$ are contracted with each other. Thus the graphs contributing to $\mathcal{Z}[\Psi]$ have no external legs: all the arrows start from some vertex and finish in another. 
To obtain the contribution we are looking for, we should integrate over all the variables of the internal legs. Moreover there are some additional symmetry factors. 
Since we must perform all the possible contractions, in principle we can get several times the same graph from different contractions and we must take care of this degeneracy. The factorial terms of (\ref{constr17}) are there to automatically take care of most of the possibilities, but often they produce an over-counting and each contribution must be divided by the number of permutations of vertices that do not change the graph. Then we must also divide by the number of permutations of internal lines that leave the graph unchanged. We give some examples in Figure \ref{figgrtool3}.

Exactly as happens in the usual Feynman dyagrams theory \cite{Ma}, the symmetry factors are such that $\mathcal{Z}[\Psi]$ can be exponentiated in terms of the connected graphs:
\begin{equation}
\mathcal{Z}[\Psi]=\sum_{\text{graphs}}[\text{graph}]=\exp\left[\sum_{\text{connected graphs}}[\text{connected graph}]\right]\label{zexp}
\end{equation}
The connected graphs are simply the graphs that cannot be split into two or more disconnected pieces.
Taking the logarithm of (\ref{zexp}) we discover that the free energy can be computed as the sum over the connected blocks:
\begin{equation}
\log\mathcal{Z}[\Psi]=\sum_{\text{connected graphs}}[\text{connected graph}]
\end{equation}
Once we understood the graphical rules to compute the free energy, it is easy to generalize them to the correlation functions. Thus suppose we want to evaluate a generic correlator:
\begin{equation}
\braket{\psi^{\dagger}(\vec{x}_{1})..\psi^{\dagger}(\vec{x}_{n})\psi(\vec{y}_{1})..\psi(\vec{y}_{m})}=\frac{\bra{\Psi}\psi^{\dagger}(\vec{x}_{1})..\psi^{\dagger}(\vec{x}_{n})\psi(\vec{y}_{1})..\psi(\vec{y}_{m})\ket{\Psi}}{\braket{\Psi|\Psi}}\label{grtool3}
\end{equation}

\begin{figure}
\begin{center}
\includegraphics[scale=0.3]{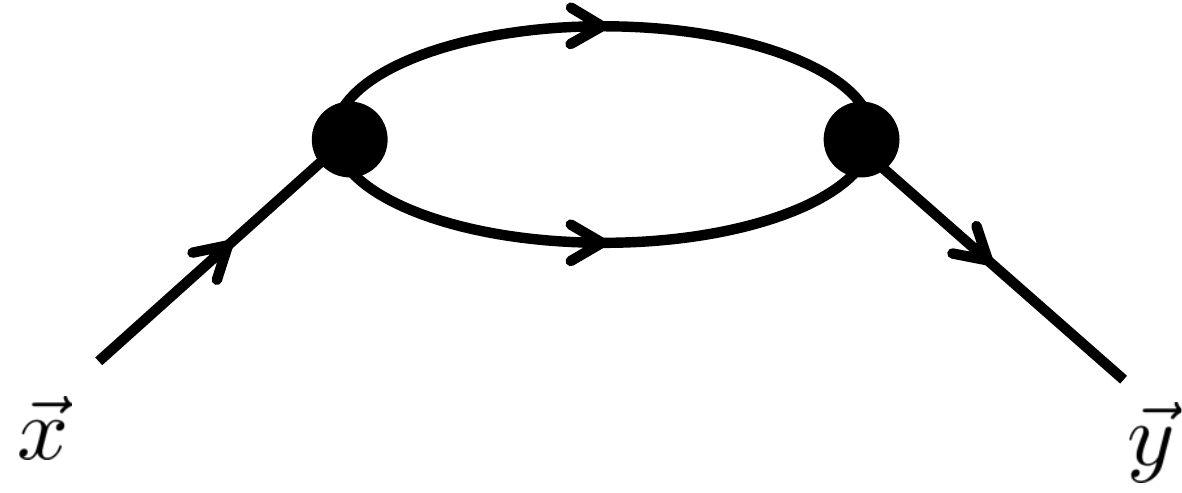}
\caption{Contribution to $\psi^{\dagger}(\vec{x})\psi(\vec{y})$ with value $\frac{1}{2}\int d^{d}x_{1}d^{d}x_{2} [W_{3}^{c}(\vec{x}_{1},\vec{x}_{2},\vec{x})]^{*}W_{3}^{c}(\vec{x}_{1},\vec{x}_{2},\vec{y})$}\label{2correlatorex}
\end{center}
\end{figure}

When we perform the contractions in the numerator, we must take into account also the possible contractions with the fields of the correlators. 
To do so we introduce the possibility of having external legs, that is to say, arrows that do not link two vertices. Also in this case the arrow notation is pretty useful: since a $\psi$ field can be contracted only with an element of the ket, an arrow departing from a vertex will be associated to a contraction with the $\psi$ field. Similarly an arrow that ends on a vertex will be associated to $\psi^{\dagger}$.
The coordinates associated to external legs are fixed by the field in the correlator and we should not integrate over them. The remaining rules are the same as for $\mathcal{Z}[\Psi]$: a simple example is given in Figure \ref{2correlatorex}.

With these rules, it is quite easy to formally compute the numerator of (\ref{grtool3}), moreover, as happens in the usual Feynman diagrams \cite{Ma}, all the blocks in the graphs of the numerator of (\ref{grtool3}) that do not contain external legs can be exponentiated, giving exactly $\braket{\Psi|\Psi}$.
Thus, at the end of the story, correlators such as (\ref{grtool3}) can be evaluated simply summing the contributions of all the graphs in which each block has at least one external leg.

\begin{center}
\begin{framed}
\centering
\hspace{1pt} \ \\
\textbf{Diagrammatic rules to compute the free energy (coordinate space):}
\begin{enumerate}
\item Draw a connected graph, that is to say, it does not contain any disconnected pieces. The vertices are connected with arrows and each vertex has either all arrows outgoing or all ingoing; mixed configurations are not allowed. External legs are not allowed.
\item The contribution of each graph can be computed as follows: assign to a vertex with $n$ outgoing arrows the function $W_{n}^{c}$, to a vertex with $n'$ incoming arrows assign $[W_{n'}^{c}]^{*}$. Take the product of all the vertices functions. Vertexes joined with an arrow have a coordinate in common. Integrate over all the free coordinates.
\item Divide by the symmetry factor of the graph, that is to say, the number of permutations of vertices and legs that leave the graph unchanged.
\item Sum over all possible connected graphs.
\end{enumerate}
\hspace{1pt} \ \\
\textbf{Diagrammatic rules to compute the correlators (coordinate space):}
\begin{enumerate}
\item Draw a graph: disconnected graphs are allowed, but each connected block must have at least one external leg. Assign a coordinate to each external leg. The remaining rules are the same as for the free energy. Notice that, if by chance we want to compute the connected part of a correlator, we must consider only connected graphs.
\item The correlator to be computed is determined by the external legs. Suppose we have $n$ outgoing external legs with coordinates $\vec{y}_{i}$ and $m$ incoming external legs with coordinates $\vec{x}_{i}$, then we are computing $\braket{\prod_{i=1}^{m}\psi^{\dagger}(\vec{x}_{i})\prod_{i=1}^{n}\psi(\vec{y}_{i})}$.
\item Integrate over all the free coordinates, that is to say, the coordinates associated to internal legs.
\item Divide by the symmetry factor obtained exactly as in the free energy case. Notice that the external legs are fixed and they cannot be interchanged.
\item Sum over all the graphs so constructed. This means summing over the distinct topological configurations, but also on the possible permutations of the external momenta that change the graph.
\end{enumerate}
\end{framed}
\end{center}

Equivalent rules hold in the momentum space. They are valid in the thermodynamic limit where we can substitute summations in the momentum space with integrals. 
Translational invariance of $W_{n}^{c}$ implies a conservation law of the momenta at each vertex.

\begin{center}
\begin{framed}
\centering
\hspace{1pt} \ \\
\textbf{Diagrammatic rules to compute the free energy (momentum space):}
\begin{enumerate}
\item Draw the same graphs as in the coordinate space case.
\item The contribution of each graph can be computed as follows: assign to a vertex with $n$ outgoing arrows the function $\mathcal{K}_{n}$, to a vertex with $n'$ incoming arrows assign $[\mathcal{K}_{n}]^{*}$. Take the product of all the vertices functions. Vertexes joined with an arrow have one momentum in common. 
\item Impose the conservation of momenta at each vertex. Integrate over all the remaining free momenta, the integration measure for a momentum $\vec{k}$ is $d^{d}k/(2\pi)^{d}$.
\item Divide by the symmetry factor of the graph, that is to say, the number of permutations of vertices and legs that leave the graph itself. Multiply the obtained value by $L^{d}$: this factor comes from the invariance under translation of the entire graph.
\item Sum over all possible connected graphs.
\end{enumerate}
\hspace{1pt} \ \\
\textbf{Diagrammatic rules to compute the correlators (momentum space):}
\begin{enumerate}
\item Draw a graph exactly as in the coordinate space case. The interpretation of vertices and arrows is the same as in the computation of the free energy in momentum space. Assign a momentum to each external leg. For simplicity consider connected graphs: the value of disconnected graphs will be simply the sum of its connected parts.
\item Suppose we have $n$ outgoing external legs with momenta $\vec{k}_{i}$ and $m$ incoming external legs with momenta $\vec{q}_{i}$, then we are computing $L^{d(n+m-2)/2}\braket{\prod_{i=1}^{m}a^{\dagger}_{\vec{q}_{i}}\prod_{i=1}^{n}a_{\vec{k}_{i}}}$. The $L$ factor is necessary because of the normalization of (\ref{construct12}). Impose the global conservation law of momenta:
 $\sum_{i=1}^{m}\vec{q}_{i}=\sum_{i=1}^{n}\vec{k}_{i}$, otherwise the correlator is zero.
\item Integrate over all the free momenta, that is to say, the momenta associated to internal legs. A momentum $\vec{k}$ has as integration measure $d^{d}k/(2\pi)^{d}$.
\item Divide by the symmetry factor obtained exactly as in the free energy case. Notice that the external legs are fixed and they cannot be exchanged between them. 
\item  Sum over all the graphs so constructed. This means summing over the distinct topological configurations, but also on the possible ways of permutations of the external momenta that change the graph.
\end{enumerate}
\end{framed}
\end{center}

\begin{figure}
\begin{center}
\begin{minipage}[c]{0.4\textwidth}
\begin{center}
\includegraphics[scale=0.3]{Figure3.pdf}
\end{center}
\end{minipage}
\begin{minipage}[c]{0.4\textwidth}
\begin{center}
\includegraphics[scale=0.3]{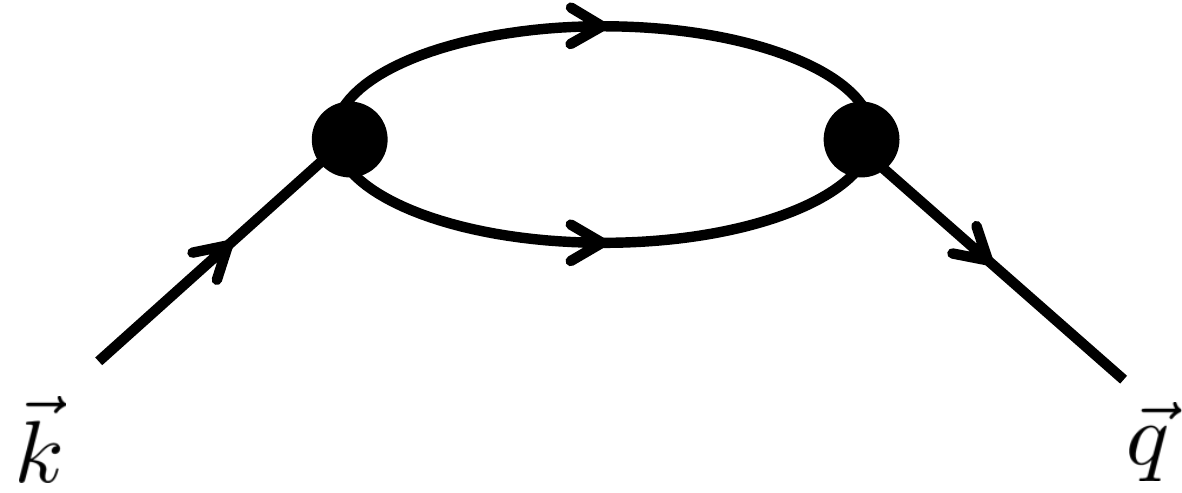}
\end{center}
\end{minipage}
\end{center}
\caption{Examples of computations of graphs in the momentum space, the first on the left is a graph of the free energy with value $\frac{L^{d}}{2}\int \frac{d^{d}k}{(2\pi)^{d}} |K_{2}(\vec{k},-\vec{k})|^{2}$, the second graph enters in the computation of $\braket{a^{\dagger}_{\vec{k}}a_{-\vec{q}}}$ and has value $\int \frac{d^{d}p}{2(2\pi)^{d}} [K_{3}(\vec{k},\vec{p},-\vec{p})]^{*}K_{3}(\vec{q},\vec{p},-\vec{p})$}\label{momenta1}
\end{figure}

A couple of examples of the computations in momentum space are given in Figure \ref{momenta1}.
The rules in coordinate space hold without any particular hypotheses on $W_{n}^{c}$, in particular we do not need to suppose (\ref{suffcond}). We will use them in Section \ref{sectech} to explore the consequences of Properties 1, 2 and 3 on the analytic properties of $W_{n}^{c}$. Instead the rules in the momentum space can be more suitable to some explicit calculations, as we will see in Section \ref{perturb} and Section \ref{dynamic}.
Notice that we can also generalize these rules to compute overlaps between cluster expansions.

\begin{framed}
\hspace{1pt} \
\begin{center}\textbf{Diagrammatic rules to compute overlaps:} \end{center}
\ 
Consider $\ket{\Psi}$ and $\ket{\Psi'}$ two cluster expansions and we want to compute $\braket{\Psi|\Psi'}$, or even insert operators between these states. The rules we explained in the case of a single state keep always distinct the vertices of the bra and the vertices of the ket by means of the direction of the arrows. If we want to differentiate the two we must simply associate to the vertices with outgoing arrows the cluster amplitudes of $\ket{\Psi'}$ and to the vertices with ingoing arrows the functions describing $\bra{\Psi}$, once we take the complex conjugate, of course. All the remaining rules are exactly the same.
\ 
\hspace{1pt} \ \\
\end{framed}

\subsection{Thermodynamic properties of the graphical expansion}
\label{sectech}

In this section we will analyze the relations between the Properties 1, 2, 3 and the features of the cluster amplitudes. In principle we would like to find necessary and sufficient conditions on the cluster amplitudes in order to have the desired thermodynamic properties, but we are able to properly address only one implication and provide a \emph{sufficient condition} for the validity of Properties 1, 2, 3 at each order of the expansion.
Assuming (\ref{suffcond}) i.e.
\begin{equation}
\int d^{d(n-j)}y\hspace{5pt} \left|W_{n}^{c}(\vec{x}_{1},..,\vec{x}_{j},\vec{y}_{1},..,\vec{y}_{n-j})\right| <M_{n}<\infty\hspace{2pc}\forall j\ge1,\hspace{5pt}\forall \vec{x}_{i}, \hspace{3pt}i\in\{1,...,j\}\label{suffcond1}
\end{equation}
and that $W_{n}^{c}$ has no explicit dependence on the system size is sufficient in order to satisfy Properties 1, 2, 3 at any order of the expansion.
It could be questioned whether or not the graphical expansion converges and the thermodynamic properties of each graph imply Properties 1, 2 and 3 on the true correlators and free energy. If the expansion does not converge any longer, it could be that Properties 1, 2 and 3 are still valid, but not respected individually by each graph. Indeed, as we argue in Appendix \ref{criticality}, in critical systems the expansion becomes divergent, even though Properties 1, 2, 3 are expected to be still valid on the correlators and free energy; nevertheless we argue that (\ref{suffcond1}) should be still valid.

\subsubsection{Property 1}
\label{secpropc2}

\begin{figure}
\begin{center}
\includegraphics[scale=0.3]{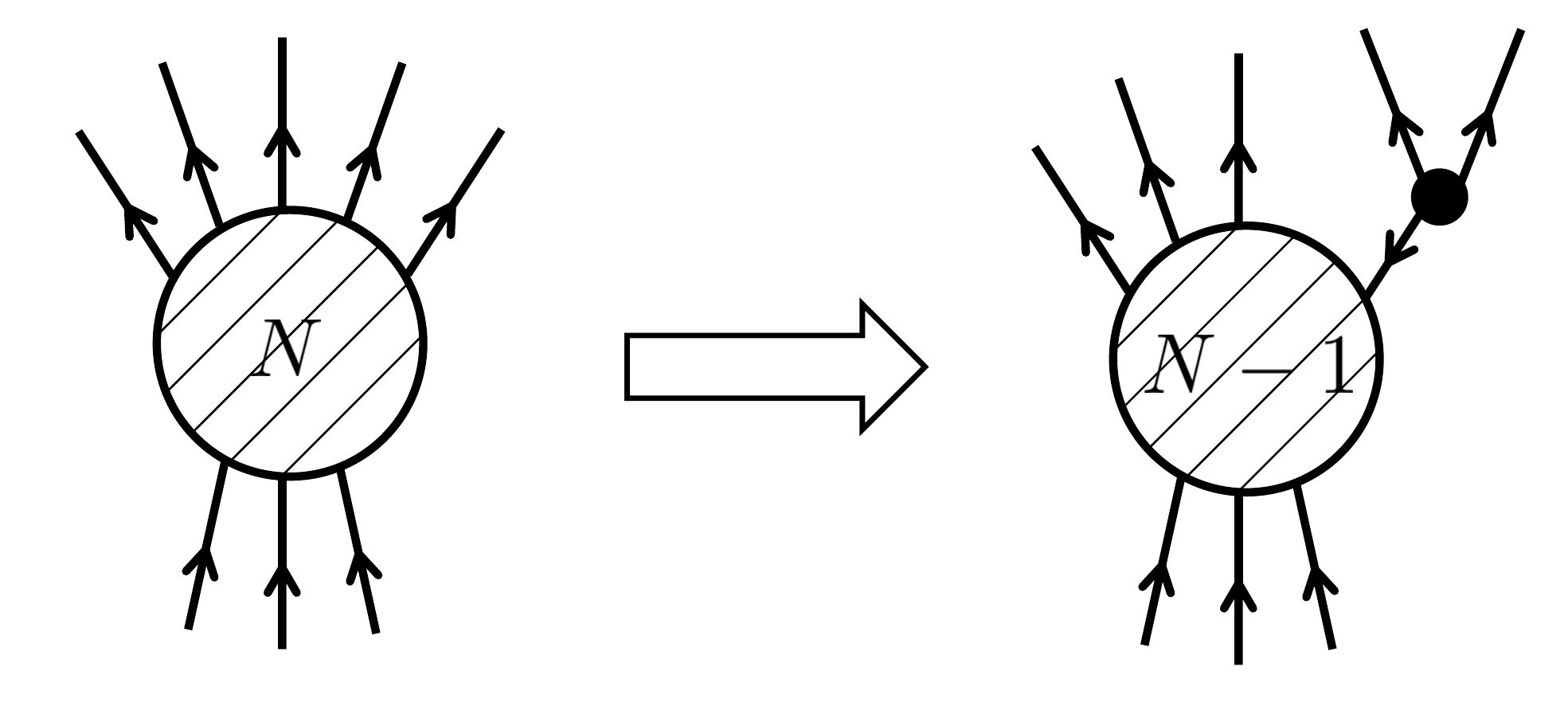}
\end{center}
\caption{Each graph with $N$ vertices can be thought as a subgraph with $N-1$ vertices eventually attached to a single vertex that has always at least one external leg. In this figure is drawn a typical example, the number of legs of the figure has not a precise meaning and want to represent a typical situation, also the nature of the selected vertex can change and have only ingoing arrows instead of outgoing. It could be also that the vertex and the subgraphs with $N-1$ are not really attached and form two disconnected pieces, what really matters is that we can always select a vertex with at least one external leg.}\label{figsuffcond}
\end{figure}

We now show that (\ref{suffcond1}) guarantees that any graph that appears in the expansion of correlation functions is well defined in the infinite size limit. 
Note that a generic graph for the correlators is nothing but a complicated convolution of the $W_{n}^{c}$ functions, therefore we need to study such an operation.
The proof is by induction on the number of vertices that contribute to a graph: in each graph with $N$ vertices we can identify a subgraph with $N-1$ vertices linked to a single $W_{n}^{c}$ vertex (or its complex conjugate) with some external legs attached to it (Figure \ref{figsuffcond}).

Then, some of the external legs should depart from $W_{n}^{c}$. Suppose $\{\vec{x}_{1},..,\vec{x}_{j}\}$ are associated with external legs of $W_{n}^{c}$, then the remaining $\vec{x}_{i}$ coordinates will be associated to external legs departing from the subgraph with $N-1$ vertices. The internal legs connecting the vertex with the subgraph will be denoted with coordinates $\vec{y}_{i}$: in Figure \ref{figsuffcond} we drew the case of only one leg connecting them, but there could be more than one or even none. We denote with $G_{N-1}(\vec{y}_{1},..,\vec{y}_{n-j},\vec{x}_{j+1},..,\vec{x}_{l})$ the value of the subgraph with $N-1$ vertices.  The function $G_{N}$ is nothing but a convolution between $W_{n}^{c}$ and $G_{N-1}$:
\begin{equation}
G_{N}(\vec{x}_{1},..,\vec{x}_{l})=\mathcal{S}\int d^{d(n-j)}y\hspace{3pt}W_{n}^{c}(\vec{x}_{1},..,\vec{x}_{j},\vec{y}_{1},..,\vec{y}_{n-j})G_{N-1}(\vec{y}_{1},..,\vec{y}_{n-j},\vec{x}_{j+1},..,\vec{x}_{l})\label{suffcond35}
\end{equation}

Above, we inserted an unimportant symmetry factor $\mathcal{S}$, that depends on the details of the connection between $W_{n}^{c}$ and the subgraph.
Now, we show that if $G_{N-1}$ is uniformly bounded, then provided (\ref{suffcond1}), $G_{N}$ exists and it is also uniformly bounded.  Note that the graphs with only one vertex are simply the $W_{n}^{c}$ functions that are uniformly bounded because of (\ref{suffcond1}). Then, by induction we can say that each graph of the correlators is well-defined in the infinite size limit.
Therefore, suppose $\left|G_{N-1}\right|<M$ with $M$ an unknown, but finite number, then:
\begin{equation}
\left|G_{N}(\vec{x}_{1},..,\vec{x}_{l})\right|<\mathcal{S}M\int d^{d(n-j)}y\hspace{3pt}\left|W_{n}^{c}(\vec{x}_{1},..,\vec{x}_{j},\vec{y}_{1},..,\vec{y}_{n-j})\right|
\end{equation}

Because of (\ref{suffcond1}) the integral in the above expression is finite and uniformly bounded for any of the remaining $\vec{x}$ coordinates. Then $G_{N}$ exists, because the integral in (\ref{suffcond35}) is absolutely convergent and moreover $G_{N}$ uniformly bounded.
This guarantees that each graph in the expansion of the correlators is finite in the $L\to\infty$ limit.

\subsubsection{Property 2}
\label{robust2}

We now show that (\ref{suffcond1}) is enough to guarantee Property 2 (i.e. cluster property) at any order of the expansion, therefore we will show that each connected graph decays to zero whenever the distance of two of its coordinates is sent to infinity.
Actually, we will prove a stronger statement: any connected graph satisfies a similar condition to (\ref{suffcond1}). Consider $G_{N}(\vec{x}_{1},...,\vec{x}_{l})$ a generic connected graph formed by $N$ vertices, then
\begin{equation}
\int_{\vec{x}_{a}\in A} |G_{N}(\vec{x}_{1},...,\vec{x}_{l})|<\text{const.}<\infty, \hspace{3pc}\forall \vec{x}_{a}\in B, \hspace{1pc}A\cup B=\{\vec{x}_{1},..,\vec{x}_{l}\}\label{seccorr30}
\end{equation}
where the integral is meant to be over the coordinates in the subset $A$. The convergence of the integral holds for any bipartition of the set of coordinates such that $B$ has at least one element, therefore there is always at least one coordinate that is not integrated. Of course the convergence of (\ref{seccorr30}) implies that the value of the connected graph decays to zero whenever the distance of two coordinates is sent to infinity, apart from a possible zero-measure subset. The proof of (\ref{seccorr30}) is by induction: note that any connected graph can always be seen as two smaller connected graphs joined together (apart the graphs with only one vertex formed by only $W_{n}^{c}$ or the complex conjugate vertex).
Therefore, assume $G_{N}$ can be divided in two connected graphs $G_{\tilde{N}}$ and $G_{N-\tilde{N}}$ with $\tilde{N}< N$; by re-labeling the coordinates we can assume that the first $j$ coordinates $\vec{x}_{1},..,\vec{x}_{j}$ belong to $G_{\tilde{N}}$. The value of $G_{N}$ can be written similarly to (\ref{suffcond35})
\begin{equation}
G_{N}(\vec{x}_{1},..,\vec{x}_{l})=\mathcal{S}\int d^{d(n-j)}y\hspace{3pt}G_{\tilde{N}}(\vec{x}_{1},..,\vec{x}_{j},\vec{y}_{1},..,\vec{y}_{n-j})G_{N-\tilde{N}}(\vec{y}_{1},..,\vec{y}_{n-j},\vec{x}_{j+1},..,\vec{x}_{l})
\end{equation}
where $\mathcal{S}$ is an unimportant symmetry factor, the coordinates $\vec{y}$ are associated to the internal lines that connect the two subgraphs: since the graph $G_{N}$ is connected, the two subgraphs must be linked through at least one arrow.
We can now write the inequality:
\begin{equation}
\int_{\vec{x}_{a}\in A} |G_{N}(\vec{x}_{1},...,\vec{x}_{l})|\le\mathcal{S}\int_{\vec{x}_{a}\in A}\int d^{d(n-j)}y\hspace{3pt}|G_{\tilde{N}}(\vec{x}_{1},..,\vec{x}_{j},\vec{y}_{1},..,\vec{y}_{n-j})||G_{N-\tilde{N}}(\vec{y}_{1},..,\vec{y}_{n-j},\vec{x}_{j+1},..,\vec{x}_{l})|\\ \label{corrsec33}
\end{equation}
Since the integration domain $A$ is never the whole set of coordinates $\vec{x}$, at least one among the two graphs $G_{\tilde{N}}$ and $G_{N-\tilde{N}}$ possesses a $\vec{x}$ coordinate that is not in $A$. Without loss of generality we can suppose that at least one of the $\vec{x}$ coordinates in $G_{\tilde{N}}$ is not an integration variable, then consider $G_{N-\tilde{N}}$. Because of the inductive step, $G_{N-\tilde{N}}$ satisfies (\ref{seccorr30}):
\begin{equation}
\int_{\vec{x}_{a}\in A, a\ge j}|G_{N-\tilde{N}}(\vec{y}_{1},..,\vec{y}_{n-j},\vec{x}_{j+1},..,\vec{x}_{l})|<M<\infty , \hspace{2pc}\forall \;\vec{y},\forall\; \vec{x}_{a}\in B\;\text{s.t.}\; a\ge j
\end{equation}
for some constant $M$. 
Note that in the above we are considering only the integration over the $\vec{x}$ coordinates that are both in $A$ and variables of $G_{N-\tilde{N}}$, instead the $\vec{y}$ coordinates are not integrated. Note that we can use (\ref{seccorr30}) because, since $G_{N}$ is connected, there is at least one internal leg that connects $G_{\tilde{N}}$ and $G_{N-\tilde{N}}$, therefore there is at least one coordinate $\vec{y}$. Using this inequality in (\ref{corrsec33}) we have:
\begin{equation}
\int_{\vec{x}_{a}\in A} |G_{N}(\vec{x}_{1},...,\vec{x}_{l})|\le M\mathcal{S}\int_{\vec{x}_{a}\in A, a\le j}\int d^{d(n-j)}y\hspace{3pt}|G_{\tilde{N}}(\vec{x}_{1},..,\vec{x}_{j},\vec{y}_{1},..,\vec{y}_{n-j})|
\end{equation}

Now we can use that $G_{\tilde{N}}$ satisfies (\ref{seccorr30}), since in the above integration $G_{\tilde{N}}$ possesses at least one $\vec{x}$ coordinate that is not an integration variable.
Using (\ref{seccorr30}) on $G_{\tilde{N}}$  it follows that also $G_{N}$ fulfills (\ref{seccorr30}). Therefore we have shown the inductive step and this concludes the proof.
Note that we can also say something about the Fourier transform of $G_{N}$. At the end of Section \ref{LQS} we showed that the Fourier transform of the cluster amplitudes that satisfy (\ref{suffcond1}) are non-singular, analogously the condition (\ref{seccorr30}) implies that the Fourier transform of $G_{N}$ has no singularity (after we factorize the overall $\delta$ function caused by the translation invariance). This simply means that if (\ref{suffcond1}) is valid then none of the graphs of the correlators computed in the momentum space has any singularity.

\subsubsection{Property 3}
\label{A3}
\begin{figure}
\begin{center}
\includegraphics[scale=0.25]{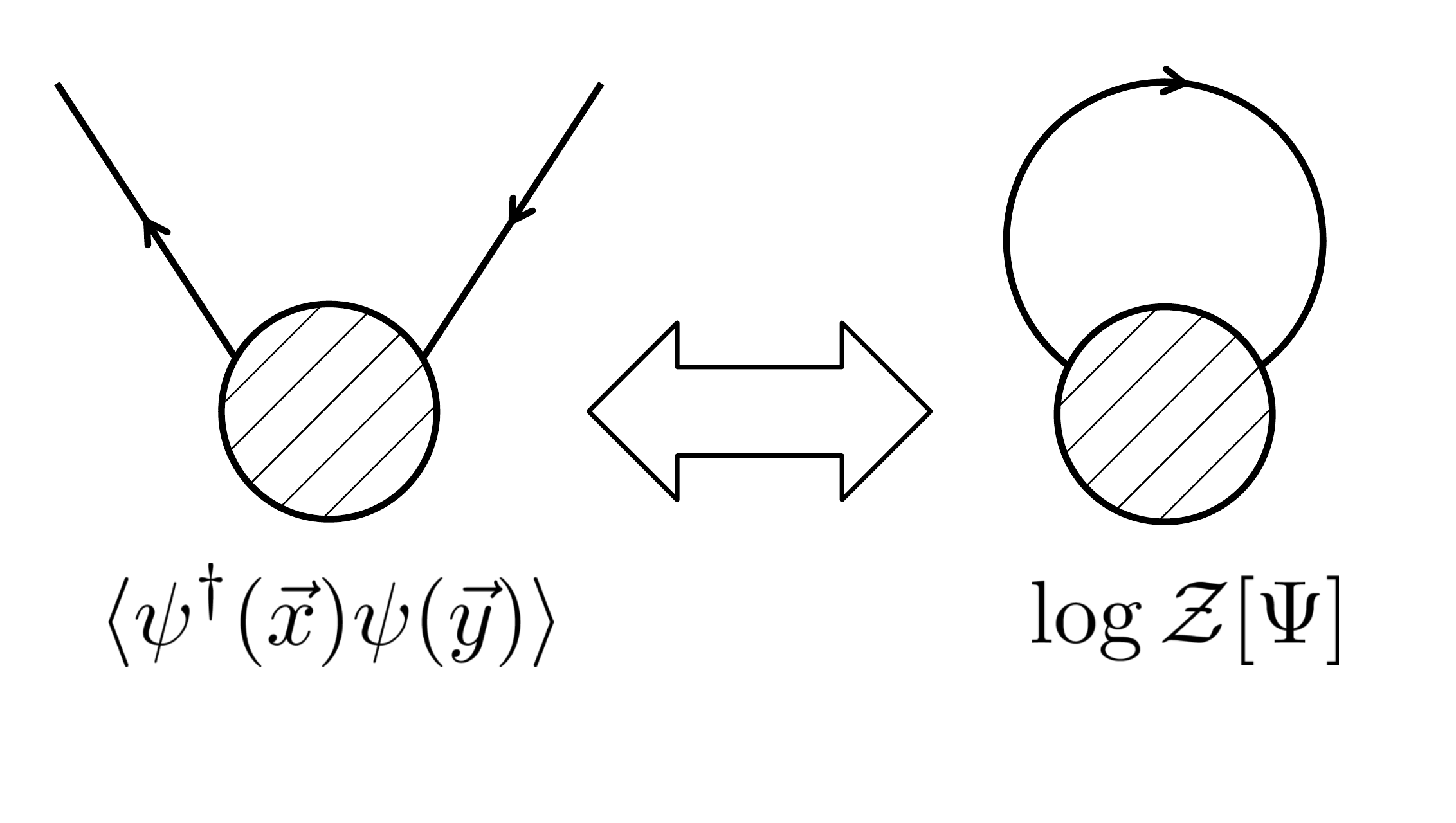}
\end{center}
\caption{The graph on the left is a contribution of the two point correlator, call $G(\vec{x},\vec{y})$ its general contribution. Gluing together the external legs we can construct a graph of the expansion of the free energy: the contribution of the latter is simply $\int d^{d}x\hspace{3pt}G(\vec{x},\vec{x})$.}\label{figpropc2}
\end{figure}

The proof of this implication is simple and relies on our graphical expansion.
We start with an interesting mapping between the graphs that appear in the free energy and the graphs appearing in the two point correlator $\braket{\psi^{\dagger}(\vec{x})\psi(\vec{y})}$. Pictorially it is very simple: consider a graph of the two point correlator $\braket{\psi^{\dagger}(\vec{x})\psi(\vec{y})}$. Its external legs must be an outgoing arrow and an ingoing one. If we join these two legs, we get a graph for the free energy. We can also proceed in the reverse direction taking a graph of $\log\mathcal{Z}[\Psi]$ and cutting an internal line to get a graph of $\braket{\psi^{\dagger}(\vec{x})\psi(\vec{y})}$. This is shown in Figure \ref{figpropc2}.  
It should not be forgotten that this mapping does not properly take in account the symmetry factors that can change during the gluing procedure, but since we are merely interested in the system size dependence of each graph it is sufficient for our purpose.
This correspondence is enough to show that any graph of $\log \mathcal{Z}[\Psi]$ is extensive. Because of the invariance under translation, each graph of the two point correlator can depend only on the difference between the two points, thus the integration in Figure \ref{figpropc2} becomes simply $L^{d}G(0,0)$. Now, using the fact that each graph of the correlators does not depend explicitly on the system size, we have that $G(0,0)$ is independent of $L$, thus each contribution to the free energy is extensive. 
It must be said that $G(0,0)$ may be a divergent quantity when the UV cut off is removed, but this does not change the extensivity since the latter refers only to the scaling with the system size, i.e. the infrared cut off.

\section{Explicit calculation of the cluster amplitudes of ground states}
\label{perturb}

In the previous sections we have studied the general properties that a cluster expansion of a local state should respect. Now we shall link our abstract discussion to some explicit models. Even if the general construction can be applied also to condensed matter systems with a finite number of particles (Appendix \ref{finitep}), from now on we will focus mainly on relativistic theories, in which we avoid the subtlety of dealing with a finite number of particles. In particular we will consider a real bosonic field $\phi$ whose dynamics is governed by a relativistic Lagrangian with local self-interactions.
\begin{equation}
\mathcal{L}=\int d^{d}x\left[\frac{1}{2}(\partial_{\mu}\phi\partial^{\mu}\phi-m^{2}\phi^{2})-\lambda\sum_{n}\frac{\beta_{n}}{n!}\phi^{n}\right]\label{gzlandau}
\end{equation}

Above, $\beta_{n}$ are the coupling constants of the theory, $\lambda$ is a parameter to tune the strength of the interaction, that is inserted to make eventual perturbative calculations more straightforward.
As we stated in the introduction and further investigated in Section \ref{thprop}, the ground state (or ground states, if many of them are present) of such a theory is supposed to satisfy some locality requirements encoded in Properties 1, 2 and 3. 
Even if most of the content of this section can be extended to the case of many degenerated ground states, in the following we will imagine $\beta_{n}$ to be tuned so that there is only one ground state.
Our physical properties have been expressed in terms of some abstract fields $\psi$ $\psi^{\dagger}$: the first step is to correctly identify these fields. In the canonical quantization scheme, the field $\phi$ is divided in terms of the modes that diagonalize the free Hamiltonian. In finite volume this decomposition is expressed by:
\begin{equation}
\phi(\vec{x})=\frac{1}{L^{d/2}}\sum_{\vec{k}}\frac{e^{i\vec{k}\vec{x}}}{\sqrt{2 E(\vec{k})}}\left(a^{\dagger}_{\vec{k}}+a_{-\vec{k}}\right)\hspace{2pc} E(\vec{k})=\sqrt{\vec{k}^{2}+m^2}\hspace{3pc} \label{newperturb39}
\end{equation}

Where $a_{\vec{k}}$ and its conjugate follow standard commutation rules. The natural thing to do is to \emph{define} the $\psi$ field from the $a_{\vec{k}}$ field. Therefore we invert relation (\ref{construct12}) and use it as the definition of $\psi$:
\begin{equation}
\psi(\vec{x})\equiv\frac{1}{L^{d/2}}\sum_{\vec{k}}e^{-i\vec{k}\vec{x}}a_{\vec{k}}\label{newperturb40}
\end{equation}

Proceeding in this way, the cluster expansion in the momentum space (\ref{expk}) (\ref{momentaexp}), will be simply an expansion in the Fock space of the theory constructed acting with the creation operators of the free theory on the bare vacumm.
The Hamiltonian derived from (\ref{gzlandau}) is:
\begin{equation}
H=\sum_{\vec{k}}E(\vec{k})a^{\dagger}_{\vec{k}}a_{\vec{k}}+\lambda\sum_{n}\frac{\beta_{n}}{n!L^{d(n-2)/2}}\sum_{\vec{k}_{1}+..+\vec{k}_{n}=0}\prod_{q=1}^{n}\left[\frac{1}{\sqrt{2E(\vec{k}_{q})}}\left(a^{\dagger}_{\vec{k}_{q}}+a_{-\vec{k}_{q}}\right)\right]\label{perturbh}
\end{equation}

Above, we ignored the zero point energy by normal ordering the free part.
A cautionary remark must be made about Properties 1 and 2: the locality requirements are written in terms of the fields $\psi$ $\psi^{\dagger}$, but the physical fields that should respect them are $\phi$ and its conjugated momentum. It turns out that, at least in the massive case $m\ne 0$, it is pretty simple to show that if $\psi$ satisfies Properties 1 and 2, then also $\phi$ and its conjugated momentum satisfy them. Also the inverse implication is true.
As a matter of fact, combining (\ref{newperturb39}) and (\ref{newperturb40}) we get:
\begin{equation}
\phi(\vec{x})=\int d^{d}y\hspace{2pt} D(\vec{x}-\vec{y})\left(\psi^{\dagger}(\vec{y})+\psi(\vec{y})\right); \hspace{3pc}
 D(\vec{x}-\vec{y})=\frac{1}{L^{d}}\sum_{\vec{k}}\frac{1}{\sqrt{2 E(\vec{k})}}e^{i\vec{k}(\vec{x}-\vec{y})}\label{kernel}
\end{equation}

The kernel is exponentially decaying $D(\vec{x}-\vec{y})\sim e^{-m|\vec{x}-\vec{y}|}$ from which it is straightforward to verify our claim.
From our reasoning, ground states of these theories should have a cluster expansion as (\ref{expk}), with regular $\mathcal{K}_{n}$ functions. 
In Section \ref{scatmatsec} we verify that the ground state possesses indeed a proper cluster expansion through standard perturbative calculations and describe the suitable method to calculate $\mathcal{K}_{n}$ in a perturbative way. At least in the massive case, such a computation confirms the absence of singularities in $\mathcal{K}_{n}$ at each order in perturbation theory. In the massless case ($m=0$) the naive perturbation theory does not work and we attempt a non-perturbative check of the absence of singularities in Appendix \ref{failure}, confirming that at least $\mathcal{K}_{2}$ has no singularities. In Section \ref{sectionexacteq} a set of exact functional equations for the cluster amplitudes $\mathcal{K}_{n}$ is presented: they provide a consistency check about the analytic properties of $\mathcal{K}_{n}$, reproduce the results of standard perturbation theory of Section \ref{scatmatsec} and give access to other approximation schemes or even to some exact predictions, as in Appendix \ref{failure}.

\subsection{Perturbative computation with Feynman graphs}
\label{scatmatsec}

This section is devoted to the construction of a perturbative method to compute the ground state of an Hamiltonian such as (\ref{perturbh}). In the following we will suppose that the  ground state is unique and `perturbatively close' to $\ket{0}$, that is to say the ground state in the limit $\lambda\rightarrow 0$ reduces to $\ket{0}$. These assumptions give constraints on the allowed form of the interaction, but it is still a rather general theory.
The Hamiltonian of the system can be split in a free and an interacting part:
\begin{equation}
H=H_{0}+\lambda V;\hspace{3pc}H_{0}=\sum_{\vec{k}}E(\vec{k})a^{\dagger}_{\vec{k}}a_{\vec{k}}\label{hamilt}
\end{equation}

Given the ground state $\ket{G}$, we can project it on a state with $n$ particles at definite position $\vec{x}_{i}$ and find the wavefunction $W_{n}$:
\begin{equation}
W_{n}(\vec{x}_{1},..,\vec{x}_{n})=\bra{0}\prod_{i=1}^{n}\psi(\vec{x}_{i})\ket{G}
\end{equation}

Now we should extract from both sides their connected part, which is the cluster amplitude that enters in (\ref{constr17}).
Going to Fourier transform, we find:
\begin{equation}
\delta_{\sum_{i=1}^{n}\vec{k}_{i},0}\hspace{3pt}\mathcal{K}_{n}(\vec{k}_{1},..,\vec{k}_{n})=L^{d(n-2)/2}\left(\bra{0}\prod_{i=1}^{n}a_{\vec{k}_{i}}\ket{G}\right)_{\text{connected}}\label{knconncorr}
\end{equation}

The $L$ factors come from (\ref{expk}) and are such that when $L\rightarrow\infty$ the right hand side becomes $L$ independent.
The Kronecker delta is due to the translation invariance of the state. Instead, the physical requirement of Property 1, 2 and 3 are completely encoded in the analytic properties of $\mathcal{K}_{n}$. In particular there should be no extra $\delta$ factors, and no other singularities are expected: we will check this claim in the next computations.
For simpler notation, it is better to compute the complex conjugate of (\ref{knconncorr}):
\begin{equation}
\delta_{\sum_{i=1}^{n}\vec{k}_{i}}\mathcal{K}^{*}_{n}(\vec{k}_{1},..,\vec{k}_{n})=L^{d(n-2)/2}\left(\bra{G}\prod_{i=1}^{n}a^{\dagger}_{\vec{k}_{i}}\ket{0}\right)_{\text{connected}}\label{normalization34}
\end{equation}

The next steps will bring the right side of the above expression in a form more suitable to perturbation theory. 
To compute the right term side we can use the same trick as in (\ref{thproj}) to obtain the ground state. As a matter of fact, applying a thermal ensemble on the bare vacuum and then taking the zero temperature limit we are able to extract the ground state.
\begin{equation}
e^{-\beta H}\ket{0}\xrightarrow{\beta\to\infty}\frac{e^{-\beta E_{G}}}{\braket{G|G}}\ket{G}
\end{equation}

Above, we explicitly used our normalization $\braket{0|G}=1$. We can use this trick to compute the right side of  (\ref{normalization34}).
\begin{equation}
\bra{G}\prod_{i=1}^{n}a^{\dagger}_{\vec{k_{i}}}\ket{0}=\lim_{\beta\rightarrow \infty}e^{\beta E_{G}}\mathcal{N}_{G}^{-1}\bra{0}e^{-\beta H}\prod_{i=1}^{n}a^{\dagger}_{\vec{k_{i}}}\ket{0}\label{vaccumextr}
\end{equation}

Above, $\mathcal{N}_{G}=\braket{G|G}$. Instead of using the Hamiltonian $H$, it is more suitable to switch to the interaction picture defined as:
\begin{equation}
e^{\beta H_{0}}e^{-\beta H}=\mathcal{T} \exp\left[-\lambda\int_{0}^{\beta}d\tau V_{I}(\tau)\right]\label{intpic}
\end{equation}
where $\mathcal{T}\exp$ is the $\tau$ ordered exponential and $V_{I}$ is the potential in the interacting picture:
\begin{equation}
V_{I}(\tau)=e^{\tau H_{0}}V e^{-\tau H_{0}}\label{vI}
\end{equation}

Using the fact that $\bra{0}H_{0}=0$ we arrive at:
\begin{equation}
\bra{G}\prod_{i=1}^{n}a^{\dagger}_{\vec{k_{i}}}\ket{0}=\lim_{\beta\rightarrow \infty}e^{\beta E_{G}}\mathcal{N}_{G}^{-1}\bra{0}\mathcal{T} \exp\left[-\lambda\int_{0}^{\beta}d\tau V_{I}(\tau)\right]\prod_{i=1}^{n}a^{\dagger}_{\vec{k_{i}}}\ket{0}\label{scatmat36}
\end{equation}

To proceed further, it is useful to define the field $\phi$ in the interaction picture and in Fourier space:
\begin{equation}
\phi_{I}(\vec{k},\tau)\equiv \frac{1}{\sqrt{2 E(\vec{k})}}\left(a^{\dagger}_{\vec{k}}\hspace{2pt}e^{\tau E(\vec{k})}+a_{-\vec{k}}\hspace{2pt}e^{-\tau E(\vec{k})}\right)\label{phiI}
\end{equation}
so that we can write:
\begin{equation}
V_{I}(\tau)=\lambda\sum_{n}\frac{\beta_{n}}{n!L^{dn/2-d}}\sum_{\vec{k}_{1}+..+\vec{k}_{n}=0}\prod_{q=1}^{n}\phi_{I}(\vec{k}_{q},\tau)
\end{equation}

We can express (\ref{scatmat36}) in terms of $\phi_{I}$:
\begin{equation}
\bra{G}\prod_{i=1}^{n}\left(\frac{a^{\dagger}_{\vec{k_{i}}}}{\sqrt{E(\vec{k}_{i})}}\right)\ket{0}=\lim_{\beta\rightarrow \infty}\frac{e^{\beta E_{G}}}{\mathcal{N}_{G}}\bra{0}\mathcal{T} \exp\left[-\lambda\int_{0}^{\beta}d\tau V_{I}(\tau)\right]:\prod_{i=1}^{n}\phi_{I}(\vec{k}_{i},0):\ket{0}
\end{equation}
where $:\hspace{3pt}:$ stands for normal order. Notice that from the relation above we can read:
\begin{equation}
1=\braket{G|0}=\lim_{\beta\rightarrow \infty}e^{\beta E_{G}}\mathcal{N}_{G}^{-1}\bra{0}\mathcal{T} \exp\left[-\lambda\int_{0}^{\beta}d\tau V_{I}(\tau)\right]\ket{0}
\end{equation}

Therefore we arrive at:
\begin{equation}
\bra{G}\prod_{i=1}^{n}\left(\frac{a^{\dagger}_{\vec{k_{i}}}}{\sqrt{E(\vec{k}_{i})}}\right)\ket{0}=\lim_{\beta\rightarrow \infty}\frac{\bra{0}\mathcal{T} \exp\left[-\lambda\int_{0}^{\beta}d\tau V_{I}(\tau)\right]:\prod_{i=1}^{n}\phi_{I}(\vec{k}_{i},0):\ket{0}}{\bra{0}\mathcal{T} \exp\left[-\lambda\int_{0}^{\beta}d\tau V_{I}(\tau)\right]\ket{0}}\label{scatmat40}
\end{equation}

Notice the strong resemblance of the expression above with the starting point of standard \emph{adiabatic perturbation theory} \cite{Peskin}, but there are some important differences. Apart from being in euclidean time, notice that the integral runs from $0$ to $\beta$, not from $-\beta$ to $\beta$ as usual. This fact has strong consequences in the definition of the diagrams, as we will see.

Now we should expand the exponential in (\ref{scatmat40}) and perform the Wick contractions. To do this, it is useful to express the time ordered propagator via an auxiliary momentum along the euclidean time direction:
\begin{equation}
\bra{0}\mathcal{T}\left[\phi(\vec{k},\tau)\phi(\vec{q},\tau')\right]\ket{0}=\delta_{\vec{k},-\vec{q}}\int\frac{d k^{0}}{2\pi}\frac{ e^{ik^{0}(\tau-\tau')}}{[k^{0}]^{2}+[E(\vec{k})]^{2}}\label{tprop}
\end{equation}

Now we can evaluate (\ref{scatmat40}) via Feynman diagrams. The rules of the game are quite the same as those we encounter in standard textbooks \cite{Peskin,Weinberg,Amit,Ma} to compute correlators, with some crucial differences. Since the derivation of these Feynman rules is a standard  computation, we merely stress what changes in comparison with the standard references and then state the rules.
We can anticipate that, as usual, the denominator of (\ref{scatmat40}) has the effect of eliminating the disconnected diagrams with no external legs in the numerator. Moreover, relation (\ref{normalization34}) tells us that, in order to obtain $\mathcal{K}_{n}$, we should compute only the connected part of (\ref{scatmat40}) which, at the end of the story, will be represented by connected Feynman graphs. 
Notice that, since we are working in an `euclidean time' formulation, the propagators have no poles (at least in the massive case) and we do not need to introduce any pole prescription. 
So far nothing is different from from usual Feynman graphs, but here is  the crucial difference: we do not have conservation of energy at each vertex. 
Conservation of energy in \cite{Peskin} comes out from the integration over the whole time axis in the analogue of (\ref{scatmat40}), while in our case the integration is over the half line $(0,\beta\rightarrow\infty)$ and this has crucial consequences. 
Imagine that we expand (\ref{scatmat40}) and contract the fields; then using (\ref{tprop}) we can separately integrate along the time direction of each vertex and we encounter terms of the form:
\begin{equation}
\int_{0}^{\beta}d\tau \hspace{2pt}e^{i\sum_{i=1}^{s}\sigma_{i}k^{0}_{i}\tau}=\frac{i}{\sum_{i=1}^{s}\sigma_{i}k_{i}^{0}+i\epsilon} \quad , \; \sigma_i =\pm 1 \label{conserven}
\end{equation}

\begin{figure}
\begin{center}
\includegraphics[scale=0.3]{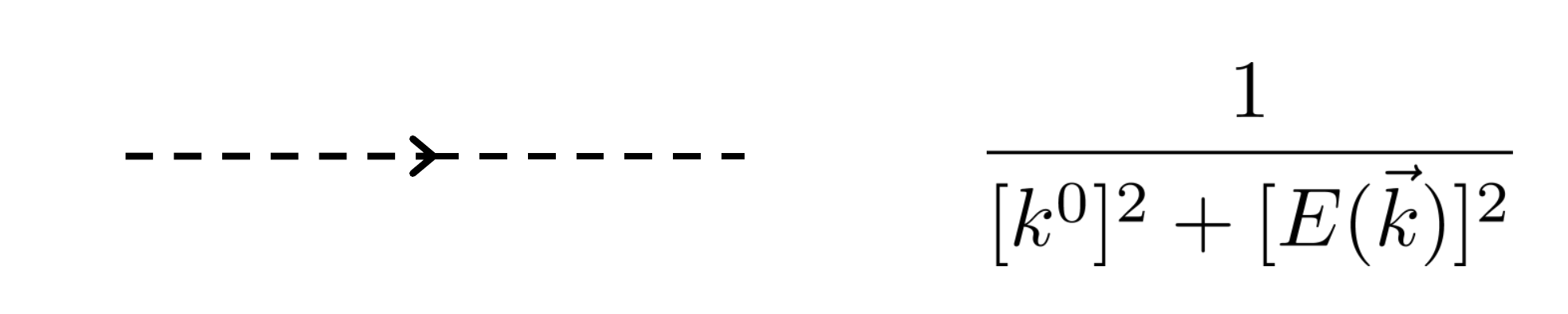}
\caption{Diagrammatic representation of the propagator with its value}\label{pert1}
\end{center}
\end{figure}

The last equality is true in the distribution sense, in the limit $\beta\rightarrow\infty$, $\sigma_{i}$ depends on the contraction we are doing, but for now their value is not important. Notice that, if in (\ref{conserven}) instead of integrating between $(0,\beta)$ we integrate between $(-\beta,\beta)$, then we would get a $\delta$-function instead of (\ref{conserven}): this is what happens in \cite{Peskin}.
Having in mind these differences from \cite{Peskin}, we can write down the Feynman rules: for simplicity we implement the $L\rightarrow\infty$ limit substituting summation in the momentum space with the proper integrations.
To avoid confusion with the graphs of Section \ref{grtool} the propagators will be indicated by dotted arrows (Figure \ref{pert1}) and the vertices as simple points.

\begin{center}
\begin{framed}
\centering
\hspace{1pt} \ \\
\textbf{Diagrammatic rules to compute $\mathcal{K}_{n}^{*}$:}

\begin{enumerate}

\item To compute (\ref{scatmat40}) draw a graph with $n$ external outgoing arrows and assign $(k^{0}_{i},\vec{k}_{i})_{i=1,...,n}$ to them. It is important to draw only outgoing arrows, otherwise we would spoil the vertices convention. 
If we are interested only in $\mathcal{K}_{n}$, we draw only connected graphs in order to extract the connected part of (\ref{scatmat40}). If instead we want the full expression (\ref{scatmat40}), disconnected graphs are allowed, but each disconnected block must have at least one external leg. Each term in (\ref{scatmat40}) has a global $L^{-d(n-2)/2}$ factor that exactly cancels with the same factor in (\ref{normalization34}), as it should. Therefore, we can ignore it.

\item Each vertex with $s$ external lines labeled by $s$ momenta $\{(q^{0}_{j},\vec{q}_{j})\}_{j\in\{1,s\}}$ counts as: 
\begin{equation}
-\frac{i\lambda\beta_{s}}{\sum_{j=1}^{s}\sigma_{j}q_{j}^{0}+i\epsilon}
\end{equation}
with $\epsilon\rightarrow 0$ and $\sigma_{i}$ is chosen $-1$ if the arrow is ingoing and $+1$ otherwise.

\item For each vertex of $s$ legs we must impose a momentum conservation law $\sum_{i=1}^{s}\vec{k}_{i}=0$ where $\sigma_{i}$ is chosen as at the previous point. Notice that the conservation refers to the spatial momentum while the $k^{0}_{j}$ momenta remain completely free.

\item Integrate over all the free momenta with integration measure for a component $q^{j}$ equal to $dq^{j}/(2\pi)$. Notice that we must also integrate over the zero component of the momenta of the external legs: only the spatial external momenta are fixed.

\item Divide each graph by the corresponding symmetry factor, that is the number of permutations of vertices and of internal legs that leave the graph unchanged. 

\item Imposing the conservation laws at each vertices we will end up with a global conservation law over the external momenta, as expected.

\item Sum over all the connected graphs to obtain $\mathcal{K}^{*}_{n}(\vec{k}_{1},..,\vec{k}_{n})/\prod_{i=1}^{n}\sqrt{2E(\vec{k}_{i})}$. Notice that a connected graph has only a global conservation law $\sum_{i=1}^{n}\vec{k}_{i}=0$, therefore the momenta are not conserved in `separate groups'. Since in (\ref{knconncorr}) we already took into account the global conservation law, $\mathcal{K}_{n}$ does not contain any additional $\delta$ factor, as expected from our locality constraints.
\end{enumerate}
\end{framed}
\end{center}

\begin{figure}
\begin{center}
\includegraphics[scale=0.3]{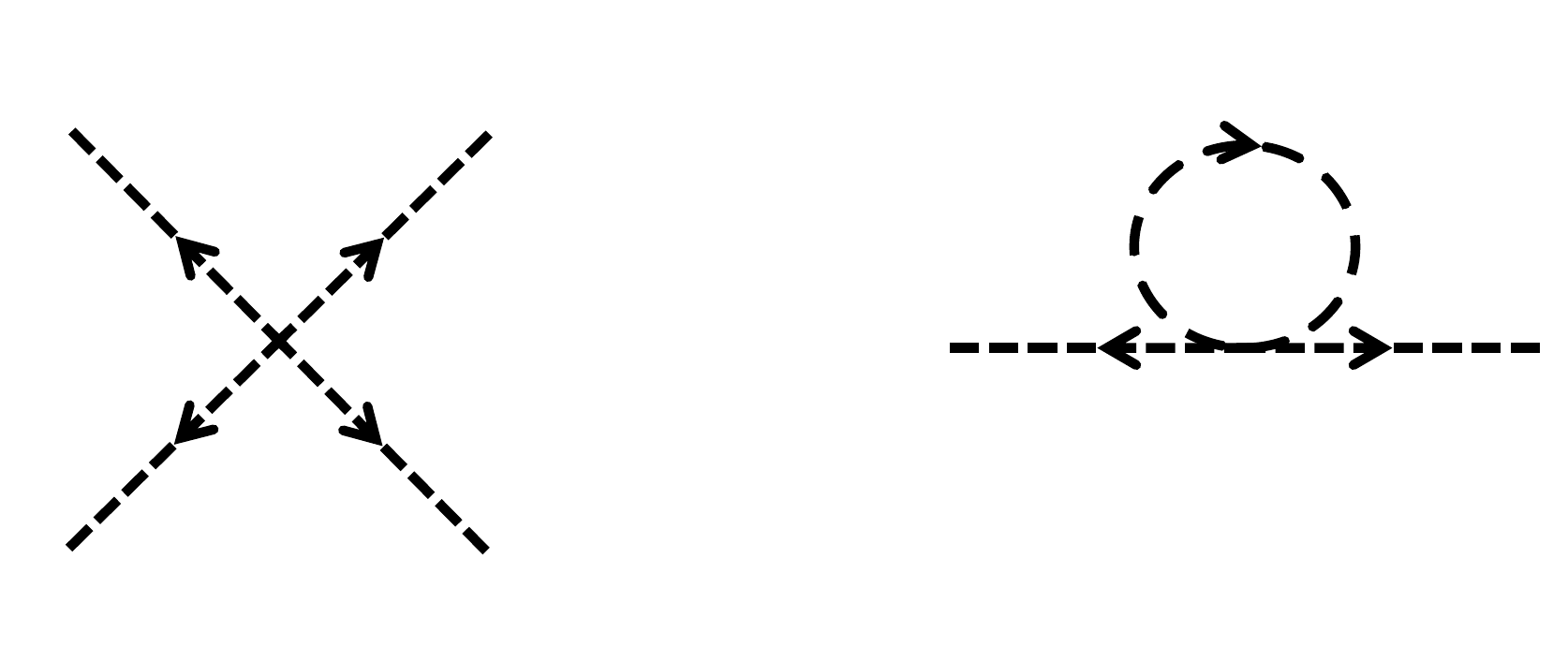}
\caption{Graphs for $\mathcal{K}_{4}$ and $\mathcal{K}_{2}$ at first order in $\lambda$ in $\phi^{4}$ theory. Their values are given in (\ref{kphi460}) (\ref{kphi461}).}\label{kphi4}
\end{center}
\end{figure}

These rules are best suited for findings the functions $\mathcal{K}_{n}$ as power series in the coupling constant. As an example, we focus on the $\phi^{4}$ theory:
\begin{equation}
\mathcal{L}=\int d^{d}x\frac{1}{2}(\partial_{\mu}\phi\partial^{\mu}\phi-m^{2}\phi^{2})-\frac{\lambda}{4!}\phi^{4}\label{phi4action}
\end{equation}
We can readily see that the graphs that contribute to $\mathcal{K}_{n}$ at first order in $\lambda$ are those of Figure \ref{kphi4}. 
\begin{equation}
\frac{\mathcal{K}_{4}(\vec{k}_{1},\vec{k}_{2},\vec{k}_{3},\vec{k}_{4})}{\prod_{j=1}^{4}\sqrt{2E(\vec{k}_{j})}}=-\lambda\int \prod_{j=1}^{4}\left[\frac{d^{4}k^{0}_{j}}{2\pi}\frac{1}{[k^{0}_{j}]^{2}+[E(\vec{k}_{i})]^{2}}\right]\frac{i}{\sum_{j=1}^{4}k^{0}_{j}+i\epsilon}\label{kphi460}
\end{equation}
\begin{equation}
\frac{\mathcal{K}_{2}(\vec{k},-\vec{k})}{2E(\vec{k})}=-\frac{\lambda}{2} \int \frac{d^{2}k^{0}dq^{0}d^{d}q}{(2\pi)^{3+d}}\frac{1}{\prod_{j=1}^{2}\left([k^{0}_{j}]^{2}+[E(\vec{k})]^{2}\right)}\frac{1}{[q^{0}]^{2}+[E(\vec{q})]^{2}}\frac{i}{k_{1}^{0}+k_{2}^{0}+i\epsilon}\label{kphi461}
\end{equation}

We can easily perform the integration over the momenta along the time direction to get:
\begin{equation}
\mathcal{K}_{4}(\vec{k}_{1},\vec{k}_{2},\vec{k}_{3},\vec{k}_{4})=-\frac{\lambda}{\prod_{j=1}^{4}\sqrt{2E(\vec{k}_{j})}}\frac{1}{\sum_{j=1}^{4}E(\vec{k}_{j})}+\mathcal{O}(\lambda^{2})\label{k4}
\end{equation}
\begin{equation}
\mathcal{K}_{2}(\vec{k},-\vec{k})=-\frac{\lambda}{[4E(\vec{k})]^{2}} \int \frac{d^{d}q}{(2\pi)^{d}}\frac{1}{E(\vec{q})}+\mathcal{O}(\lambda^{2})\label{k2}
\end{equation}

All other $\mathcal{K}_{n}$ with $n>4$ are of order higher than $\mathcal{O}(\lambda)$. All $\mathcal{K}_{n}$ with odd $n$ vanish due to the symmetry of the interaction under $\phi\to-\phi$. We can easily count the minimum number of vertices needed in order to construct the graphs that contribute to the computation of $\mathcal{K}_{2n}$: they correspond to a connected graph with no closed loops, i.e. a tree-like graph with $2n$ external legs. Thus the order of $\mathcal{K}_{n}$ in perturbation theory is:
\begin{equation}
\mathcal{K}_{2n+2}\sim \lambda^{n}  \hspace{3pc}n\ge 1
\end{equation}

More generally, for an interaction of the form $\phi^{j}$, there is a topology relation between the number of external legs $n$, the number of the interaction vertices $V$ and the number of internal loops $\mathfrak{L}$:
\begin{equation}
n=2-2\mathfrak{L}+(j-2)V
\end{equation}

This can be easily derived from the Euler's polyhedral formula and can be found in \cite{Peskin}.
Using  this formula it is not difficult to pin down the first order in $\lambda$ for each $\mathcal{K}_{n}$, since it is simply the number of vertices of the graph. We get $\mathcal{K}_{n}\sim \lambda^{(n+2\mathfrak{L}_{min}-2)/(j-2)}$ where $\mathfrak{L}_{min}$ is the minimal number of loops needed to draw the graph with $n$ external legs. 
Notice that, if $m$ is even, then we cannot construct any graph with an odd number of external legs.
Truncation in some order of $\lambda$ means  truncation in the functions $\mathcal{K}_{n}$: at order $\mathcal{O}(\lambda^{\alpha})$, only $\mathcal{K}_{n}$ with 
\begin{equation}
n\le n_{max} = (j-2)\alpha+2
\end{equation}
are non-vanishing. 
As a self-consistency check of this diagrammatic technique, it can be verified that the $\phi$ field correlators computed using the above results for $\mathcal{K}_{n}$ and the rules of Section \ref{grtool}, agree with the zero temperature correlators derived from standard perturbation theory. This is the content of Appendix \ref{check}.

As expected, some of the diagrams may involve integrals that diverge for large momenta,  as for example in (\ref{k2}): this is due to the ultraviolet divergences of the theory and in order to make sense of them we should implement a renormalization scheme, which must be compatible with the computation of the field correlators (see for example \cite{Peskin}). This procedure is implemented with the insertion of the proper renormalization counterterms in the Lagrangian with the net effect of modifying the definition of the coupling constants.
As we already stated, we will not proceed along this path and we rather assume that integrals are finite because of a UV cutoff.
Notice that, as long as $m\ne 0$, we do not have any singularity in each diagram we can draw, therefore $\mathcal{K}_{n}$ have no singular behavior: this is a non trivial check of the condition (\ref{suffcond}).

\subsection{Exact functional equations from the Schr\"oedinger equation}
\label{sectionexacteq}

The method of Section \ref{scatmatsec} is a perturbative method mostly suited for an expansion of $\mathcal{K}_{n}$ in terms of $\lambda$,  even thought a careful treatment of analogous expansion could lead to non perturbative results, for example the mass renormalization in quantum field theory \cite{Peskin}. To obtain such information at least a partial summation of the graphs is needed, but this is not an easy task, mostly because the of lack of energy conservation at the vertices.
Actually, there exists another way to implement a non perturbative study of the coefficients $\mathcal{K}_{n}$: we can derive a set of exact equations that $\mathcal{K}_{n}$ must satisfy in order for $\ket{G}$ to be the ground state of the Hamiltonian.
It is worth to anticipate that the general structure of these equations is consistent with the expected analytic properties of the $\mathcal{K}_{n}$ functions. Moreover these equations allow us to find the $\lambda$ expansion of $\mathcal{K}_{n}$ in a systematic way, avoiding the graphical technique of Section \ref{scatmatsec}, but consistently with it.
Moreover, if in Section \ref{scatmatsec} we must ask that the ground state is unique in order to extract it from the vacuum through (\ref{vaccumextr}), the equations we are going to use do not use this fact, therefore can be used in principle to study systems with degenerated ground states: of course, the hypothesis in which (\ref{intro0}) can be written must still be valid, therefore we need a non zero overlap of the state with the free vacuum of the theory.

We start with the observation that the ground state is simply an eigenstate of $H$:
\begin{equation}
H\ket{G}=E_{G}\ket{G}\label{exacteq55}
\end{equation}

To solve the eigenstate equation we can directly impose a solution $\ket{G}$ in the form of the cluster expansion (\ref{constr17}), so that we would obtain a set of equations for the unknown functions $\mathcal{K}_{n}$. 
Since each translational invariant state can be written as (\ref{constr17}), among the solutions of (\ref{exacteq55}) there are all the translational invariant eigenstates, but since we are interested in the ground state we can look for a solution only in the set of local quantum states with the precise analytic properties of the cluster amplitudes we identified in Section \ref{clusterexp}.
At this point we could have more than one solution: it would be interesting to study which eigenstates different from the ground state are local quantum states. In principle, to select the ground state we should compute the energy of all of them and choose the lowest, but if we make in the same hypothesis as in Section \ref{scatmatsec} we can isolate the correct ground state with the requirement that, for $\lambda\rightarrow 0$, it becomes the ground state of the free theory, i.e. the vacuum.

As before we assume an Hamiltonian of the form (\ref{perturbh}), but without loss of generality it is easier to suppose that it is normal-ordered, defining new coupling constants $\tilde{\beta}_{n}$:

\begin{equation}
H=\sum_{\vec{k}}E(\vec{k})a^{\dagger}_{\vec{k}}a_{\vec{k}}+\lambda\sum_{n}\frac{\tilde{\beta}_{n}}{n!L^{d(n-2)/2}}\sum_{\vec{k}_{1}+..+\vec{k}_{n}=0}:\prod_{q=1}^{n}\left[\frac{1}{\sqrt{2E(\vec{k}_{q})}}\left(a^{\dagger}_{\vec{k}_{q}}+a_{-\vec{k}_{q}}\right)\right]:\label{hnormal}
\end{equation}

In the above $:\hspace{4pt}:$ means that we consider the normal-ordered expression, with all the creation operators on the left of the annihilation operators.
The action of $H$ on $\ket{G}$ can be computed using the observation that $a_{\vec{k}}$, because of the commutation rules, acts as a differentiation with respect to $a^{\dagger}_{\vec{k}}$. Since $\ket{G}$ is an exponential state, acting on it with as many operators $a_{\vec{k}_{i}}$ we want, we will always obtain a state of the form $\Theta\ket{G}$, with $\Theta$ an operator containing only $a^{\dagger}$ operators. For example, if we consider $a_{-\vec{k}}\ket{G}$ we have:
\begin{equation}
a_{-\vec{k}}\ket{G}=\frac{\delta}{\delta a^{\dagger}_{-\vec{k}}}\ket{G}
\end{equation}
\begin{equation}
a_{-\vec{k}}\ket{G}=\sum_{n}\frac{1}{(n-1)!}\frac{1}{ L^{d(n-2)/2}}\left[\sum_{\vec{k}_{1}+..+\vec{k}_{n-1}=\vec{k}}\mathcal{K}_{n}(\vec{k}_{1},..,\vec{k}_{n-1},\vec{k})a^{\dagger}_{\vec{k}_{1}}..a^{\dagger}_{\vec{k}_{n-1}}\right]\ket{G}\label{exacteq56}
\end{equation}

With this, we can see immediately that:

\begin{equation}
\sum_{\vec{k}}E(\vec{k})a^{\dagger}_{\vec{k}}a_{\vec{k}}\ket{G}=\sum_{n}\frac{1}{n!}\frac{1}{ L^{d(n-2)/2}}\left[\sum_{\vec{k}_{1}+..+\vec{k}_{n}=0}\left(\sum_{i=1}^{n}E(\vec{k}_{i})\right)\mathcal{K}_{n}(\vec{k}_{1},..,\vec{k}_{n})a^{\dagger}_{\vec{k}_{1}}..a^{\dagger}_{\vec{k}_{n}}\right]\ket{G}\label{exacteq}
\end{equation}

Dealing with the interaction term of the Hamiltonian is more involved, because we do not have only a derivative:
\begin{equation}
:\prod_{q=1}^{n}\left(a^{\dagger}_{\vec{k}_{q}}+a_{-\vec{k}_{q}}\right):\ket{G}=:\prod_{q=1}^{n}\left(a^{\dagger}_{\vec{k}_{q}}+\frac{\delta}{\delta a^{\dagger}_{-\vec{k}_{q}}}\right):\ket{G}\label{uderivative}
\end{equation}

The combinatorics of such an expression is difficult, because $\ket{G}$ contains all the cluster amplitudes $\mathcal{K}_{n}$ and having more than one derivative links them in a non trivial way. 
In Section \ref{grexacteqsec} we show a graphical representation to deal with this complicated combinatorics.
After this cumbersome computation, we can write $H\ket{G}$ as:

\begin{eqnarray}
\nonumber&&H\ket{G} =\sum_{n=0}^{\infty}\frac{1}{n!}\frac{1}{ L^{d(n-2)/2}} \sum_{\vec{k}_{1}+...+\vec{k}_{n}=0} 
\Bigg[ \left(\sum_{i=1}^{n}E(\vec{k}_{i})\right)\mathcal{K}_{n}(\vec{k}_{1},...,\vec{k}_{n}) + \\
&&  +\lambda \frac{\tilde{\beta}_{n}}{\prod_{i=1}^{n}\sqrt{2E(\vec{k}_{i})}} +\lambda \mathfrak{F}_{n}(\vec{k}_{1},...,\vec{k}_{n}) \Bigg] 
a^{\dagger}_{\vec{k}_{1}}...a^{\dagger}_{\vec{k}_{n}}\ket{G}\label{exact58}
\end{eqnarray}

In the second line we have explicitly written the simplest part of (\ref{uderivative}) that does not involve derivatives and all the remaining action of the interaction is contained in
$\mathfrak{F}_{n}$. This is a complicated object involving sums and convolutions of $\mathcal{K}$ with the interaction terms as we are going to explain in Section \ref{grexacteqsec}.
Now we should impose equation (\ref{exacteq55}) and we obtain:
\begin{eqnarray}
\nonumber&&\sum_{n=0}^{\infty}\frac{1}{n!}\frac{1}{ L^{d(n-2)/2}} \sum_{\vec{k}_{1}+...+\vec{k}_{n}=0} 
\Bigg[ \left(\sum_{i=1}^{n}E(\vec{k}_{i})\right)\mathcal{K}_{n}(\vec{k}_{1},...,\vec{k}_{n}) + \\
&&  +\lambda \frac{\tilde{\beta}_{n}}{\prod_{i=1}^{n}\sqrt{2E(\vec{k}_{i})}} +\lambda \mathfrak{F}_{n}(\vec{k}_{1},...,\vec{k}_{n}) -\frac{E_{G}}{L^{d}}\delta_{n,0}\Bigg] 
a^{\dagger}_{\vec{k}_{1}}...a^{\dagger}_{\vec{k}_{n}}\ket{G}=0\label{newexacteq79}
\end{eqnarray}

Note that we used a factor $L^{d}$ to put the ground state energy inside the summation: what is contained in square brakets is now well-defined in the $L\to\infty$ limit, since the ground state energy is extensive.
From these equations we can extract directly equations for the cluster amplitudes $\mathcal{K}_{n}$:

\begin{equation}
\left(\sum_{i=1}^{n}E(\vec{k}_{i})\right)\mathcal{K}_{n}(\vec{k}_{1},..,\vec{k}_{n})=-\frac{\lambda\tilde{\beta}_{n}}{\prod_{i=1}^{n}\sqrt{2E(\vec{k}_{i})}} -\lambda\hspace{2pt}\mathfrak{F}_{n}(\vec{k}_{1},..,\vec{k}_{n})+\frac{E_{G}}{L^{d}} \delta_{n,0}\label{excteq57}
\end{equation}

Notice that, if equations (\ref{excteq57}) are satisfied, then also (\ref{newexacteq79}) is true. It is also possible to show the inverse using suitable test states. The argument is this: applying $\bra{0}$ to (\ref{newexacteq79}), all the terms with $n>0$ will automatically cancel; only the $n=0$ term survives.

\begin{equation}
\left(\lambda\tilde{\beta}_{0} +\lambda\hspace{2pt}\mathfrak{F}_{0}-\frac{E_{G}}{L^{d}}\right)\braket{0|G}=0
\end{equation}

Since $\braket{0|G}=1$, we have that such an equation is satisfied only if the equation $n=0$ of (\ref{excteq57}) is satisfied. Now, move on with another test state containing one particle, that is to say $\bra{0}a_{\vec{k}}$: applying it on the left of (\ref{newexacteq79}), knowing that the term $n=0$ is zero and using the commutation rules, we have that only the term with $n=1$ is not annihilated by this test state. This time, we would find that the equation  (\ref{excteq57}) with $n=1$ must be satisfied. At this point using a test state with two particles, we will find $n=2$ out of  (\ref{excteq57}), then proceed step by step increasing the number of particles: with this idea, it is immediate to see by induction the equivalence of (\ref{newexacteq79}) and (\ref{excteq57}).
Notice that in (\ref{excteq57}) there are no extra $\delta-$like contributions, which would be inconsistent (there are no extra $\delta$ hidden in $\mathfrak{F}$, as it will be clear in Section \ref{grexacteqsec}). Moreover, since there are no explicit singularities in the equations (at least for $m\ne 0$), we should expect a smooth behavior for $\mathcal{K}$, without any singularity. Some care must be taken in the massless case in which $E(\vec{k})=|\vec{k}|$ and therefore poles arise; we discuss about this fact in Appendix \ref{failure}.
These features of the equations can be regarded as a consistency check for the analytic properties of the cluster expansion.
It should also be explicitly said that these equations could also been used to compute the ground state energy, simply taking the $n=0$ term in (\ref{excteq57}):
\begin{equation}
\frac{E_{G}}{L^{d}}=\lambda\tilde{\beta}_{0}+\lambda \mathfrak{F}_{0}\label{newGenergy}
\end{equation}

In the above, $\mathfrak{F}_{0}$ is completely fixed once we determine $\mathcal{K}_{n}$ through the other equations: this relation is particularly important if we have more than one solution of (\ref{excteq57}), since in order to find the ground state we must choose the solution that minimizes $E_{G}$.
Notice that starting from the solution for $\lambda=0$, these equations can be solved recursively to obtain an expansion of $\mathcal{K}$ in powers of $\lambda$: the first order solution can be obtained simply neglecting $\mathfrak{F}$.
\begin{equation}
\mathcal{K}_{n}(\vec{k}_{1},..,\vec{k}_{n})=-\frac{1}{\sum_{i=1}^{n}E(\vec{k}_{i})}\frac{\lambda\tilde{\beta}_{n}}{\prod_{i=1}^{n}\sqrt{2E(\vec{k}_{i})}}+\mathcal{O}(\lambda^{2})\label{apprkn}
\end{equation}

In this way we have that $\mathcal{K}$ is of first order in $\lambda$ and $\lambda\mathfrak{F}$ is of order at least $\lambda^{2}$, since $\mathfrak{F}$ contains itself at least one $\mathcal{K}$. Inserting (\ref{apprkn}) in $\mathfrak{F}$ we can update the right hand side of (\ref{excteq57}) and find new $\mathcal{K}$ corrected up to $\lambda^{2}$. We can then iterate the procedure: this is a systematic way to obtain exactly the same perturbative expansion as Section \ref{scatmatsec}.
As a matter of fact in the $\phi^4$ model, once we normal order the Hamiltonian obtained from (\ref{phi4action}) we get the coupling constants:
\begin{equation}
\tilde{\beta}_{4}=1\hspace{4pt};\hspace{2pc}
\tilde{\beta}_{2}=\frac{1}{2} \int \frac{d^{d}q}{(2\pi)^{d}}\frac{1}{2E(\vec{q})}\hspace{4pt};\hspace{2pc}
\tilde{\beta}_{0}=\frac{1}{8}\left(\int\frac{d^{d}q}{(2\pi)^{d}}\frac{1}{2E(\vec{q})}\right)^{2}\label{u2}
\end{equation}

The other coupling constants $\tilde{\beta}_{n}$ are zero. Inserting (\ref{u2}) in (\ref{apprkn})  we obtain again (\ref{k4}) and (\ref{k2}), as expected.

\subsubsection{Diagrammatic representation of the functional equations}
\label{grexacteqsec}

\begin{figure}
\begin{center}
\includegraphics[scale=0.24]{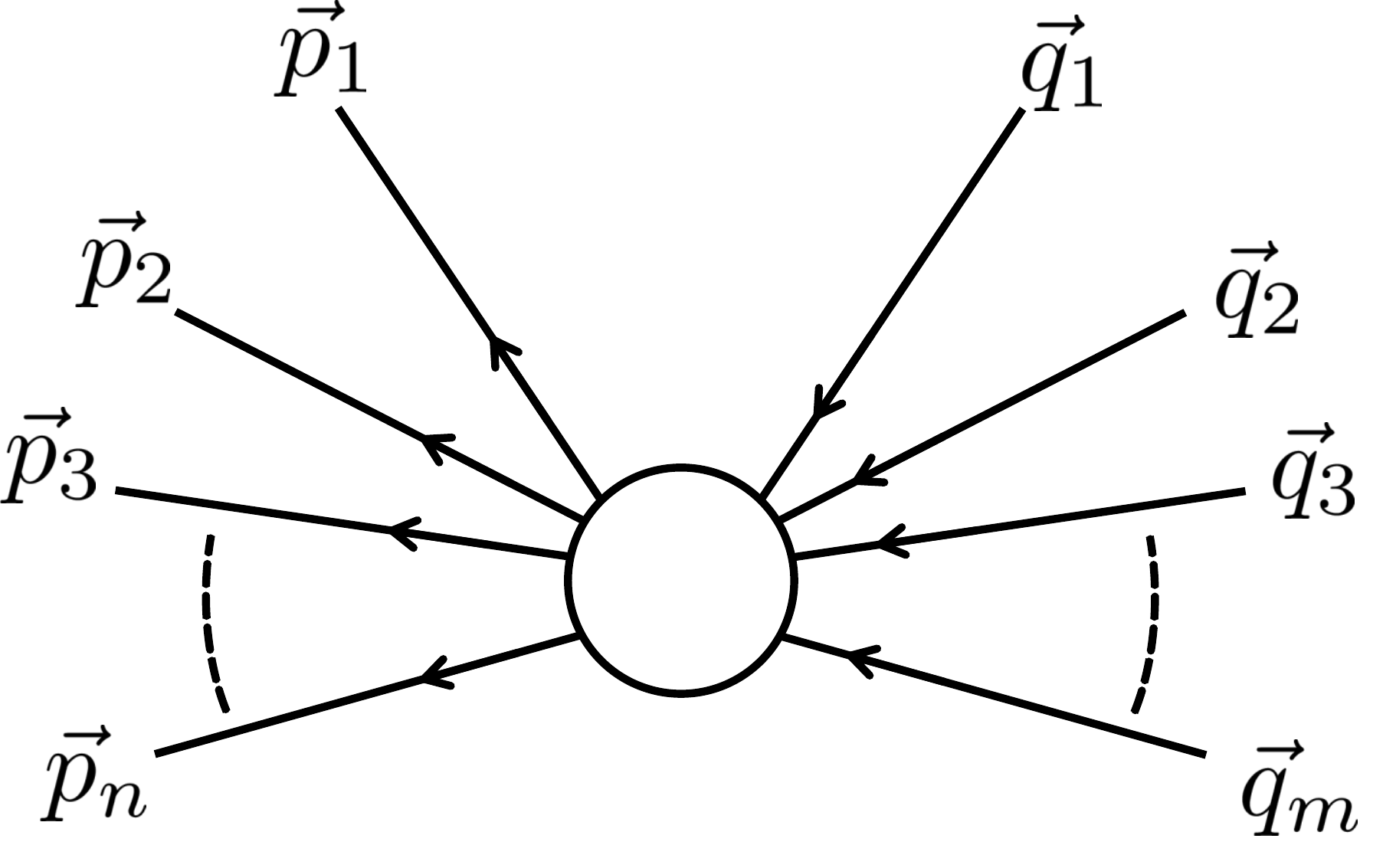}
\end{center}
\caption{With a blank circle we denote the interaction vertex between the Hamiltonian and the cluster amplitudes, its value is $\left(\prod_{i=1}^{n}\sqrt{2E(\vec{p}_{i})}\prod_{i=1}^{m}\sqrt{2E(\vec{q}_{i})}\right)^{-1}\tilde{\beta}_{n+m}$}\label{figexacteq1}
\end{figure}

We now want to present a simple graphical way to deal with the combinatorics of (\ref{uderivative}) and compute $\mathfrak{F}_{n}$ of (\ref{exact58}). We invite the reader to check its consistency with the explicit computation of some terms through (\ref{uderivative}). 
We still use arrows to connect vertices and in this graphical approach we need two  types of vertices, one to represent $\mathcal{K}_{n}$ and one to represent $\tilde{\beta}_{n}$: the vertices related to $\mathcal{K}_{n}$ are simple dots, instead $\tilde{\beta}_{n}$ is represented by a circle (Figure \ref{figexacteq1}). The following rules hold in the thermodynamic limit, where we can replace sums with integrals.

\begin{center}
\begin{framed}
\centering
\hspace{1pt} \ \\
\textbf{Diagrammatic rules to compute $\mathfrak{F}_{n}$:}

\begin{enumerate}
\item A graph associated to $\mathfrak{F}_{n}$ is made by one and only one circle representing the interaction vertex and a number of dots representing the $\mathcal{K}_{n}$ amplitudes. 
\item Each dot must be connected to the circle by an arrow, which automatically implies that the graph is connected. Arrows can only point outwards from dots, but at the circle they can be either ingoing or outgoing. Following this recipe, the dots related to $\mathcal{K}$ can never be directly connected.
\item To get $\mathfrak{F}_{n}(\vec{k}_{1},..,\vec{k}_{n})$ we must have $n$ external legs with assigned momenta $\{\vec{k}_{i}\}_{i=1,..,n}$. All the external arrows must point outwards: this is necessary to respect the sign convention we adopt.
\item Vertices joined with an arrow have the same momenta; an internal arrow with momentum $\vec{q}$ enters as $\vec{q}$ in the respective $\mathcal{K}$.
\item At each vertex, dot or circle, we must impose the momentum conservation. Notice that regarding the $\tilde{\beta}$ circle we must take in account also the direction of the arrows. (For example in Figure \ref{figexacteq1} the requirement is $\sum_{i=1}^{n}\vec{p}_{i}=\sum_{i=1}^{m}\vec{q}_{i}$.)
\item After we impose all the conservation laws we have a global conservation law on the momenta and no other restriction over them, as it should. We must integrate over all remaining free momenta. A free momentum $\vec{q}$ has integration measure $d^{d}q/(2\pi)^{d}$.
\item Divide by the symmetry factor, that is to say, the number of permutations of vertices and internal lines that leave the graph (topologically) unchanged. External lines are fixed ant they cannot be exchanged.
\item Sum over all possible graphs. This means sum over all the possible topologies and over all the possible permutations of the external momenta that change the graph. To avoid confusion, once we fix the topology of the graph we will always mean that we are also summing over all non equivalent permutations of external legs.
\end{enumerate}
\hspace{1pt} \ \\
\end{framed}
\end{center}

\begin{figure}
\begin{center}
\includegraphics[scale=0.35]{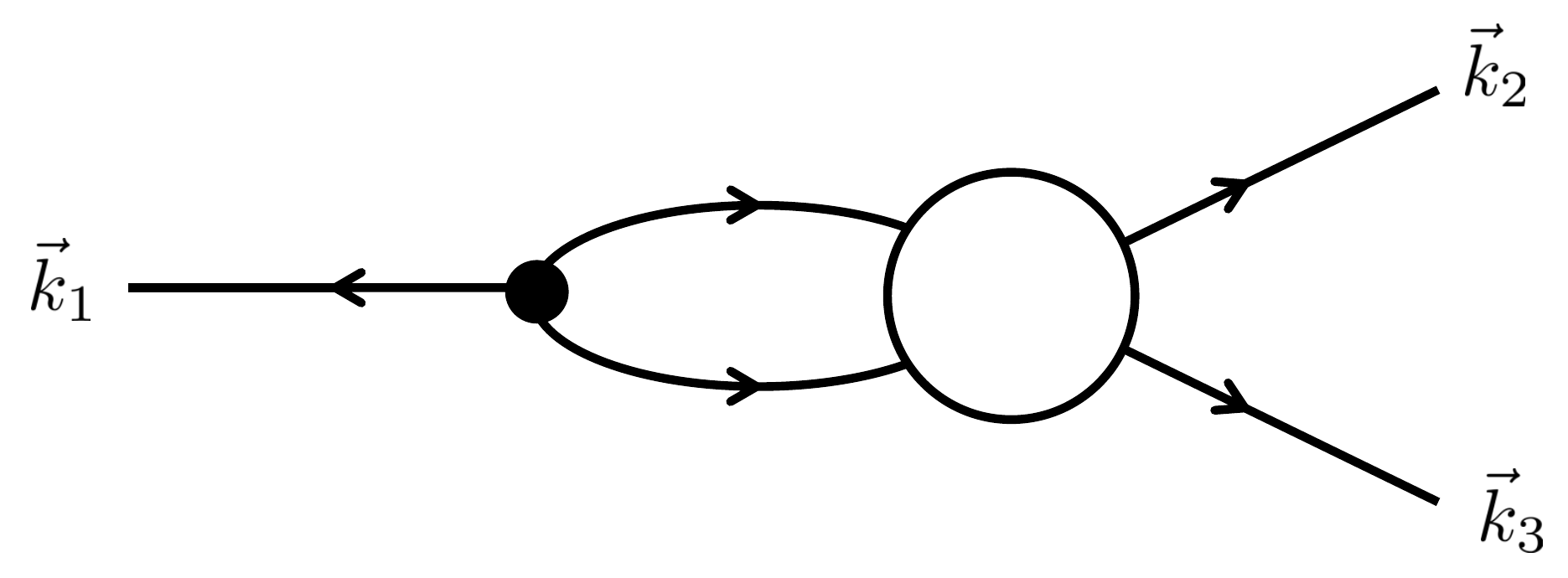}
\caption{Example: this graph enters in the computation of $\mathcal{F}_{3}$, when the coupling constant $\tilde{\beta}_{4}$ is present, therefore it appears in the $\phi^{4}$ theory.}\label{exacteqexample}
\end{center}
\end{figure}

As an example, we are going to see in detail the calculation of  Figure \ref{exacteqexample}, that contributes to $\mathfrak{F}_{3}$. We assign the external momenta and arrived at point 5 we have:
\begin{equation}
\frac{\mathcal{K}_{3}(\vec{q},-\vec{q}-\vec{k}_{1},\vec{k}_{1})}{\sqrt{4E(\vec{p
}_{2})E(\vec{k}_{3})}}\frac{\tilde{\beta}_{4}}{\sqrt{4E(\vec{q})E(\vec{q}-\vec{k}_{2}-\vec{k}_{3})}}
\end{equation}

The free momentum is $\vec{q}$ and it must be integrated out; the symmetry factor is $1/2$ because we can exchange the two internal lines. At point 7 we get:
\begin{equation}
f_{3}(\vec{k}_{1},\vec{k}_{2},\vec{k}_{3})\equiv\frac{1}{2}\int \frac{d^{d}q}{(2\pi)^{d}}\frac{\mathcal{K}_{3}(\vec{q},-\vec{q}-\vec{k}_{1},\vec{k}_{1})}{\sqrt{2E(\vec{k}_{2})}\sqrt{2E(\vec{k}_{3})}}\frac{\tilde{\beta}_{4}}{\sqrt{4E(\vec{q})E(\vec{q}-\vec{k}_{2}-\vec{k}_{3})}}
\end{equation}

Now we must sum over the distinct permutations of the external momenta, therefore after point 8 we get:
\begin{equation}
f^{Sym}_{3}(\vec{k}_{1},\vec{k}_{2},\vec{k}_{3})=f_{3}(\vec{k}_{1},\vec{k}_{2},\vec{k}_{3})+f_{3}(\vec{k}_{2},\vec{k}_{3},\vec{k}_{1})+f_{3}(\vec{k}_{3},\vec{k}_{2},\vec{k}_{1})
\end{equation}

Now the quantity $f^{Sym}_{3}(\vec{k}_{1},\vec{k}_{2},\vec{k}_{3})$ is completely symmetric in its argument and it is the contribution for $\mathfrak{F}_{3}(\vec{k}_{1},\vec{k}_{2},\vec{k}_{3})$ generated by such a graph.
The expressions for $\mathfrak{F}_{n}$ are quite involved and usually analytically intractable; there is only one case in which we can analytically solve them and it is the case of a gaussian state.
This is something expected, since we know that the gaussian state can be explicitly solved by other methods (Appendix \ref{gaussian}) and we can use this special case as a check for our method.
The quadratic case is such that in (\ref{hnormal}) we have only $\tilde{\beta}_{2}$ and $\tilde{\beta}_{0}$ different from zero, without loss of generality we can redefine the $\lambda$ parameter in order to have $\tilde{\beta}_{2}=1$. With this choice, the Lagrangian we are considering is simply
\begin{equation}
\mathcal{L}=\int d^{d}x \hspace{3pt}\frac{1}{2}(\partial_{\mu}\phi\partial^{\mu}\phi-m^{2}\phi)-\frac{\lambda}{2}\phi^{2}
\end{equation}
while the normal ordered Hamiltonian is: 
\begin{equation}
H= \tilde{\beta}_{0}+\sum_{\vec{k}} \left [ E(\vec{k})a^{\dagger}_{\vec{k}}a_{\vec{k}}+\frac{\lambda}{2} \frac{1}{2E(\vec{k})} \left(a^{\dagger}_{\vec{k}}a^{\dagger}_{-\vec{k}}+a_{\vec{k}}a_{-\vec{k}}+2a^{\dagger}_{\vec{k}}a_{\vec{k}}\right) \right ] 
\end{equation}

$\tilde{\beta}_{0}$ does not matter for what follows, since an additive constant does not change the ground state, but only its energy.
In Appendix \ref{gaussian} we study such an Hamiltonian by standard methods and find its ground state is a squeezed state with $\mathcal{K}_{2}$ given by (\ref{k2mass}); here we want to obtain the same result through our equations.
From our equations, it is not difficult to understand that only $\mathcal{K}_{2}$ will be non zero. Consider the perturbative solution described in Section \ref{sectionexacteq}: at first order the only non zero term is $\mathcal{K}_{2}$, as it is clear from (\ref{apprkn}). Using now this solution in the next iterative step, we should compute $\mathfrak{F}_{n}$ with the first order solution (\ref{apprkn}). Using the diagrammatic rules to compute $\mathfrak{F}_{n}$ we see immediately that, thanks to the fact that only $\mathcal{K}_{2}$ is not zero at first order, only $\mathfrak{F}_{2}$ and $\mathfrak{F}_{0}$ are not zero at the same order. At the next order, $\mathfrak{F}_{2}$ corrects $\mathcal{K}_{2}$ as it is clear from (\ref{excteq57}), but all $\mathcal{K}_{n\ne 2}$ remain zero: repeating the same reasoning we see all $\mathcal{K}_{n\ne 2}$ remain zero at any perturbative order.
Thanks to this reasoning, we know that the solution we are looking for possesses only $\mathcal{K}_{2}$ different from zero. Using this fact we can now write the exact equations that $\mathcal{K}_{2}$ must satisfy, then exactly solve them.
From the topology of the graphs it is clear that only $\mathfrak{F}_{2}$ will be non zero and has two different contributions (Figure \ref{figexacteq2}).
Using the rules we explained, we can write $\mathfrak{F}_{2}$ as:
\begin{figure}
\begin{center}
\includegraphics[scale=0.3]{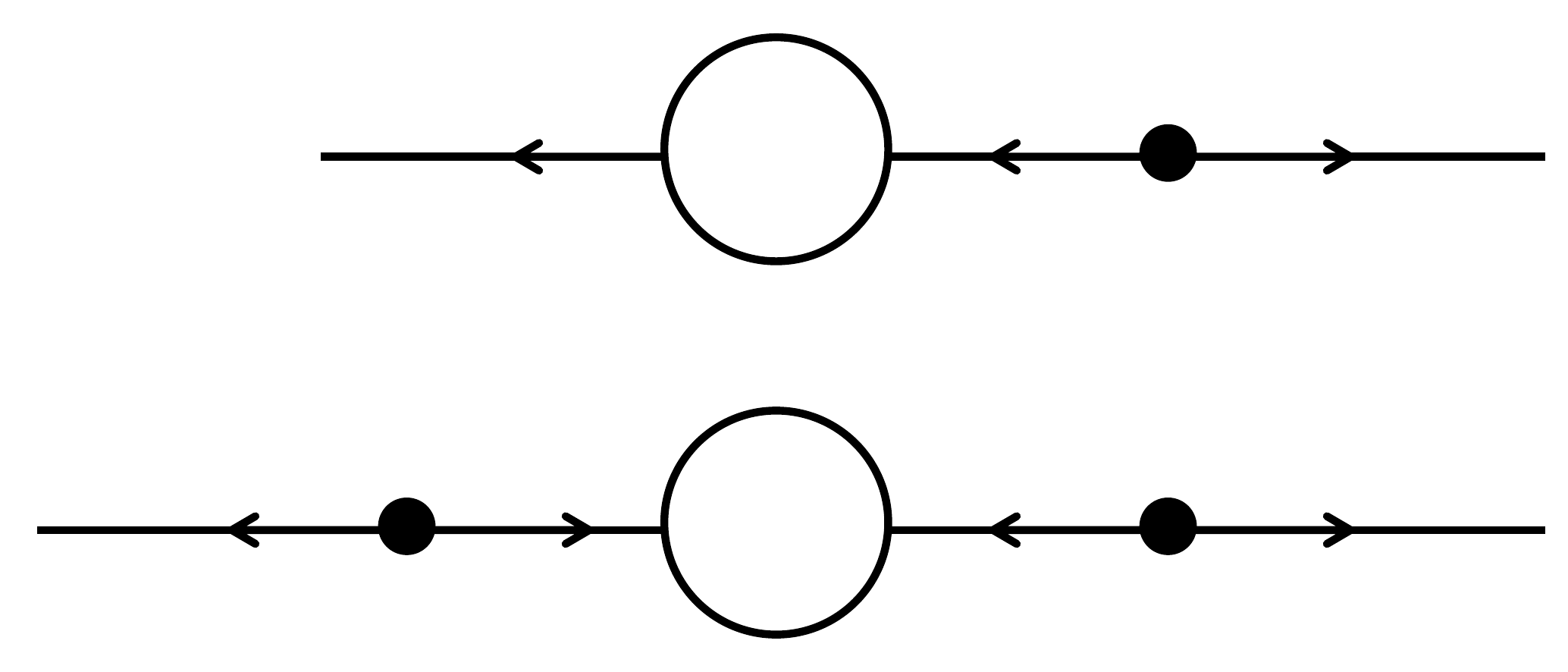}
\caption{Graphs representing the two contributions to $\mathfrak{F}_{2}$ keeping only $\mathcal{K}_{2}\ne 0$, to obtain $\mathfrak{F}_{2}$ we must sum them.}\label{figexacteq2}
\end{center}
\end{figure}
\begin{equation}
\mathfrak{F}_{2}(\vec{k},-\vec{k})=\frac{1}{E(\vec{k})}\mathcal{K}_{2}(\vec{k},-\vec{k})+\frac{1}{2E(\vec{k})}[\mathcal{K}_{2}(\vec{k},-\vec{k})]^{2}
\end{equation}

Notice that in the first graph of Figure \ref{figexacteq2} we get two different graphs if we exchange the momenta of the two external legs, but the value is unchanged: this gives a multiplicative factor of $2$ to the linear term in $\mathcal{K}_{2}$, correctly counted in the expression above.
Now we should insert this quantity in (\ref{excteq57}) and we finally get the equations for $\mathcal{K}_{2}$:
\begin{equation}
2E(\vec{k})\mathcal{K}_{2}(\vec{k},-\vec{k})+\lambda\frac{1}{2E(\vec{k})}+\lambda\frac{1}{E(\vec{k})}\mathcal{K}_{2}(\vec{k},-\vec{k})+\lambda\frac{1}{2E(\vec{k})}[\mathcal{K}_{2}(\vec{k},-\vec{k})]^{2}=0\label{eqgaussian}
\end{equation}

These have two solutions:
\begin{equation}
\mathcal{K}_{2}(\vec{k},-\vec{k})=-2\lambda^{-1}[E(\vec{k})]^{2}-1\pm\sqrt{\left(2\lambda^{-1}[E(\vec{k})]^{2}+1\right)^{2}-1}\label{solquadk2}
\end{equation}

We can identify the correct solution requiring that, for $\lambda\to 0$, the ground state becomes the vacuum, therefore we must have $\mathcal{K}_{2}\rightarrow 0$ and so we must choose the plus sign. With a little bit of algebra, the expression can be rearranged as:
\begin{equation}
\mathcal{K}_{2}(\vec{k},-\vec{k})=\frac{\sqrt{\vec{k}^{2}+m^{2}}-\sqrt{\vec{k}^{2}+m^{2}+\lambda}}{\sqrt{\vec{k}^{2}+m^{2}}+\sqrt{\vec{k}^{2}+m^{2}+\lambda}}\label{new88}
\end{equation}

This is exactly expression (\ref{k2mass}) of Appendix \ref{gaussian}, as it should. As we anticipated at the beginning of this section, we have found also another solution to the eigenstate equations, but note that the state we would have obtained choosing the minus sign in (\ref{solquadk2}) is such that $|\mathcal{K}_{2}|\ge 1$ and this is an unphysical state, as we explain in Appendix \ref{gaussian}. 

After this check in the gaussian case we would like to use these equations to study a real interacting field theory, like the $\phi^4$ model. The equations in this case become involved and we do not attempt an analytical solution of them, nevertheless they are suited for different approximation schemes, as we are going to discuss, and we can even acquire precious, non perturbative information through them, as we will do in Appendix \ref{failure}.

\subsection{Approximation schemes}
\label{approxscheme}
In this brief section we want to sketch some possible approximation schemes that can be used to get the cluster amplitudes of a ground state, having in mind the example of a relativistic field theory (such as (\ref{gzlandau})): we do not want to be quantitative and simply comment on the possible validity and problems of these approximated methods, leaving their development for further studies.

\begin{enumerate}
\item \textbf{Naive perturbation theory:}

This is the simplest method that we can think of: we can use the perturbation theory of Section \ref{scatmatsec} or, equivalently, solve recursively in $\lambda$ the functional equations of Section \ref{sectionexacteq}. Solving up to order $\mathcal{O}(\lambda^{\alpha})$ we get an approximation to $\mathcal{K}_{n}$ that provides a natural truncation in the cluster expansion, as commented at the end of Section \ref{scatmatsec}. The appealing fact is that such an approximate method is straightforward to perform, but we know that often naive perturbation theory lacks of predictive power and we need to go beyond it. This truncation scheme, from the viewpoint of the functional equations, can be generalized also to the dynamical case: in Section \ref{dynamic} we will see that the evolution of a local quantum state with a local Hamiltonian follows a simple generalization of the functional equations.

\item \textbf{Perturbation theory starting from Hartree-Fock approximation:}

This method represents the first improvement of the naive perturbation theory: we focus on the $\phi^{4}$ theory as an example, but can be done in general.
The idea is simply to separated in the action a quadratic part that captures as much information as possible, in a consistent way.
To do this, we add and subtract terms in the $\phi^{4}$ Lagrangian (\ref{phi4action}) and write:
\begin{equation}
\mathcal{L}=\int d^{d}x\hspace{2pt}\frac{1}{2}\left[\partial_{\mu}\phi\partial^{\mu}\phi-\left(m^{2}+\frac{\lambda\braket{\phi^{2}}}{6}\right)\phi^{2}\right]-\frac{\lambda}{4!}(\phi^{2}-\braket{\phi^{2}})^{2}+\frac{\lambda}{4!}\braket{\phi^{2}}^{2}
\end{equation}

This is exactly the same Lagrangian as (\ref{phi4action}), in which we added and subtracted $\braket{\phi^{2}}$ in a proper way: now, in square brakets, it appears a quadratic Lagrangian with a shifted mass. The unknown expectation value $\braket{\phi^{2}}$ must be computed selfconsistently. With this trick we have a better control on the interaction term: now the small parameter is not the bare $\lambda$ parameter anymore, but rather the variance of the square of the field $\braket{(\phi^{2}-\braket{\phi^{2}})^{2}}$. The crudest approximation is obtained neglecting the latter term in the action and is known as Hartree-Fock approximation \cite{HF}. Since the latter term would be zero if the ground state were gaussian, such a method is valid as long as the gaussian state is a good approximation. To proceed further we should write the eigenmodes of $\phi$ using the shifted mass; we call $\tilde{a}_{\vec{k}}$ these new operators {which are} related to the old $a_{\vec{k}}$ and $a^{\dagger}_{\vec{k}}$ simply through a linear transformation, known as Bogoliubov rotation.
\begin{equation}
\phi(\vec{x})=\frac{1}{L^{d/2}}\sum_{\vec{k}}\frac{e^{i\vec{k}\vec{x}}}{\sqrt{2 \tilde{E}(\vec{k})}}\left(\tilde{a}^{\dagger}_{\vec{k}}+\tilde{a}_{-\vec{k}}\right)\hspace{2pc} \tilde{E}(\vec{k})=\sqrt{\vec{k}^{2}+\tilde{m}^2}
\end{equation}
In the above, the shifted mass $\tilde{m}$ is defined as:
\begin{equation}
\tilde{m}^{2}=m^{2}+\frac{\lambda\braket{\phi^{2}}}{6}
\end{equation}

The Bogoliubov transformation in terms of these parameters is:

\begin{equation}
a_{\vec{k}}=\frac{1}{2}\left(\sqrt{\frac{E(\vec{k})}{\tilde{E}(\vec{k})}}+\sqrt{\frac{\tilde{E}(\vec{k})}{E(\vec{k})}}\right)\tilde{a}_{\vec{k}}+\frac{1}{2}\left(\sqrt{\frac{E(\vec{k})}{\tilde{E}(\vec{k})}}-\sqrt{\frac{\tilde{E}(\vec{k})}{E(\vec{k})}}\right)\tilde{a}^{\dagger}_{-\vec{k}}
\end{equation}

The Hilbert space will be constructed out of a new vacuum $\ket{\tilde{0}}$ that is annihilated by all the $\tilde{a}_{\vec{k}}$. Then all the construction of Section \ref{perturb} can be repeated simply replacing $a_{\vec{k}}$ and $m$ with the new operators and mass. We can use Section \ref{scatmatsec} or Section \ref{sectionexacteq} to perform naive perturbation theory as explained before, but notice that such an expansion is not perturbative in $\lambda$ due to the shifted mass: this approximation goes beyond naive perturbation theory. After we obtain the proper truncated approximation of $\mathcal{K}_{n}$, the expectation value $\braket{\phi^{2}}$ must be computed in a self consistent way. We can formally compute $\braket{\phi^{2}}$ with the methods of Section \ref{grtool} and Appendix \ref{grtoolperturb}: the graphical expansion of $\braket{\phi^{2}}$ must be truncated to the proper order, consistently with the approximation of $\mathcal{K}_{n}$. In the latter expression $\braket{\phi^{2}}$ would enter as a parameter and therefore we would have obtained a set of self-consistent equations for $\braket{\phi^{2}}$ that have to be solved.
Such an approximation scheme has the nice feature of going beyond the perturbative approach; moreover through the generalization of the functional equations presented in Section \ref{dynamic} it can be also applied to study the time evolution of local states.
Unfortunately, this method has two main drawbacks: the self consistency equations for $\braket{\phi^{2}}$ will become more and more involved as soon as we proceed further in perturbation theory in the $\tilde{a}_{\vec{k}}$ basis. Moreover, the cluster expansion will be expressed in terms of $\tilde{a}_{\vec{k}}$: if, by chance, we need the cluster expansion in terms of $a_{\vec{k}}$ (for example, in Section \ref{secworkstat} we focus on quenches from an interacting to a free Hamiltonian and the time evolution is trivial in terms of the $a_{\vec{k}}$ operators) it is not an easy task to switch from one language to another.

\item \textbf{Variational approach:}

The above methods use the already found equations. Here instead we want to describe a way to derive new equations from a variational principle.
The ground state is the state of minimum energy, therefore it is the state that minimizes the expectation value $\braket{H}$. As explained in standard books \cite{Sakurai}, one way to approximate the ground state is to minimize the energy over a selected set of  trial states. Within these trial states, the state with the lowest energy is the best approximation for the ground state.
At this point, we can use our knowledge about the generic properties of a  ground state: it is a local quantum state and we know the form of its cluster expansion. Let $\ket{\Psi}$ be a generic local quantum state with cluster expansion (\ref{intro0}); the ground state $\ket{G}$ is identified by the set of functional equations:
\begin{equation}
\frac{\delta \braket{H}_{\Psi}}{\delta W_{n}^{c}(\vec{x}_{1},..,\vec{x}_{n})}=0\label{vareq}
\end{equation}

In the above, $\braket{H}_{\Psi}$ is the expectation value of the Hamiltonian evaluated on the local quantum state $\ket{\Psi}$, the derivative is a functional derivative where $W_{n}^{c}$ and $[W_{n}^{c}]^{*}$ must be thought of as independent quantities. Equations (\ref{vareq}) (that can be written also in momentum space with $\mathcal{K}_{n}$ as well) express only the stationarity of the energy with respect to variations of the state: if more solutions are present, in order to find the ground state we should keep the solution with the lowest energy.
So far if we were able to perform the above procedure, we would obtain the exact ground state, since it belongs to the set of local quantum states: the approximation comes out because we cannot compute $\braket{H}_{\Psi}$ in an exact way.
For example, we can focus on the general Hamiltonian (\ref{perturbh}): computing its average reduces to computing multipoint correlators of $a_{\vec{k}}$ and $a_{\vec{k}}^{\dagger}$, that we can compute only through a graphical expansion. Therefore the best we can do is to approximate $\braket{H}_{\Psi}$ through a truncation of the set of its graphs.
Actually, there could be different meaningful truncation schemes; the easiest one we can imagine is to implement corrections to the gaussian state: if the gaussian state is a good approximation, then $\mathcal{K}_{n>2}$ must be small. Using the rules of Appendix \ref{grtoolperturb} we can organize the graphs of $\braket{H}_{\Psi}$ in terms of how many $\mathcal{K}_{n>2}$ vertices they contain and truncate the expansion. Notice that we can exactly take into account the two point correlators through the resummation of one particle irreducible graphs: of course, the sum of irreducible graphs appearing in (\ref{irrgaussian}) must be truncated as well.
Using this approximation for $\braket{H}_{\Psi}$ we would find an approximation to the equations (\ref{vareq}): once we find the solution, we should check self consistently that $\mathcal{K}_{n>2}$ are actually small, as well as the neglected terms in $\braket{H}_{\Psi}$.
Such a method has the appealing feature that it can give non perturbative information; moreover if by chance we know some further properties of $\mathcal{K}_{n}$ we can use them from the beginning in our trial local state. As a drawback we must understand the proper way to truncate the expansion $\braket{H}_{\Psi}$ and to solve the set of complicated equations coming from its variation.
Another appealing fact is that such a variational procedure can be applied also to systems whose ground state does not have a truly cluster expansion, but the latter represents only a faithful thermodynamic description of it as we explained at the end of Section \ref{LQS}:  such a variational method could improve our knowledge of the ground states of many interesting condensed matter systems, for example the interacting Bose gas \cite{castin}.

\end{enumerate}

\section{Dynamics of the cluster expansion and application to quenches}
\label{dynamic}

So far we have studied only static properties of local quantum states; we shall now turn to the study of dynamics. 
In this section we ask ourselves what happens to a system that evolves with a local Hamiltonian, provided it is initialized in a local quantum state, which is the case in many out of equilibrium situations. 

More specifically, we consider a thermodynamically large closed system lying on the ground state $|\Psi\rangle$ of some local Hamiltonian $H_0$. At time $t=0$ the Hamiltonian changes abruptly from $H_0$ to $H$ and the system is let to evolve for a long time under the new Hamiltonian. This process has been termed ``quantum quench'' \cite{cc06}. Even though the system remains always in a pure state, for large times expectation values of local observables may become stationary. This is a consequence of the interference of incoherent quasiparticles produced at the time of the quench and coming from distant points. Any finite subsystem of the infinite system exhibits such stationary behaviour, meaning that the reduced density matrix of any subsystem becomes stationary. The complement of the subsystem can therefore be seen as a `bath' in contact to the subsystem, which drives it to equilibrium. 

Such equilibration is not however necessarily thermal. If the evolution is constrained by extra conserved quantities, apart from the total energy, then the system is conjectured to equilibrate to a `Generalised Gibbs Ensemble' (GGE) \cite{thr1a} which maximises the entropy subject to all conserved quantities \cite{GGEold1,GGEold2}. This is the case when the evolution is done under an integrable Hamiltonian, which is characterised by an infinite set of local conserved charges. Free (i.e. Gaussian) Hamiltonians are the simplest examples of integrable models. Non-trivial integrable models are the one-dimensional Bethe Ansatz solvable models. The lack of thermalisation in a one-dimensional integrable system was demonstrated in a seminar experiment in ultra-cold atoms \cite{exp1}, while the experimental observation of a GGE was recently achieved \cite{exp8}. The GGE hypothesis has been successfully tested in a large number of quenches with integrable evolution, while in some cases `quasi-local' conserved charges \cite{Prosen1,Prosen2,Panfil} must be included apart from the local ones, in order for the GGE to describe correctly the large time limit \cite{GGEfail1,GGEfail2,thr6}.

The role of clustering properties of the initial state in equilibration has been first recognised in \cite{thr3,thr3b} (see also the more recent \cite{FE13,thr3c}) in the context of lattice systems, where it is shown that the system equilibrates locally to a Gaussian ensemble. Cluster expansions have been applied to quantum quenches in continuous models in \cite{SotiriadisCalabrese,SM15} where it is shown that any quench in a bosonic field theory (both in one and in higher dimensions) from any initial Hamiltonian to a massive free Hamiltonian leads to the GGE, as a consequence of the cluster decomposition property of the initial state. This is not so in the case of massless evolution and interacting initial Hamiltonian in one dimension \cite{S15} where more memory of the initial state survives at large times. Recently, the cluster property has been used to rigorously study the relaxation of a  general lattice system after a quantum quench and the locality properties of the charges that should be included in the GGE \cite{Doyon}.
Numerical studies based on cluster expansions have also been fruitful, even in the case of evolution under interacting Hamiltonians \cite{Rigol}. 

In most of the above studies the calculation is done at the level of local observables. Here instead we present an analysis of the evolution of the quantum state itself. We derive time-dependent equations for the cluster amplitudes for the time evolution under a general Hamiltonian and solve them in the case of a Gaussian Hamiltonian before and after the quench. The starting point of our general analysis is to understand what happens with the locality-related Properties 1, 2 and 3 of Section \ref{thprop} when we evolve the system with an Hamiltonian $H$ that is supposed to have local or short range interactions. As we will show, evolution under such Hamiltonians preserves the local character of local quantum states. Before we present the mathematical argument, we first give a physically intuitive picture.
We can start analyzing Property 2, i.e. the cluster decomposition property. At the initial time, points at large distance to each other are completely uncorrelated.
Once we start to evolve the system, excitations will start to travel from one point to another, eventually building up correlations between them. It is plausible that far distant points require a large time to be correlated, since we must wait until an excitation starting from one point reaches the other one. This should be true if the interaction is short ranged: long range Hamiltonians correlate distant points instantaneously. 
This statement can be made more precise if the evolution has a light-cone form, which is certainly true in relativistic field theories but also in short-range lattice models (Lieb-Robinson bound \cite{LiebRobinson}) (Figure \ref{lightcone}). When this reasoning is correct, it means that two points at infinite distance cannot be correlated at any finite time, therefore the cluster property is preserved during the time evolution. Note that this has already been established in \cite{Doyon} (Theorem 6.3) where it is shown that the cluster property of a quantum state is preserved by the evolution under a local Hamiltonian. We will see below how we can naturally reach this conclusion also in our formalism.

For the same reasons, also Property 1 i.e. the finiteness of correlation functions in the thermodynamic limit, should not be spoiled: a point deep in the bulk cannot be affected by what happens at the boundaries placed at infinite distance, at least not at finite time. 
These considerations suggest that a local quantum state, when evolved with a short range Hamiltonian, remains a local quantum state, therefore we expect that:

\begin{figure}
\begin{center}
\includegraphics[scale=0.4]{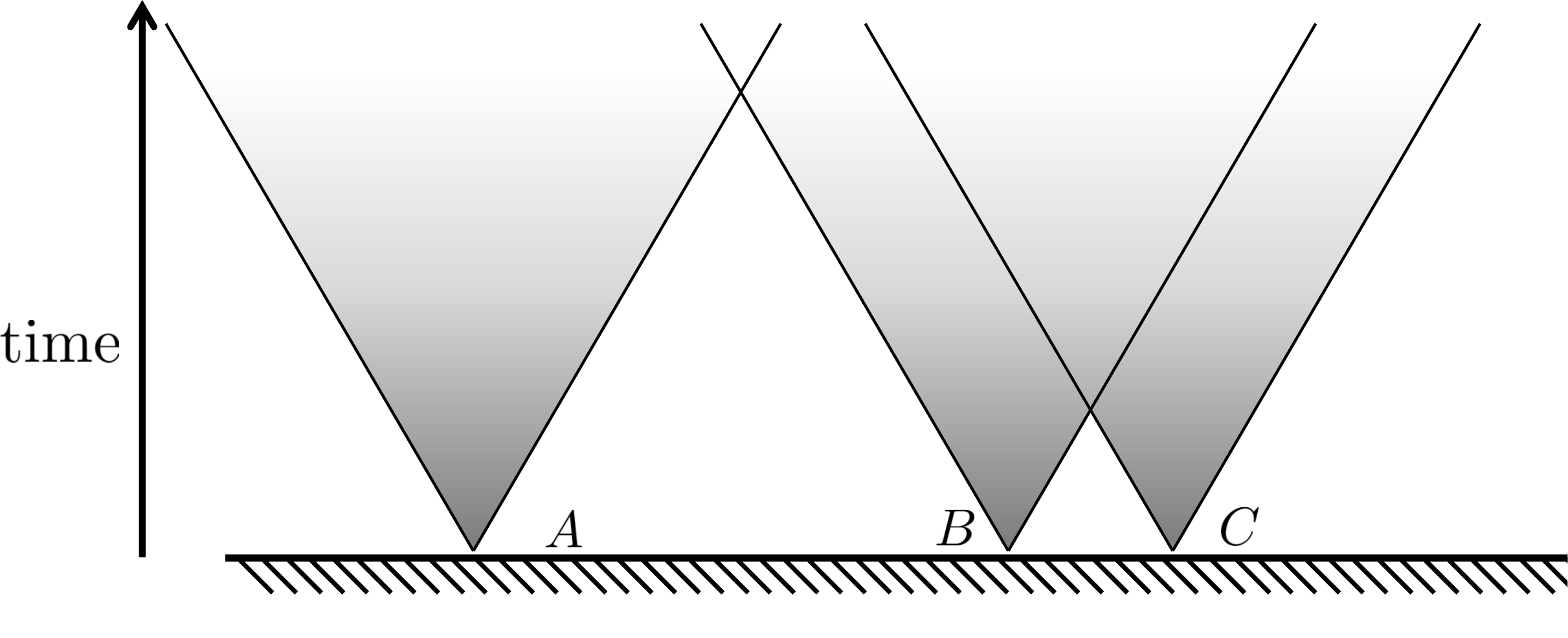}
\caption{Pictorial representation of spreading of correlation from some points in presence of lightcone: $A$ and $B$ require a larger time to become correlated than $B$ and $C$. At any finite time, infinitely distant points remain uncorrelated and cluster property is ensured.}\label{lightcone}
\end{center}
\end{figure}
\begin{equation}
\ket{\Psi_{t}}\equiv e^{-iHt}\ket{\Psi}=e^{L^{d}\mathcal{K}_{0}(t)}\exp\left[\sum_{n=1}\frac{1}{ L^{d(n-2)/2}}\frac{1}{n!}\sum_{\vec{k}_{1}+..+\vec{k}_{n}=0}\mathcal{K}_{n}(\vec{k}_{1},..,\vec{k}_{n},t)a^{\dagger}_{\vec{k}_{1}}..a^{\dagger}_{\vec{k}_{n}}\right]\ket{0} \label{evolvedc}
\end{equation}
In other words, the effect of the evolution is simply to change $\mathcal{K}_{n}$ in time, without spoiling their analytic properties.
Note that the zero-order cluster amplitude $\mathcal{K}_{0}(t)$ must be included, since it is not guaranteed that the time evolution preserves our conventional normalization $\braket{0|\Psi}=1$. The notation is chosen in such a way that we can absorb $\mathcal{K}_{0}$ in the cluster expansion including $n=0$ in the summation: we explicitly put a factor $L^{d}$ because we will see in a while that, with this choice, $\mathcal{K}_{0}$ is $L$ independent.

The above statement, which is based on physical arguments, can be made much more solid and, in order to be quantitative,  we will focus on systems that evolve with an Hamiltonian of the same form (\ref{hnormal}) we used in Section \ref{sectionexacteq}.
Remember that the cluster expansion can be written for any state such that $\braket{0|\Psi_{t}}\ne 0$, therefore (\ref{evolvedc}) can be written without supposing that $\ket{\Psi_{t}}$ is a local quantum state. What we should do is to write the equations that $\mathcal{K}_{n}$ must satisfy during the time evolution and, assuming that the analytic properties of the initial condition are those of a local quantum state, to check whether they are respected also at any later time. 
To do this, we can write the time dependent Schr\"odinger equation:
\begin{equation}
i\partial_{t}\ket{\Psi_{t}}=H\ket{\Psi_{t}}
\end{equation}

We already studied the time independent version of the Schr\"oedinger equation in Section \ref{sectionexacteq}; including time derivatives to obtain the time dependent version is straightforward and with the same steps we arrive at:
\begin{equation}
i\partial_{t}\mathcal{K}_{n}(\vec{k}_{1},..,\vec{k}_{n},t)=\left(\sum_{i=1}^{n}E(\vec{k}_{i})\right)\mathcal{K}_{n}(\vec{k}_{1},..,\vec{k}_{n},t)+\frac{\lambda\tilde{\beta}_{n}}{\prod_{i=1}^{n}\sqrt{2E(\vec{k}_{i})}} +\lambda\hspace{2pt}\mathfrak{F}_{n}(\vec{k}_{1},..,\vec{k}_{n},t)\label{evolutionkn}
\end{equation}

Now $\mathfrak{F}$ is time dependent because the cluster amplitudes $\mathcal{K}_{n}$ that enter in its definition evolve in time. Notice that the time independent equation (\ref{excteq57}) is readily obtained from the above if we look for eigenvectors: since eigenvectors evolve with phases $e^{-iEt}$, therefore we should require from (\ref{evolutionkn}) $\partial_{t}\mathcal{K}_{n>0}=0$ and $\mathcal{K}_{0}=-iE/L^{d}$ and (\ref{excteq57}) is obtained.

Even though these equations are rather involved, we see that at least in the massive case $m\ne 0$, they respect the expected analytical properties of $\mathcal{K}_{n}$ for a local state: if the initial conditions for $\mathcal{K}_{n}$ are non-singular functions, then in the equations of motion do not appear any singularities, which implies that $\mathcal{K}_{n}$ itself has no singularity at any finite time $t$.
Additional care should be taken when we consider the massless case $m=0$ due to the presence of singularities in (\ref{evolutionkn}): also in the static case we had this problem and in Appendix \ref{failure} we see how the singularity disappears, but we do not attempt a similar analysis in the time dependent case.

Notice that equations (\ref{evolutionkn}) are valid even when the Hamiltonian is taken time dependent, simply allowing for time dependent $\tilde{\beta}_{n}$: this has important consequences regarding generic out of equilibrium problems, where the interaction can be varied in time. It does not matter how we change the coupling constants of the theory: the state preserves its form with time dependent $\mathcal{K}_{n}$.

Equations (\ref{evolutionkn}) are non-trivial, but they present two simple special cases that can be explicitly solved:

\begin{itemize}
\item Time evolution governed by the free Hamiltonian ($\lambda=0$). In absence of interaction the equations become trivial and we simply have:
\begin{equation}
\mathcal{K}_{n}(\vec{k}_{1},..,\vec{k}_{n},t)=e^{-it\sum_{j=1}^{n}E(\vec{k}_{j})}\mathcal{K}_{n}(\vec{k}_{1},..,\vec{k}_{n},0)\label{freekn}
\end{equation}

This is obvious, since if we take a generic cluster expansion and let it evolve freely, we simply have to multiply each creation operator by an oscillating phase factor $a^{\dagger}_{\vec{k}}\to e^{-iE(\vec{k})t}a^{\dagger}_{\vec{k}}$: in Schr\"oedinger picture, these phases are attached to the $\mathcal{K}_{n}$ rather than $a^{\dagger}$.

\item Quadratic Hamiltonian and gaussian initial state ($\mathcal{K}_{n\ne2}=0$). In this case the interaction preserves the gaussianity and the equations are exactly the same as (\ref{eqgaussian}) with the addition of the time derivative part:
\begin{equation}
\left(2E+\lambda\frac{\tilde{\beta}_{2}}{E}\right)\mathcal{K}_{2}+\lambda\frac{\tilde{\beta}_{2}}{2E}+\lambda\frac{\tilde{\beta}_{2}}{2E}[\mathcal{K}_{2}]^{2}=i\partial_{t}\mathcal{K}_{2}
\end{equation}
\end{itemize}

where for brevity we suppressed the explicit dependence on time and momenta. If $\tilde{\beta}_{2}$ is chosen time independent, we can rescale $\lambda$ to have $\tilde{\beta}_{2}=1$. The solution can be readily written as:
\begin{equation}
\mathcal{K}_{2}(\vec{k},-\vec{k},t)=-\alpha(\vec{k})+\sqrt{\alpha^{2}(\vec{k})-1}\hspace{3pt}\frac{1+C(\vec{k})e^{-i2\mathcal{E}(\vec{k})t}}{1-C(\vec{k})e^{-i2\mathcal{E}(\vec{k})t}}\label{evolvedk}
\end{equation}

where:
\begin{equation}
\alpha(\vec{k})=1+2\lambda^{-1}[E(\vec{k})]^{2}\hspace{2pc}\mathcal{E}(\vec{k})=\sqrt{\vec{k}^{2}+m^{2}+\lambda}
\end{equation}

\begin{equation}
C(\vec{k})=\frac{\mathcal{K}_{2}(\vec{k},-\vec{k},0)+\alpha(\vec{k})-\sqrt{\alpha^{2}(\vec{k})-1}}{\mathcal{K}_{2}(\vec{k},-\vec{k},0)+\alpha(\vec{k})+\sqrt{\alpha^{2}(\vec{k})-1}}
\end{equation}

Notice that if we search for time independent solutions, we must impose $C(\vec{k})=0$: this ensures $\mathcal{K}_{2}$ is constant in time and it is exactly equal to (\ref{new88}), as it should.

We can briefly comment on the result above, keeping in mind the discussion of Appendix \ref{gaussian}. As we show there, $|\mathcal{K}_{2}|\le1$ is a necessary requirement in order to have a meaningful gaussian initial state.
Note that $\mathcal{E}(\vec{k})$ is the energy of the modes of the full Hamiltonian with $\lambda\ne 0$.
Therefore, in order to have a meaningful theory we should impose $\mathcal{E}(\vec{k})$ to be a real quantity, which implies $\alpha^{2}>1$. 
Then, with this hypothesis, it is a matter of simple algebraic manipulations to show that $|\mathcal{K}_{2}(\vec{k},-\vec{k},0)|\le1$ implies $|C|<1$. 
Knowing that $|C|<1$ we see that $\mathcal{K}_{2}(\vec{k},-\vec{k},t)$ never acquires singularities during the time evolution, and moreover $|\mathcal{K}_{2}(\vec{k},-\vec{k},t)|\le1$, as it should.

It must be said that the evolution of $\mathcal{K}_{2}$ can be obtained also through the method of Appendix \ref{gaussian}: (\ref{gaussiancorrelators}) remains valid also if we replace the initial $\mathcal{K}_{2}$ with the evolved counterpart and the initial correlators $\braket{aa^{\dagger}}_{S}$ with the evolved ones $\braket{aa^{\dagger}}_{S}(t)$. At this point from (\ref{gaussiancorrelators}) we see that we can find $\mathcal{K}_{2}$ through the correlators:

\begin{equation}
\mathcal{K}_{2}(\vec{k},-\vec{k},t)=\frac{\braket{a_{-\vec{k}}a_{\vec{k}}}_{S}(t)}{\braket{a_{\vec{k}}a^{\dagger}_{\vec{k}}}_{S}(t)}
\end{equation}

Then the time dependent correlators can be evaluated using the suitable Bogoliubov transformation (\ref{bfunca}) that diagonalizes the full Hamiltonian and have a trivial evolution. We do not report these calculations since, even though straightforward, they are lengthy, but the result would match (\ref{evolvedk}) as it should.
We will not proceed further in the general discussion of these equations and for the time being we are going to see two possible applications of this method to quench problems.

\subsection{Relaxation to GGE in quenches from interacting to free theories}

As it is clear from the above, the cluster expansion is especially convenient for the study of quench dynamics in which the evolution is done according to a Hamiltonian diagonal in terms of the creation and annihilation operators $a_{\vec{k}}^\dagger,a_{\vec{k}}$. This can be achieved if the evolution Hamiltonian is Gaussian (free) in terms of local fields and we choose to expand the initial state as a cluster expansion in terms of the corresponding creation and annihilation operators. In such an interacting to free quench, our approach gives a particularly simple proof of validity of the GGE conjecture. 

The original GGE conjecture \cite{thr1a} explicitly claims that in thermodynamically large systems and for a large class of initial states $|\Psi\rangle$, including (but not necessarily limited to) ground states of local Hamiltonians, the evolution under an integrable Hamiltonian is such that the large time value of any local observable $\mathcal{O}$ equilibrates to a stationary point, whose value can be computed through a Generalised Gibbs Ensemble density matrix $\rho_{\text{GGE}}$ that maximises the entropy subject to all local conserved charges $Q_n$ that characterise the integrable system. Regarding these charges, it is common a little abuse of notation: we talk about local charges as conserved quantities that can be written as an integral over all the space of some point wise operator, constructed out of the field and its derivatives.
In a more quantitative way, the GGE conjecture says:

\begin{equation}
\mathcal{O}(t\to\infty) := \lim_{t\to\infty} \braket{ \mathcal{O}(t) }_{\Psi} = \text{Tr} \, \{\mathcal{O}(t) \rho_{\text{GGE}}  \}
\end{equation}
where $\rho_{\text{GGE}}$ is defined as
\begin{equation}
\rho_{\text{GGE}} = \frac{e^{-\sum_n \lambda_n Q_n}}{\text{Tr}\{e^{-\sum_n \lambda_n Q_n}\}}
\end{equation}
and the Lagrange multipliers $\lambda_n$ are defined by the condition
\begin{equation}
\text{Tr}\{Q_n \, \rho_{\text{GGE}}  \} = \braket{ Q_n }_{\Psi}\label{GGE105}
\end{equation}
for all the local charges $Q_n$. 
By local observable $\mathcal{O}$ we mean any observable defined within a finite size subsystem, i.e. including also multi-point correlation functions. Thus the GGE conjecture equivalently claims that the reduced density matrix $\rho^{\mathcal{A}}$ of any finite size subsystem $\mathcal{A}$ at $t\to\infty$ is equal to the reduced density matrix corresponding to the GGE
\begin{equation}
\rho^{\mathcal{A}}(t\to\infty) =  \rho_{\text{GGE}}^{\mathcal{A}}.
\end{equation}

Notice that integrable field theories possess an infinite set of local conserved charges, apart from the Hamiltonian itself, therefore the GGE prediction preserves lot of information about the initial state, but it is still economic. As a matter of fact, any quantum system possesses the maximal set of conserved quantities constructed as the projectors on the eigenstates of the Hamiltonian. The size of this set increases exponentially in the  system size, but the GGE conjecture states that only local conserved charges matter in the long time limit: this latter set increases only linearly in the system size.
Locality of the charges is a desirable feature, since statistical ensembles are expected to satisfy locality requirements which are linked, as we explained, to the extensivity of thermodynamic quantities, but using only these conserved quantities in the construction of the GGE has been proven to be a too tight constraint: \emph{quasi local charges} must also be used \cite{Panfil}.
Quasi local charges are a relaxation of the locality constraint, that is to say such charges, instead of being integrals of a point wise operator, are constructed from short range operators integrated on the whole space.
The interacting-to-free quench is the simplest situation in which the post-quench Hamiltonian is integrable and constitutes an important benchmark both to test the GGE prediction  itself, but also to study the locality properties of the conserved charges that enter in its construction.
It has been proven in \cite{SotiriadisCalabrese,SM15} that in relativistic quenches in free theories, the expectation values of local observables indeed equilibrates and the right GGE density matrix can be written in terms of the occupation modes $a_{\vec{k}}^{\dagger}a_{\vec{k}}$:

\begin{equation}
\rho_{\text{GGE}} = \frac{e^{-\int d^dk \, \beta(k) E(\vec{k})a^{\dagger}_{\vec{k}}a_{\vec{k}}}}{\text{Tr}\{e^{-\int d^dk \, \beta(k) E(\vec{k}) a^{\dagger}_{\vec{k}}a_{\vec{k}}}\}}
\label{gge2}
\end{equation}

The GGE density matrix is gaussian and closely resembles a thermal ensemble, but each mode equilibrates at a different, momentum dependent, temperature. In this case, (\ref{GGE105}) is rephrased as:
\begin{equation}
\text{Tr}\{a^{\dagger}_{\vec{k}}a_{\vec{k}} \, \rho_{\text{GGE}}  \} = \braket{ a_{\vec{k}}^{\dagger}a_{\vec{k}}}_{\Psi}
\label{gge3}
\end{equation}

This result can be rephrased as follows: computing large time correlators of the relativistic field $\phi$ and its conjugate field $\pi$, only the two point connected correlators survive and, once we have expressed the latter in terms of the mode operators, only $\braket{a^{\dagger}_{\vec{k}}a_{\vec{k}}}$  gives a non vanishing contribution to the observable. The key information that was used in \cite{SotiriadisCalabrese,SM15} to obtain this result, is that the initial state respects the cluster property: this assumption was used to keep under control the analytic properties of the correlator in momentum space, similarly to what we will see in a while.
However long calculations were necessary in order to use directly the $\phi$ and $\pi$ operators, while we know that the cluster property, in a massive theory, can be expressed in terms of $\psi$ and $\psi^{\dagger}$ as well. Testing the GGE prediction on these operators is very simple, consider for example the connected two point correlator function for the field $\psi$: since the evolution is free, we can obtain the evolved fields in Heisenberg representation simply replacing $a_{\vec{k}}\to e^{-iE(\vec{k})t}a_{\vec{k}}$.
\begin{equation}
\braket{\psi(\vec{x},t)\psi(\vec{y},t)}_{c}=\int \frac{d^{d}k}{(2\pi)^{d}}e^{-i\vec{k}(\vec{x}-\vec{y})}e^{-2iE(\vec{k})t}\braket{a_{\vec{k}}a_{-\vec{k}}}_{c}\label{GGE109}
\end{equation}

At this point in the limit $t\rightarrow\infty$ we can apply the stationary phase method: if $\braket{a_{\vec{k}}a_{-\vec{k}}}_{c}$ is smooth enough, then the integral becomes zero at $t\to\infty$.
In this context, smooth enough means that we require that $\braket{a_{\vec{k}}a_{-\vec{k}}}_{c}$ and the other higher multipoint connected correlators have not extra $\delta$ singularities: if we would had a delta placed somewhere, then even at large times we would retain an oscillating time dependent term.
As long as the cluster property for the $\psi$ and $\psi^{\dagger}$ fields works, multipoint correlators in the momentum space have only one $\delta$ singularity on the overall momentum, due to translational invariance in the coordinate space.
In principle, from the cluster property (Property 2), other singularities in the momentum space are allowed, but they must be mild enough to have that, transforming back in the coordinate space, the correlator is well-defined in the infinite size limit (Property 1). Having under control the singularities, we can straightforwardly apply the stationary phase argument and (\ref{GGE109}) vanishes. 
With the same steps, it is easy to see that all the higher multipoint connected correlators vanish when $t\to\infty$, instead one out of the two point correlators and the one point correlators survive:

\begin{equation}
\braket{\psi^{\dagger}(\vec{x},t)\psi(\vec{y},t)}_{c}=\int \frac{d^{d}k}{(2\pi)^{d}}e^{i\vec{k}(\vec{x}-\vec{y})}\braket{a^{\dagger}_{\vec{k}}a_{\vec{k}}}_{c}\label{GGE110}
\end{equation}

\begin{equation}
\braket{\psi(\vec{x},t)}_{c}=e^{iE(0)t}L^{-d/2}\braket{a_{0}}\hspace{3pc}\braket{\psi^{\dagger}(\vec{x},t)}_{c}=e^{-iE(0)t}L^{-d/2}\braket{a^{\dagger}_{0}}\label{GGE111}
\end{equation}

Notice that from our rules of Section \ref{grtool} $\braket{a_{0}}$ has embedded in it a $L^{d/2}$ factor, therefore $\braket{\psi}$ is actually $L$ independent, as it should. 
On these correlators we cannot apply the stationary phase argument: the delta structure in momentum space of (\ref{GGE110}) makes it time independent, instead in (\ref{GGE111}) we have no momentum to integrate over.
If we have $\braket{\psi}=0$ because of symmetries of the initial state (for example if $\ket{\Psi}$ is the ground state of the $\phi^{4}$ model this is the case), then the long time behavior of local observables constructed with $\psi$ and $\psi^{\dagger}$ can be completely described in terms of a time independent gaussian ensemble ruled by the two point correlator in the momentum space, i.e. the GGE construction applied to the free case.
Actually, there is a subtlety when we study relativistic field theories: in these theories, the $\psi$ field is somewhat artificial and the real physical fields are the bosonic field $\phi$ and its conjugated momentum. As we stated at the beginning of Section \ref{perturb}, we can switch from the $\psi$ field to $\phi$ with (\ref{kernel}): as long as $m\ne 0$ it is easy to show that cluster property for $\psi$ and $\psi^{\dagger}$ holds if and only if it holds for $\phi$ and the conjugated momentum, this is thanks to the exponential damping of the kernel in  (\ref{kernel}).
This implies that, as long as we require the cluster property on the physical fields, we have it on $\psi$ and $\psi^{\dagger}$, therefore we have the same control on the singularities of the correlators in the momentum space.
At this point, we can study the long time limit of  connected correlators of the field $\phi$ exactly as we have just done for the $\psi$ fields. Their expression, thanks to (\ref{newperturb39}), can be written as the Fourier transform of multipoint correlators of $a_{\vec{k}}$, with some additional $E(\vec{k})$ factors, but these in the massive case do not introduce any additional singularity and the stationary phase argument can be applied.
This is no longer true in the massless case, but also the fact that cluster property of the field $\phi$ guarantees cluster property for $\psi$ does not hold anymore, since the kernel in (\ref{kernel}) decays too slowly to guarantee it. It can happen that, even if we have the cluster property on the physical field $\phi$, this is spoiled on the $\psi$ fields: therefore the argument we used before to study the $\delta$ singularities is spoiled and we do not expect the GGE to describe the large time behavior.
This is consistent with \cite{SotiriadisCalabrese}, where relativistic field theories are studied and the fact of having massive excitations was crucial to have the GGE prediction realized.

\subsection{Spectral expansion and work statistics}
\label{secworkstat}

Among many other possible applications of the cluster expansion to quantum quench problems, a simple and straightforward application is the computation of the work statistics. The work statistics i.e. the probability distribution of the work done in the system by a quantum quench, is a quantity of central interest for several reasons. Firstly, the study of work statistics is important for the verification of quantum fluctuation relations which constitute the thermodynamic laws that govern non-equilibrium physics \cite{QFR}. 
Second, it turns out to be related to other independently studied quantities like the Loschmidt echo and the Casimir force \cite{Silva, SotiriadisGambassiSilva}. 
Through its relation to the latter quantity it can be shown that it exhibits universal characteristics \cite{SotiriadisGambassiSilva,Palmai}. 
Third, it can be used to probe the spectral properties of the pre-quench and post-quench system. Indeed the work statistics is nothing but the probabilities of the transitions from the initial state to any possible post-quench excitation. As such its general properties can be easily deduced. The minimum work level corresponds to the transition from the pre-quench ground state to the post-quench ground state, which appears as an isolated $\delta$-peak. Above it and starting from a threshold related to the energy gap of the lowest possible excitations there exists a series of other peaks or edge singularities that correspond to any isolated bound states or multi-particle excitations respectively that are present in the post-quench system \cite{SotiriadisGambassiSilva}. From the type of the edge singularity we can extract information about the nature of the quasiparticles that describe the corresponding excitations \cite{SotiriadisPalmai,Palmai}. 
By classifying the various excitations in increasing order of particle number, the cluster expansion provides directly a spectral decomposition of the transition channels and a characterisation of the corresponding peaks and edge singularities of the work statistics. 

In general, supposing our system is initialized in a ground state $\ket{G}$ of some physical Hamiltonian, the work statistics $P(W)$ is simply defined as:
\begin{equation}
P(W)=\sum_{n}\delta(W-(E_{n}-E_{G}))|\braket{\psi_{n}|G}|^{2}\mathcal{N}_{G}^{-1}\label{defwork}
\end{equation}

Where $\ket{\psi_{n}}$ are the eigenstates of the post-quench Hamiltonian $H$ with eigenvalues $E_{n}$, $E_{G}$ is the energy of the pre-quench state. We introduce explicitly the norm $\mathcal{N}_{G}=\braket{G|G}$ to match the notation used so far for cluster expansions.
Actually, instead of computing $P(W)$, it is often easier to compute its generating function $g(t)$:
\begin{equation}
g(t)=\int_{-\infty}^{\infty} dW \hspace{3pt}e^{-iWt}P(W)\label{generating}
\end{equation}

Once we have it, we can compute the moments $\braket{W^{n}}$ or the cumulants $\kappa_{n}$:
\begin{equation}
g(t)=1+\sum_{n=1}^{\infty}\frac{(-it)^{n}}{n!}\braket{W^{n}}\quad ,\hspace{3pc}\log g(t)=\sum_{n=1}^{\infty}\frac{(-it)^{n}}{n!}\kappa_{n}\label{cumulants}
\end{equation}

Notice that once we know all the cumulants we can reconstruct the moments, simply comparing the analytic expansion of the two expressions above. In particular, we will focus on the computation of cumulants.

The appealing feature of $g(t)$ is that it acquires the physical meaning of a particular overlap, called Loschmidt echo \cite{Silva, SotiriadisGambassiSilva}. Inserting (\ref{defwork}) in (\ref{generating}) we simply get:
\begin{equation}
g(t)=\sum_{n}e^{-it(E_{n}-E_{G})}|\braket{\psi_{n}|G}|^{2}\mathcal{N}_{G}^{-1}=e^{it E_{G}}\bra{G}e^{-itH}\ket{G}\mathcal{N}_{G}^{-1}=\frac{\bra{G}e^{itH_{0}}e^{-itH}\ket{G}}{\braket{G|G}}\label{overlap62}
\end{equation}
The Loschmidt Echo for a state $\ket{G}$ is defined to be the last quantity in (\ref{overlap62}), where $H_{0}$ is the pre-quench Hamiltonian. 
From the above, we can trivially notice that the momenta of the work distribution are associated with the expectation value of the powers of the post quench Hamiltonian on the initial state:
\begin{equation}
\braket{W^{n}}=\frac{\braket{G|(H-E_{G})^{n}|G}}{\braket{G|G}}\label{newsp118}
\end{equation}
Of course, the cumulants $\kappa_{n}$ are nothing else than the connected part of $\braket{G|(H-E_{G})^n|G}_{G}$. Momenta and cumulants can be computed with standard perturbative methods, since they are simply expectation values of some observables (powers of the post quench Hamiltonian). Nevertheless through the cluster expansion we can directly compute the generating  function $g(t)$, since it can be expressed as an overlap of local quantum states (\ref{overlap62}).
The logarithm of $g(t)$ is often more convenient for computations:
\begin{equation}
\log g(t)=itE_{G}+\log \left(\bra{G}e^{-iHt}\ket{G}\right)-\log\braket{G|G}\label{loggt}
\end{equation}

The method of Section \ref{grtool} is pretty suitable for evaluating the above expression. 
If $\ket{G}$ has the structure of a local quantum state and we define $\ket{G_{t}}=e^{-iHt}\ket{G}$, from the discussion of Section \ref{dynamic} we expect also $\ket{G_{t}}$ to be a local quantum state, whose cluster amplitudes are changing in time. Therefore, we can also apply our graphical tool to evaluate the logarithm of the overlap $\log\braket{G|G_{t}}$, as we explained at the end of Section \ref{grtool}, once we know the proper cluster expansion. There are different possible ways to compute the ground state energy, for example we can use (\ref{newGenergy}). Notice that  $\log g(t)$ is certainly an extensive quantity: this implies that all cumulants of the energy are extensive, as expected from the discussion of Section \ref{thprop}.

To simplify our presentation we perform a computation in the simpler case in which $H$ is free: in this case the $\mathcal{K}_{n}$ factors evolve with simple phases (\ref{freekn}) and the problem reduces to determining the right initial conditions and  dealing with our graphical expansion.
In this section we will suppose the state initialized as the ground state of an Hamiltonian like (\ref{hnormal}), in particular we focus on the exactly solvable case of a quadratic Hamiltonian and then on the prototypical example of $\phi^4$ theory.
Actually, it is quite easy to have exact information about $P(W)$ when $W$ is tuned very close to $-E_{G}$, simply inserting the lowest eigenstates of the free Hamiltonian, constructed acting with $a^{\dagger}$ on the bare vacuum, whose energy is chosen as reference energy (therefore zero).
Increasing $W$ we have no contribution to (\ref{defwork}) until we reach the energy $-E_{G}$: this is simply the gap between the pre-quench ground state and the post-quench ground state. In both cases, when $0<W+E_{G}<2m$ we cannot excite any particle, but as soon as $W+E_{G}$ reaches the threshold $2m$ a pair of opposite momentum excitations can be produced and $P(W)$ is no longer zero. Then another threshold is reached at $W+E_{G}=4m$: in this case we can produce two such pairs or, in the $\phi^{4}$ theory for example, also a cluster of $4$ particles represented by $\mathcal{K}_{4}$. Then other thresholds are reached for any even multiple of the particle mass. As long as the odd amplitudes $\mathcal{K}$ vanish, these are the only allowed thresholds; otherwise other thresholds for odd multiples of $m$ arise too.

Explicitly expressing (\ref{defwork}) in terms of the eigenstates of the free theory we have:
\begin{equation}
P(W)=\delta(W+E_{G})\mathcal{N}_{G}^{-1}|\braket{0|G}|^{2}+\sum_{l}\sum_{\vec{k}_{1},..,\vec{k}_{l}}\delta\left(W+E_{G}-\sum_{j=1}^{l}E(\vec{k}_{j})\right)\frac{\mathcal{N}_{G}^{-1}}{l!}|\bra{0}\prod_{j=1}^{l}a_{\vec{k}_{j}}\ket{G}|^{2}\label{sumwork}
\end{equation}

Notice that, since the energy of each particle is at least $m$, we can keep only the terms such that $lm\le W+E_{G}$. 
The overlaps that appear are simply the wavefunctions with $l$ particles, whose connected part is $\mathcal{K}_{l}$ (\ref{knconncorr}): therefore, a truncation in the sum of (\ref{sumwork}) up to $l$, implies that only $\mathcal{K}_{n}$ with $n\le l$ will enter in the computation of the work statistics. Therefore they do not enter as an infinite series as it happens for correlators, but as a finite sum, even though rather involved. 
In principle, the lowest terms of $P(W)$ can be easily calculated once we know $\mathcal{K}_{n}$.
For example, if the odd cluster amplitudes are zero as it happens for the ground state of the $\phi^{4}$ model, the contribution up to $W+E_{G}<4m$ is:
\begin{equation}
P(W)=\delta(W+E_{G})\mathcal{N}_{G}^{-1}+\frac{\mathcal{N}_{G}^{-1}}{2}\sum_{\vec{k}}\delta\left(W+E_{G}-2E(\vec{k})\right)\left|\mathcal{K}_{2}(\vec{k},-\vec{k})\right|^{2}\label{thr117}
\end{equation}

From the above expression it is easy to study the behavior of the work statistics at the first threshold $W+E_{G}=2m$: in this case, only the value of $\mathcal{K}_{2}$ (supposed to be not zero) at zero momentum matters. Once we replaced the summation with the proper integration, we easily find:
\begin{equation}
P(W)\sim\frac{\mathcal{N}_{G}^{-1}}{4}\frac{L^{d}m^{\frac{d}{2}}\Omega_{d}}{(2\pi)^{d}}\left|\mathcal{K}_{2}(0,0)\right|^{2}\Theta(W+E_{G}-2m)\left(W+E_{G}-2m\right)^{\frac{d}{2}-1}\label{thr118}
\end{equation}
above, $\Theta$ is the Heaviside Theta function, $\Omega_{d}$ is the $d$ dimensional solid angle. It must be noticed that this and the other thresholds are immediately washed away by the thermodynamic limit, since $\mathcal{N}_{G}$ is exponentially divergent in the system size.
We can start with the simple case of a quadratic pre-quench Hamiltonian:
\begin{equation}
H= \sum_{\vec{k}} \left [ E(\vec{k})a^{\dagger}_{\vec{k}}a_{\vec{k}}+ \frac{\lambda}{4E(\vec{k})} \left(a^{\dagger}_{\vec{k}}a^{\dagger}_{-\vec{k}}+a_{\vec{k}}a_{-\vec{k}}+2a^{\dagger}_{\vec{k}}a_{\vec{k}}\right) \right ] + \lambda L^{d}\tilde{\beta}_{0};\hspace{2pc}
\tilde{\beta}_{0}=\int\frac{d^{d}k}{(2\pi)^{d}}\frac{1}{4E(\vec{k})}
\end{equation}

The post-quench free Hamiltonian is simply obtained setting $\lambda=0$. As reported in Section \ref{grexacteqsec} and Appendix \ref{gaussian}, the initial state is a gaussian state with $\mathcal{K}_{2}$ given by (\ref{new88}): it remains only to apply our graphical rules.
This case has already been studied in literature with other methods \cite{SotiriadisGambassiSilva}, but nevertheless it is interesting to derive it again in our framework and use it as a check.

Since only $\mathcal{K}_{2}$ is non-zero, the graphs that contribute to $\log\braket{G|G_{t}}$ are simply closed chains of $2n$ vertices, $n$ vertices of the ket and $n$ of the bra, as shown in Figure \ref{dihedral}. Using the rules of Section \ref{grtool} we have that such a graph contributes as:

\begin{figure}
\begin{center}
\includegraphics[scale=0.25]{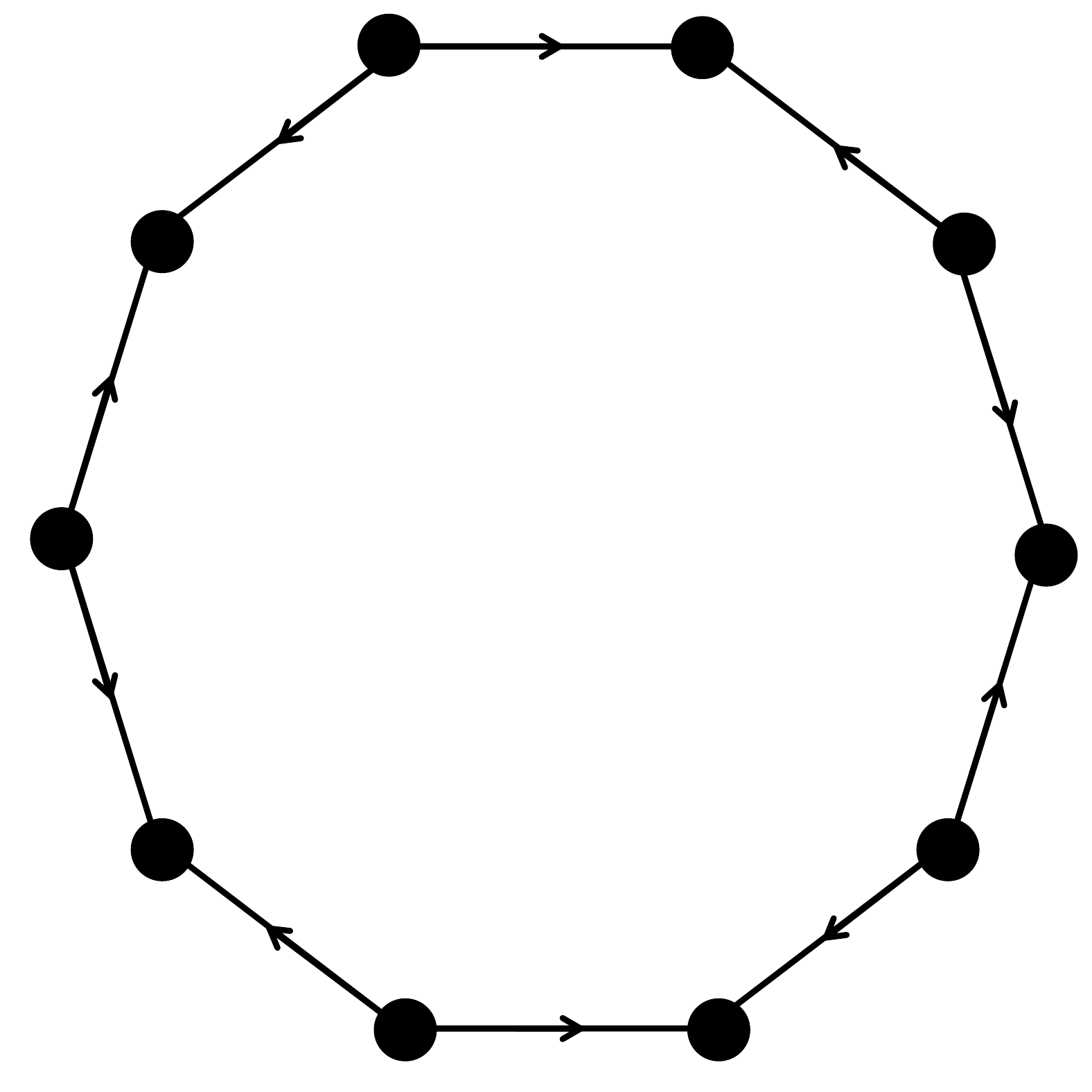}
\caption{Graphs for $\log\braket{G|G_{t}}$ when a squeezed state is considered: they are made by closed chains of $2n$ vertices, the permutations that leave the graph unchanged are given by the dihedral group $D_{n}$ that counts $2n$ elements.}\label{dihedral}
\end{center}
\end{figure}

 \begin{equation}
S_{2n}L^{d}\int \frac{d^{d}k}{(2\pi)^{d}}e^{-i2ntE(\vec{k})}|\mathcal{K}_{2}(\vec{k},-\vec{k})|^{2n}
\end{equation} 

With $S_{2n}$ the proper symmetry factor. Looking at Figure \ref{dihedral} we can easily see that a graph with $2n$ vertices has $2n$ equivalent permutations, therefore we can write:
\begin{equation}
\log\braket{G|G_{t}}=L^{d}\sum_{n=1}^{\infty}\frac{1}{2n}\int \frac{d^{d}k}{(2\pi)^{d}}e^{-i2ntE(\vec{k})}|\mathcal{K}_{2}(\vec{k},-\vec{k})|^{2n}
\end{equation}

Since we have $|\mathcal{K}_{2}|\le 1$ we can sum the series:
\begin{equation}
\log\braket{G|G_{t}}=-\frac{L^{d}}{2}\int \frac{d^{d}k}{(2\pi)^{d}}\log\left(1-e^{-i2tE(\vec{k})}|\mathcal{K}_{2}(\vec{k},-\vec{k})|^{2}\right)\label{workstatgaussian}
\end{equation}

Then, inserting this expression in (\ref{loggt}) we get:
\begin{equation}
\log g(t)=itE_{G}-\frac{L^{d}}{2}\int \frac{d^{d}k}{(2\pi)^{d}}\log\left(\frac{1-e^{-i2tE(\vec{k})}|\mathcal{K}_{2}(\vec{k},-\vec{k})|^{2}}{1-|\mathcal{K}_{2}(\vec{k},-\vec{k})|^{2}}\right)
\end{equation}

The last  step is to compute $E_{G}$. There are many standard ways of doing this, but it is rather simple to obtain it from (\ref{newGenergy}). Once we evaluate correctly $\mathfrak{F}_{0}$ with the rules of Section \ref{sectionexacteq} we get:

\begin{equation}
\frac{{E}_{G}}{L^{d}}=\lambda \tilde{\beta}_{0}+\frac{\lambda}{2}\int\frac{d^{d}k}{(2\pi)^{d}}\frac{\mathcal{K}_{2}(\vec{k},-\vec{k})}{2E(\vec{k})}
\end{equation}

We have now the exact expression of $\log g(t)$, once we insert $\mathcal{K}_{2}$ from (\ref{new88}) of Section \ref{grexacteqsec}. The expression for the Loschmidt echo agrees with the result of \cite{SotiriadisGambassiSilva}, as it should.

Now that we exactly solved the gaussian case, we can focus on the $\phi^4$ model: here the situation is much more complicated since, unlike before, the cluster amplitudes $\mathcal{K}_{n}$ are non-vanishing for any even $n$ and this makes impossible to explicitly sum all the graphs as we did before. 
To obtain an approximate result we can perform a naive perturbative expansion in $\lambda$, using the results (\ref{k4}) and (\ref{k2}) that tell us $\mathcal{K}_{2,4}\sim \lambda$. With this information, the lowest order contributions in $\lambda$ to $\log\braket{G|G_{t}}$ are given by the two graphs of Figure \ref{overlap1ord} and they are $\mathcal{O}(\lambda^{2})$:

\begin{figure}
\begin{center}
\includegraphics[scale=0.3]{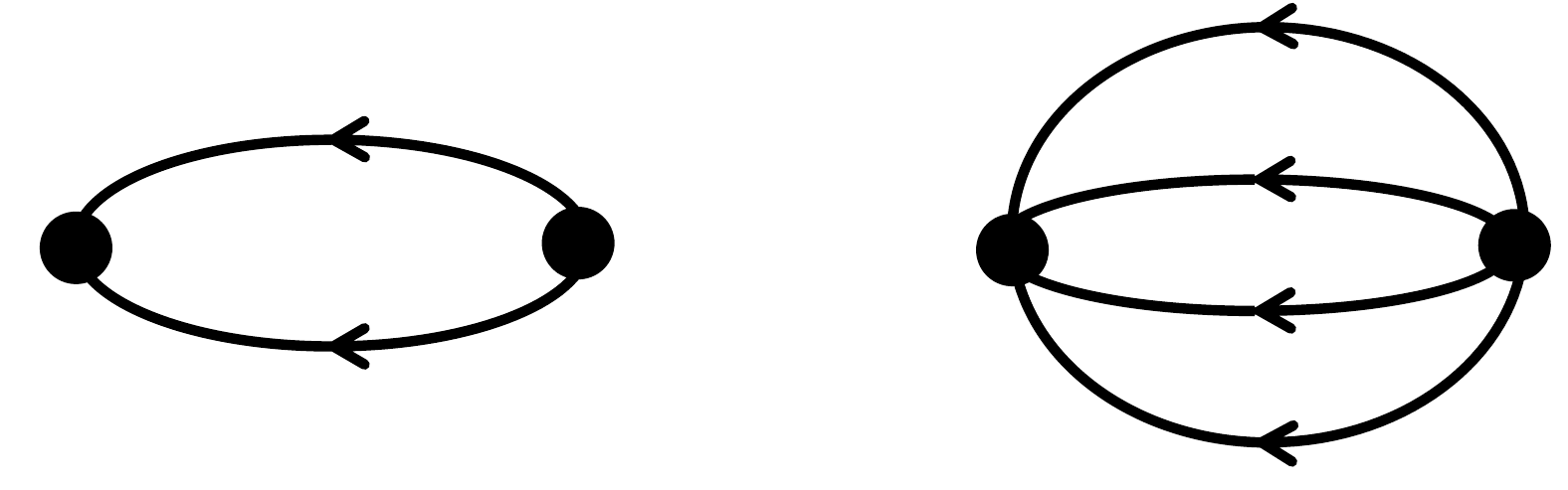}
\caption{Lowest order contributions in naive perturbation theory to $\log\braket{G|G_{t}}$ for the $\phi^4$ model: these graphs are $\mathcal{O}(\lambda^{2})$ in naive perturbation theory.}\label{overlap1ord}
\end{center}
\end{figure}
\begin{eqnarray}
\nonumber&& \log\braket{G|G_{t}}=\frac{L^{d}}{2}\int\frac{d^{d}k}{(2\pi)^{d}}e^{-2itE(\vec{k})}|\mathcal{K}_{2}(\vec{k},-\vec{k})|^{2}+\\
&&+\frac{L^{d}(2\pi)^{d}}{4!}\int \prod_{i=1}^{4}\frac{d^{d}k_{i}}{(2\pi)^{d}}\delta\left(\sum_{i=1}^{4}\vec{k}_{i}\right)e^{-it\sum_{i=1}^{4}E(\vec{k}_{i})}|\mathcal{K}_{4}(\vec{k}_{1},\vec{k}_{2},\vec{k}_{3},\vec{k}_{4})|^{2}+\mathcal{O}(\lambda^{3})
\end{eqnarray}

Consistently, also $E_{G}$ must be evaluated at the same order. Using (\ref{newGenergy}) and keeping only $\mathcal{O}(\lambda^{2})$ in $\mathfrak{F}_{0}$ we get:
\begin{equation}
\frac{E_{g}}{L^{d}}=\lambda\tilde{\beta}_{0}+\frac{\lambda\tilde{\beta}_{2}}{2}\int\frac{d^{d}k}{(2\pi)^{d}}\frac{\mathcal{K}_{2}(\vec{k},-\vec{k})}{2E(\vec{k})}+\frac{\lambda\tilde{\beta}_{4}(2\pi)^{d}}{4!}\int \prod_{i=1}^{4}\frac{d^{d}k_{i}}{(2\pi)^{d}}\delta\left(\sum_{i=1}^{4}\vec{k}_{i}\right)\frac{\mathcal{K}_{4}(\vec{k}_{1},\vec{k}_{2},\vec{k}_{3},\vec{k}_{4})}{\prod_{i=1}^{4}\sqrt{2E(\vec{k}_{i})}}+\mathcal{O}(\lambda^{3})
\end{equation}

The expression for $\tilde{\beta}_{0,2,4}$ in the $\phi^{4}$ case can be found in (\ref{u2}). At this point, putting all the pieces together we can find $\log g(t)$ up to order $\mathcal{O}(\lambda^{2})$: the expression is rather long, but notice that through (\ref{cumulants}) we can easily evaluate the cumulants, simply deriving $\log g(t)$ around zero.
\begin{eqnarray}
\nonumber&&\kappa_{n}=-E_{G}\delta_{n,1}+\frac{L^{d}}{2}\int\frac{d^{d}k}{(2\pi)^{d}}\left(2E(\vec{k})\right)^{n}|\mathcal{K}_{2}(\vec{k},-\vec{k})|^{2}+\\
\nonumber&&+\frac{L^{d}(2\pi)^{d}}{4!}\int \prod_{i=1}^{4}\frac{d^{d}k_{i}}{(2\pi)^{d}}\delta\left(\sum_{i=1}^{4}\vec{k}_{i}\right)\left(\sum_{i=1}^{4}E(\vec{k}_{i})\right)^{n}|\mathcal{K}_{4}(\vec{k}_{1},\vec{k}_{2},\vec{k}_{3},\vec{k}_{4})|^{2}+\mathcal{O}(\lambda^{3})\\
\end{eqnarray}

Once we insert (\ref{k4}) and (\ref{k2}) we get:
\begin{eqnarray}
\nonumber&&\kappa_{n}=-E_{G}\delta_{n,1}+\frac{L^{d}}{2}\frac{\lambda^{2}}{4}\left(\int \frac{d^{d}q}{(2\pi)^{d}}\frac{1}{2E(\vec{q})}\right)^{2}\int\frac{d^{d}k}{(2\pi)^{d}}\left(2E(\vec{k})\right)^{n-4}+\\
\nonumber&&+\frac{\lambda^{2}L^{d}(2\pi)^{d}}{4!}\int \prod_{i=1}^{4}\frac{d^{d}k_{i}}{(2\pi)^{d}}\delta\left(\sum_{i=1}^{4}\vec{k}_{i}\right)\left(\sum_{i=1}^{4}E(\vec{k}_{i})\right)^{n-2}\left(\prod_{i=1}^{4}2E(\vec{k}_{i})\right)^{-1}+\mathcal{O}(\lambda^{3})\\
\end{eqnarray}

The cumulants are all extensive quantities, as they should.
Notice that the above integrals are UV divergent and a UV cut-off must be used. 
This means in particular  that they are non-universal quantities. Instead universality is expected in the new edge singularity associated with the 4-particle cluster at the threshold $W=4m$. In fact as we explained these edge singularities are low-energy effects and their exponents are essentially universal.
This expression for the cumulants can be compared with those obtained from usual perturbation theory, since the momenta are nothing else than expectation values of powers of the free Hamiltonian on the $\phi^4$ ground state (\ref{cumulants}) and the expression match, as it should.
From the $\mathcal{O}(\lambda^2)$ expression of $\braket{G|G_{t}}$ we can also study the $\mathcal{O}(\lambda^2)$ the thresholds $W+E_{G}=2m$ and $W+E_{G}=4m$. The general expression for the threshold at $2m$ has already been reported in (\ref{thr118}), therefore we need simply to plug in it the expression for $\mathcal{K}_{2}$ at first order in $\lambda$ (\ref{k2}). To study the threshold at $4m$ we write $P(W)$ as Fourier transform of the generating function $g(t)$. Close to the threshold $W+E_{G}=4m$ we have:
\begin{eqnarray}
\nonumber&&P(W)=\mathcal{N}_{G}^{-1}\frac{L^{d}}{2}\int\frac{d^{d}k}{(2\pi)^{d}}\delta\left(W+E_{G}-2E(\vec{k})\right)|\mathcal{K}_{2}(\vec{k},-\vec{k})|^{2}+\\
\nonumber&&+\mathcal{N}_{G}^{-1}\frac{L^{d}(2\pi)^{d}}{4!}\int \prod_{i=1}^{4}\frac{d^{d}k_{i}}{(2\pi)^{d}}\delta\left(\sum_{i=1}^{4}\vec{k}_{i}\right)\delta\left(W+E_{G}-\sum_{i=1}^{4}E(\vec{k}_{i})\right)|\mathcal{K}_{4}(\vec{k}_{1},\vec{k}_{2},\vec{k}_{3},\vec{k}_{4})|^{2}+\mathcal{O}(\lambda^{3})\\
\end{eqnarray}
 Consider the first term in the above: its threshold is at $2m$, therefore it is smooth in correspondence with the $4$ particle threshold. The second term is not smooth at the four particle threshold and can be analyzed similarly to (\ref{thr117}) giving:
\begin{equation}
\mathcal{N}_{G}^{-1}L^{d}Q|\mathcal{K}_{4}(0,0,0,0)|^2m^{\frac{3}{2}d}\Theta(W+E_{G}-4m)\left(W+E_{G}-4m\right)^{\frac{3}{2}d-1}\label{thr130}
\end{equation}
The $\mathcal{O}(\lambda^2)$ expression for $\mathcal{K}_{4}$ is reported in (\ref{k4})
\begin{equation}
|\mathcal{K}_{4}(0,0,0,0)|^2=\frac{\lambda^2}{4(2m)^3}
\end{equation}
and $Q$ is a numerical factor.
We have therefore found that the introduction of $\phi^4$ interactions in the initial state produced an extra edge singularity at $W=4m$ that was absent in the quadratic case. This is a signature of the presence of clusters of four particles and as we see it is characterised by a different exponent $\tfrac32 d -1$ which was derived by means of an essentially phase-space calculation, while the presence of only pairs is characterized by the exponents $d-1$ and $\frac{d}{2}-1$. This can be seen considering in full generality the threshold, without restricting to the $\mathcal{O}(\lambda^{2})$ contributions.
The fact that only $\mathcal{K}_{4}$ contributes to the $4$ particles threshold is true only at $\mathcal{O}(\lambda^2)$: the threshold at $4m$ in general depends on both $\mathcal{K}_{2}$ and $\mathcal{K}_{4}$ as it can be easily understood from (\ref{sumwork}), but these terms are of higher order in $\lambda$.

\section{Conclusions and outlook}
\label{conclusions}

In this work we studied the effects of the general physical assumptions of locality on the analytic structure of a cluster expansion for local quantum states, in particular we find a wide class of states that satisfy the cluster properties for local observables (Section \ref{clusterexp}). We denoted such states as local quantum states and their cluster expansion is studied. Since ground states of physical Hamiltonians are known to satisfy the cluster property, we proposed to look for them among the general class of states we found: this ansatz was confirmed by perturbation theory at all orders (Section \ref{scatmatsec}).
The knowledge of the analytic structure of local quantum states allows us to study ground states directly from the Schr\"odinger equation (Section \ref{sectionexacteq}), changing the usual point of view of field theory. With the knowledge of the ground state itself, rather than the usual information about the expectation values computed on it, we have the possibility of studying physical quantities that are not directly linked to correlators of fields  and we sketch the study of the spectral decomposition and the work statistics.

The knowledge of the exact analytic form of the ground state opens up the possibility of implementing efficient approximation schemes over it, as we commented in Section \ref{approxscheme}. Using the cluster expansion we can naturally go beyond the well known gaussian approximation, increasing our understanding of ground states. Moreover, using approximation schemes on the cluster expansion guarantees that any approximation we are implementing will produce a state whose qualitative features do not differ from those of a ground state: for example, cluster property will be always satisfied.
This is quite important both on a conceptual level and on a practical one, since any approximation that gives a state qualitatively different from a ground state is doomed to be a very poor approximation of it.
In particular, having exact information about the state gives us the possibility of implementing variational approximation schemes: these could permit to go beyond perturbation theory and such a program will be subject of future studies.

As we already commented, we expect that such an approach could find interesting applications also in out-of-equilibrium problems: in Section \ref{dynamic} we studied the behavior under time evolution of local quantum states, showing that evolution under local Hamiltonians does not spoil their physical properties and the evolved state remains a local quantum state. The cluster expansion permits us to explicitly write the time dependent Schr\"odinger equation for the state: this is an important change of perspective, since usually the time evolution is explored in the Heisenberg picture, where are the operators that evolve.

This work is far from being exhaustive and it has been mainly thought to present the basics of the cluster expansion for ground states, giving in the meanwhile as many tools as possible to handle such a construction. During the exposition we tried to explicitly comment over all the possibilities of such a construction and they will be subject of future investigations.
\ \\
\ \\

\textbf{Acknowledgments}

This work has been carried out partially thanks to the support of the A*MIDEX project Hypathie (no. ANR-11-IDEX-0001-02) funded by the ``Investissements d'Avenir'' French Government program, managed by the French National Research Agency (ANR). SS acknowledges also the ERC for financial support under Starting Grant 279391 EDEQS.


%
\newpage
\appendix

\section{Cluster expansion for systems with a finite number of particles}
\label{finitep}

The application of the cluster expansion to many body systems with a finite number of particles deserves some further comments. 
Considering the cluster expansion (\ref{intro0}) with the constraint (\ref{suffcond}) on the decay of the cluster amplitudes, we find contributions with an arbitrary number of particles and the cluster expansion, as we commented in Section \ref{LQS}, can only be a thermodynamic description of the ground state.
Because of the conservation of particles, a naive application of the methods of Section \ref{perturb} to compute the cluster amplitudes does not work.
Nevertheless, if a condensate is present we can define new local bosonic operators in terms of which the cluster amplitudes can be computed by mean of the same methods of Section \ref{perturb}.
The idea is to use $\psi^{\dagger}$ operators that do not create a particle, rather an excitation moving a particle from a 'reference state' to another state: in this way $\psi$ does not change the number of particles. Then we will take the thermodynamic limit and get an Hamiltonian in terms of $\psi$ with not restrictions on the number of excitations and the cluster expansion can be readily derived with the same methods of Section \ref{perturb}.

We illustrate this program in the prototypical example of the $d-$dimensional interacting Bose Gas, whose Hamiltonian in second quantization is:
\begin{equation}
H=\int d^{d}x\hspace{2pt} \nabla \Psi^{\dagger}(\vec{x})\nabla 	\Psi(\vec{x})+\lambda \Psi^{\dagger}(\vec{x})\Psi^{\dagger}(\vec{x})\Psi(\vec{x})\Psi(\vec{x})\label{ref25}
\end{equation}

The operators $\Psi$ follow standard commutation rules $[\Psi(\vec{x}),\Psi^{\dagger}(\vec{y})]=\delta(\vec{x}-\vec{y})$ and the Hilbert space is a Fock space constructed acting with $\Psi^{\dagger}$ on a vacuum $\ket{0_{\Psi}}$ annihilated by $\Psi$.
The number of particles is fixed:
\begin{equation}
N=\int d^{d}x\hspace{3pt} \Psi^{\dagger}(\vec{x})\Psi(\vec{x}) \label{ref26}
\end{equation}

Note that for $d\ge 2$ the ground state of (\ref{ref25}) exhibits condensation of the zero mode, that is macroscopically occupied.
The idea is to define $\ket{0}$ as the condensate state and define $\psi^{\dagger}$ as operators that move particles from this reference state to other states, in such a way that the number of particles (\ref{ref26}) is surely conserved.
Let $\alpha_{\vec{k}}$ be the modes of the field $\Psi$, then following \cite{castin, MoraCastin} we define the reference state $\ket{0}$ as:
\begin{equation}
\ket{0}=\frac{1}{\sqrt{N!}}\alpha_{0}^{N}\ket{0_{\Psi}}
\end{equation}

The particles in $\ket{0}$ are all in the condensate at zero momentum. Now, as  mode operators $a_{\vec{k}}$ define:
\begin{equation}
a_{\vec{k}}^{\dagger}=\frac{1}{\sqrt{L^{d}n_{0}}}\alpha_{\vec{k}}^{\dagger}\alpha_{0}, \hspace{2pc}\vec{k}\ne 0
\end{equation}
where $n_{0}$ is the density of particles in the condensate and it must be determined self-consistently.
The $a^{\dagger}$ operator does not create a particle, rather excites it outside of the condensate $\ket{0}$.
It is possible to show \cite{castin} that in the thermodynamic limit $a_{\vec{k}}$ can be regarded as a canonical bosonic operator in a weak sense, when we compute correlators on the ground state. As a matter of fact:
\begin{equation}
[a_{\vec{q}},a^{\dagger}_{\vec{k}}]=\delta_{\vec{k},\vec{q}}\frac{\alpha^{\dagger}_{0}\alpha_{0}}{L^{d}n_{0}}-\frac{\alpha_{\vec{q}}\;\alpha^{\dagger}_{\vec{k}}}{L^dn_{0}}
\end{equation}
when the above is inserted in a correlator the first term is $\mathcal{O}(L^{0})$, since $\braket{\alpha_{0}^{\dagger}\alpha_{0}}=L^{d}n_{0}$, while the extra correction is suppressed in $L\to\infty$ and we obtain the usual commutation rules $[a_{\vec{q}},a_{\vec{k}}^{\dagger}]=\delta_{\vec{k},\vec{q}}$. Further details of the limit can be found in \cite{castin}.
The $\psi$ operators are defined as the Fourier transform of $a_{\vec{k}}$.
\begin{equation}
\psi(\vec{x})\equiv\frac{1}{L^{d/2}}\sum_{\vec{k}\ne 0}e^{-i\vec{k}\vec{x}}a_{\vec{k}}
\end{equation}

From the expectation value of (\ref{ref26}) we get the self consistency equations for the condensate density $n_{0}$.
\begin{equation}
\frac{N}{L^{d}}=\braket{\psi^{\dagger}\psi}+n_{0}
\end{equation}

The final step is to take the thermodynamic limit of (\ref{ref25}) in terms of $\psi$ and obtain an effective Hamiltonian directly defined in the thermodynamic limit $L\to\infty$ \cite{MoraCastin}:
\begin{equation}
H=\lambda n_{0}^{2}L^{d}+\int d^{d}x \; \nabla\psi^{\dagger}\nabla\psi+\lambda\left[n_{0}\left( \psi^{\dagger}\psi^{\dagger}+\psi\psi\right)+4n_{0}\psi^{\dagger}\psi+2\sqrt{n_{0}}\left(\psi^{\dagger}\psi\psi+\psi^{\dagger}\psi^{\dagger}\psi\right)+\psi^{\dagger}\psi^{\dagger}\psi\psi\right]\label{refA8}
\end{equation}
Note that the above is a thermodynamic limit of the Hamiltonian valid only in a weak sense, nevertheless it is enough to study the thermodynamic properties of the ground state. 
The $\psi^{\dagger}$ operators create localized excitations, therefore it is plausible that they satisfy the cluster property. Moreover, the number of excitations is not fixed by the above Hamiltonian.
The methods of Section \ref{perturb} can be easily adapted to the Hamiltonian (\ref{refA8}) and used to compute the cluster expansion of the ground state of (\ref{ref25}) in terms of $\psi$.

\section{Gaussian states and quadratic Hamiltonians}
\label{gaussian}

A well known type of state in quantum field theories is a translational invariant squeezed state that is actually a particular case of (\ref{expk}) obtained keeping only $\mathcal{K}_{1}$ and $\mathcal{K}_{2}$, but we will focus on the case in which $\mathcal{K}_{1}$ is absent. Therefore:

\begin{equation}
\ket{S}=\exp\left[\sum_{\vec{k}}\frac{\mathcal{K}_{2}(\vec{k},-\vec{k})}{2}a^{\dagger}_{\vec{k}}a^{\dagger}_{-\vec{k}}\right]\ket{0}\label{squeezedstate}
\end{equation}

Because of its simplicity many quantities (correlators, free energy..) can be computed exactly and we often use this state as a check and a guide for our more general approach, therefore we want to enlist some of its properties.
A squeezed state naturally arises considering ground states of Hamiltonians with a perturbation quadratic in the fields. As an example, consider the simple Lagrangian:
\begin{equation}
\mathcal{L}=\int d^{d}x \hspace{3pt}\frac{1}{2}(\partial_{\mu}\phi\partial^{\mu}\phi-m^{2}\phi)-\frac{\lambda}{2}\phi^{2}\label{gaussian1}
\end{equation}

 If we consider $\lambda\phi^{2}$ as a perturbation, then we will proceed defining creation and annihilation operators based on the mass $m$:
\begin{equation}
\phi(\vec{x})=\frac{1}{L^{d/2}}\sum_{\vec{k}}\frac{e^{i\vec{k}\vec{x}}}{\sqrt{2E(\vec{k})}}\left(a^{\dagger}_{\vec{k}}+a_{-\vec{k}}\right)\hspace{3pc}E(\vec{k})=\sqrt{\vec{k}^{2}+m^{2}}\label{gaussian2}
\end{equation}

Actually, we could have proceeded also in a smarter way, interpreting (\ref{gaussian1}) as a mass shift:
\begin{equation}
\mathcal{L}=\int d^{d}x \hspace{3pt}\frac{1}{2}(\partial_{\mu}\phi\partial^{\mu}\phi-M^{2}\phi)\hspace{3pc}M^{2}=m^{2}+\lambda\label{newgaussian122}
\end{equation}

Now we have a simple free theory and we can diagonalize its Hamiltonian through new bosonic operators $b$ and $b^{\dagger}$ such that:

\begin{equation}
\phi(\vec{x})=\frac{1}{L^{d/2}}\sum_{\vec{k}}\frac{e^{i\vec{k}\vec{x}}}{\sqrt{2\mathcal{E}(\vec{k})}}\left(b^{\dagger}_{\vec{k}}+b_{-\vec{k}}\right)\hspace{3pc}\mathcal{E}(\vec{k})=\sqrt{\vec{k}^{2}+M^{2}}\label{gaussian3}
\end{equation}

Therefore the ground state $\ket{S}$ (with this notation we anticipate it will be a squeezed state) is simply the vacuum of the $b$ operators, thus it is completely identified by the requirement $b_{\vec{k}}\ket{S}=0$. This last relation permits us to express $\ket{S}$ in terms of the $a$ operators and their vacuum state $\ket{0}$: as a matter of fact comparing the expressions (\ref{gaussian2}) and (\ref{gaussian3}) and the analogous expressions of the conjugated momenta of the field $\phi$ we can express the $b$ operators in terms of $a$.

\begin{equation}
b_{\vec{k}}=\cosh\theta_{\vec{k}}a_{\vec{k}}-\sin\theta_{k}a^{\dagger}_{-\vec{k}}\label{bfunca}
\end{equation}

Where:
\begin{equation}
\sinh\theta_{k}=\frac{1}{2}\left(\frac{\sqrt{E(\vec{k})}}{\sqrt{\mathcal{E}(\vec{k})}}-\frac{\sqrt{\mathcal{E}(\vec{k})}}{\sqrt{E(\vec{k})}}\right)\hspace{3pc}\cosh\theta_{\vec{k}}=\frac{1}{2}\left(\frac{\sqrt{E(\vec{k})}}{\sqrt{\mathcal{E}(\vec{k})}}+\frac{\sqrt{\mathcal{E}(\vec{k})}}{\sqrt{E(\vec{k})}}\right)
\end{equation}

With these relations we can readily obtain an equation for the ground state in terms of the $a$ operators:
\begin{equation}
b_{\vec{k}}\ket{S}=0\hspace{3pt}\Longrightarrow\hspace{3pt}a_{\vec{k}}\ket{S}=\tanh\theta_{\vec{k}}a^{\dagger}_{\vec{k}}\ket{S}
\end{equation}

The solution of this last equation is a squeezed state (\ref{squeezedstate}) with:
\begin{equation}
\mathcal{K}_{2}(\vec{k},-\vec{k})=\tanh\theta_{\vec{k}}=\frac{\sqrt{\vec{k}^{2}+m^{2}}-\sqrt{\vec{k}^{2}+m^{2}+\lambda}}{\sqrt{\vec{k}^{2}+m^{2}}+\sqrt{\vec{k}^{2}+m^{2}+\lambda}}\label{k2mass}
\end{equation}

We can use (\ref{bfunca}) also to compute correlators of the $a$ fields on the ground state, simply inverting (\ref{bfunca}). For example:
\begin{equation}
\braket{a^{\dagger}_{\vec{k}}a_{\vec{k}}}_{S}=\frac{\bra{S}(\cosh\theta_{\vec{k}}b^{\dagger}_{\vec{k}}+\sinh\theta_{\vec{k}}b_{-\vec{k}})(\cosh\theta_{\vec{k}}b_{\vec{k}}+\sinh\theta_{\vec{k}}b^{\dagger}_{-\vec{k}})\ket{S}}{\braket{S|S}}=\sinh^{2}\theta_{\vec{k}}
\end{equation}

With similar computation we can arrive at the following result for the correlators on a squeezed state as (\ref{squeezedstate}). In particular the expression below is valid for each $\mathcal{K}_{2}$, not necessary in the form (\ref{k2mass}) and also complex $\mathcal{K}_{2}$ are allowed:
\begin{equation}
\begin{pmatrix} \braket{a_{\vec{k}}a^{\dagger}_{\vec{k}}}_{S} && \braket{ a_{-\vec{k}}a_{\vec{k}}}_{S} \\  \braket{a^{\dagger}_{\vec{k}}a^{\dagger}_{-\vec{k}}}_{S} && \braket{a_{-\vec{k}}a^{\dagger}_{-\vec{k}}}_{S}\end{pmatrix}=\frac{1}{1-|\mathcal{K}_{2}(\vec{k},-\vec{k})|^{2}}\begin{pmatrix} 1 && \mathcal{K}_{2}(\vec{k},-\vec{k}) \\ \mathcal{K}^{*}_{2}(-\vec{k},\vec{k}) && 1\end{pmatrix}\label{gaussiancorrelators}
\end{equation}

Since the squeezed state (\ref{squeezedstate}) is gaussian with zero mean, all the correlators can be computed once we know the two point correlators above. Notice that, since our squeezed state is translational invariant, all the correlators with a non zero total momentum must be zero.
Notice that something odd happens in (\ref{gaussiancorrelators}) if we let $|\mathcal{K}(\vec{k},-\vec{k})|^{2}>1$: in that case the correlator $\braket{a_{\vec{k}}a^{\dagger}_{\vec{k}}}_{S}$ becomes negative.
This situation is impossible to reach when $\mathcal{K}_{2}$ describes the ground state of (\ref{gaussian1}) since (\ref{k2mass}) implies $|\mathcal{K}_{2}|\le 1$, nevertheless we can learn something important studying such a state.
Having a negative correlator is non sense, since:

\begin{equation}
\braket{a_{\vec{k}}a^{\dagger}_{\vec{k}}}_{S}=\frac{\bra{S}a_{\vec{k}}a^{\dagger}_{\vec{k}}\ket{S}}{{\braket{S|S}}}=\frac{\braket{S'|S'}}{\braket{S|S}} \hspace{2pc}\text{Where: } \ket{S'}\equiv a^{\dagger}_{\vec{k}}\ket{S}
\end{equation}

The inner product is positively defined and we must have $\braket{S'|S'}\ge 0$, therefore $\braket{a_{\vec{k}}a^{\dagger}_{\vec{k}}}_{S}\ge0$. We can solve this absurd fact if we notice:

\begin{equation}
\braket{a_{\vec{k}}a^{\dagger}_{\vec{k}}}_{S}=\frac{1}{1-|\mathcal{K}_{2}(\vec{k},-\vec{k})|^{2}}=\sum_{n=0}^{\infty}|\mathcal{K}_{2}(\vec{k},-\vec{k})|^{2n}\label{gaussianity134}
\end{equation}

The two expressions are identical whenever $|\mathcal{K}_{2}|<1$, but they are rather different as soon as this bound is violated. We can see that the right expression for $\braket{a_{\vec{k}}a^{\dagger}_{\vec{k}}}_{S}$ is indeed the series rather than the fraction. In this way whatever is the value of $\mathcal{K}_{2}$ we always have $\braket{a_{\vec{k}}a^{\dagger}_{\vec{k}}}_{S}\ge 0$ as it should and when $|\mathcal{K}_{2}|>1$ the correlator does not become negative, but it rather becomes a positive divergent quantity.
From this expression it is clear that the correct interpretation of the bound $|\mathcal{K}_{2}(\vec{k},-\vec{k})|<1$ is as the radius of convergence of the geometric series: this is a very useful observation we are going to use in Appendix \ref{criticality}.
Notice that if the inequality is saturated, we have a divergence in the correlation function and this is a signal of criticality, as we will see in Appendix \ref{criticality}.
For the moment, notice that if we are studying the ground state of a Lagrangian such as (\ref{gaussian1}), then for $\lambda>-m^{2}$ the condition $|\mathcal{K}_{2}|<1$ is always met, instead when $\lambda=-m^{2}$ we have $\mathcal{K}_{2}(0,0)=1$ and the Lagrangian (\ref{newgaussian122}) becomes massless $M=0$, therefore it is critical.
In the remaining case $\lambda<-m^{2}$ we have that the theory described by (\ref{gaussian1}) is ill defined because $M^{2}<0$, therefore the Hamiltonian is no more bounded from below and has no more a ground state.

\section{Partial resummation in the two point correlator}
\label{grtoolperturb}

In Section \ref{grtool} we have seen the basic graphical tool to compute the correlators and the free energy, but apart from formal calculations, we can really evaluate only a finite number of graphs and truncate the expansion.
In this section we are going to perform a partial resummation of graphs that can give us some insight on important, non perturbative phenomena, for example we are going to use these results to correctly insert critical systems in our framework (Appendix \ref{criticality}).
In Feynman graphs a central role is played by the one particle irreducible graphs, for example the renormalization of the masses is to due their resummation in the perturbative series \cite{Peskin}.
Also in our case we can define irreducible graphs and in this section we explain how they can be summed, exactly as it happens in the usual Feynman graphs \cite{Peskin}. There will be some complications due to the presence of two kinds of vertices, but the ideas remain the same.
The best way to introduce the irreducible graphs is looking at the two point connected correlator in momentum space:
\begin{equation}
\braket{a^{\dagger}_{\vec{k}}a_{\vec{k}}}_{c}=\braket{a^{\dagger}_{\vec{k}}a_{\vec{k}}}-\braket{a^{\dagger}_{\vec{k}}}\braket{a_{\vec{k}}};\hspace{3pc}
\braket{a_{\vec{k}}a_{-\vec{k}}}_{c}=\braket{a_{\vec{k}}a_{-\vec{k}}}-\braket{a_{\vec{k}}}\braket{a_{-\vec{k}}}
\end{equation}

\begin{figure}
\begin{center}
\includegraphics[scale=0.3]{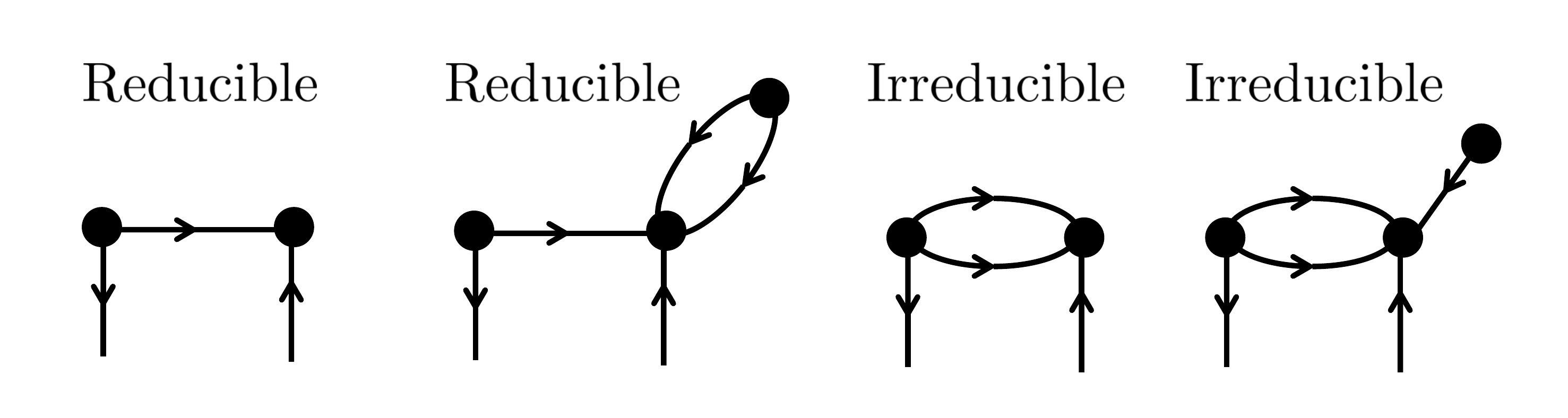}
\caption{Examples of irreducible and reducible graphs}\label{figirr}
\end{center}
\end{figure}

It is immediate to notice that the graphs that contribute to the connected correlators are all and only the connected graphs, among them we can define the irreducible graphs. A graph is called reducible if after we cut one internal line, the external legs become disconnected, irreducible otherwise. Some examples are given in Figure \ref{figirr}.
The irreducible graphs can be used as building blocks to construct general graphs of the two point connected correlators, simply joining them. This is quite obvious, since if a graph is reducible then it can be divided in two blocks, if one of them is reducible it can be divided again: we can proceed until we have divided the graph in its irreducible parts.
It must be stressed that the presence of two different kinds of vertices means some difficulties, as a matter of fact in the correlator $\braket{a^{\dagger}a}_{c}$ do not enter contributions only from its irreducible graphs, but also from the irreducible graphs of $\braket{aa}_{c}$ and so on.
With a little thought, it is possible to organize this combinatorial problem in a simple manner, if we define a $2\times 2 $ matrix $\mathcal{G}_{\text{ir}}(\vec{k})$, whose entries are defined as:

\begin{equation}
\left[\mathcal{G}_{\text{ir}}(\vec{k})\right]_{1,1}=\sum[\hspace{3pt}\text{irreducible graphs of }\braket{a^{\dagger}_{\vec{k}}a_{\vec{k}}}_{c}\hspace{3pt}]
\end{equation}
\begin{equation}
\left[\mathcal{G}_{\text{ir}}(\vec{k})\right]_{1,2}=\sum[\hspace{3pt}\text{irreducible graphs of }\braket{a_{-\vec{k}}a_{\vec{k}}}_{c}\hspace{3pt}]
\end{equation}
\begin{equation}
\left[\mathcal{G}_{\text{ir}}(\vec{k})\right]_{2,1}=\sum[\hspace{3pt}\text{irreducible graphs of }\braket{a^{\dagger}_{\vec{k}}a^{\dagger}_{\vec{k}}}_{c}\hspace{3pt}]
\end{equation}
\begin{equation}
\left[\mathcal{G}_{\text{ir}}(\vec{k})\right]_{2,2}=\sum[\hspace{3pt}\text{irreducible graphs of }\braket{a^{\dagger}_{-\vec{k}}a_{-\vec{k}}}_{c}\hspace{3pt}]
\end{equation}
\begin{figure}
\begin{center}
\includegraphics[scale=0.15]{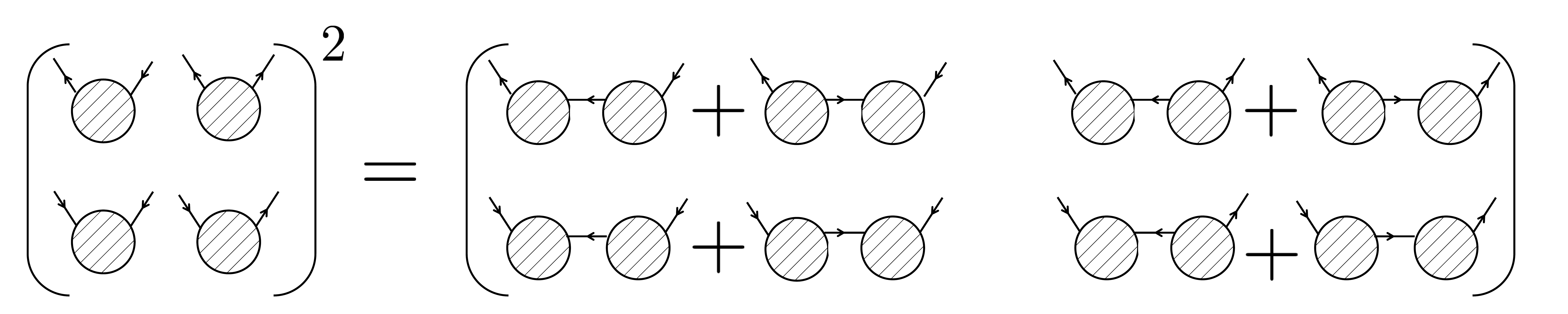}
\caption{Graphical representation of the square of $\mathcal{G}_{\text{ir}}(\vec{k})$, graphically the products of the matrix entries are simply the graphs obtained joining the arrows. In the square of $\mathcal{G}_{\text{ir}}$ appear all the graphs of the two point correlator made of two irreducible pieces.}\label{matrixprod}
\end{center}
\end{figure}

Taking the square of such a matrix is equivalent to consider the contributions of all the graphs that are made of two irreducible pieces. The key to understand this fact is to notice that for graphs with two external legs the value of the graphs obtained by joining two of them is simply the product of the two contributions. 
With this in mind consider  Figure \ref{matrixprod}, in each entry we can put the value of an irreducible graph: when we perform the matrix product, the product of the  values of graphs is simply the value of the graph obtained joining them. Therefore the square of such a matrix contains the contributions of all the graphs constructed joining two of the irreducible graphs.
If instead of considering only an irreducible graph for each entry we consider the sum over all the possible graphs, then the square produces the sum of the values of all the possible graphs made by two irreducible pieces.
In a similar way the $n^{th}$ power produces all the possible graphs made of $n$ irreducible graphs. Notice that each entry of the $n^{th}$ power is still associated to the same correlator of the same entry of $\mathcal{G}_{\text{ir}}$,  for example $[\mathcal{G}_{\text{ir}}(\vec{k})]^{n}_{1,1}$ is a contribution to $\braket{a^{\dagger}_{\vec{k}}a_{\vec{k}}}_{c}$. It is not difficult to convince ourselves of the last statements after some specific examples are considered.
At this point we can do the last step to find the two point correlators: we have simply to sum over all the possible graphs. This means that we have to sum over all the possible powers of $\mathcal{G}_{\text{ir}}$, therefore we have a geometric series.
\begin{equation}
\begin{pmatrix} \braket{a^{\dagger}_{\vec{k}}a_{\vec{k}}}_{c} && \braket{ a^{\dagger}_{\vec{k}}a^{\dagger}_{-\vec{k}}}_{c} \\  \braket{a_{-\vec{k}}a_{\vec{k}}}_{c} && \braket{a^{\dagger}_{-\vec{k}}a_{-\vec{k}}}_{c}\end{pmatrix}=\sum_{n=1}^{\infty}[\mathcal{G}_{\text{ir}}(\vec{k})]^{n}=\frac{\mathcal{G}_{\text{ir}}(\vec{k})}{\mathbb{1}-\mathcal{G}_{\text{ir}}(\vec{k})}\label{1pirr140}
\end{equation}

The last equality holds only if we are within the radius of convergence of the series and if $\mathcal{G}_{\text{ir}}$ is too big, the correlators diverge: in Appendix \ref{gaussian} we discuss the specific case of a squeezed state, but the same discussion can be easily generalized.
We can write this expression in an even more compact form if, instead of the normal order, we reverse it and use $\braket{a_{\vec{k}}a^{\dagger}_{\vec{k}}}_{c}=1+\braket{a^{\dagger}_{\vec{k}}a_{\vec{k}}}_{c}$:
\begin{equation}
\begin{pmatrix} \braket{a_{\vec{k}}a^{\dagger}_{\vec{k}}}_{c} && \braket{ a^{\dagger}_{\vec{k}}a^{\dagger}_{-\vec{k}}}_{c} \\  \braket{a_{-\vec{k}}a_{\vec{k}}}_{c} && \braket{a_{-\vec{k}}a^{\dagger}_{-\vec{k}}}_{c}\end{pmatrix}=\frac{1}{\mathbb{1}-\mathcal{G}_{\text{ir}}(\vec{k})}\label{sumirr}
\end{equation}

\begin{figure}
\begin{center}
\includegraphics[scale=0.4]{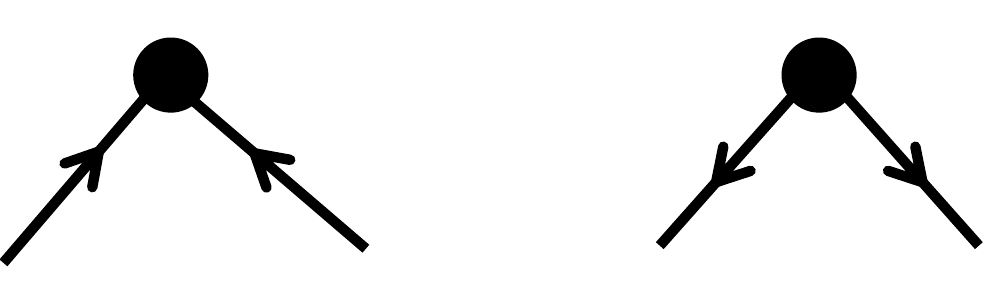}
\caption{Irreducible graphs for squeezed state}\label{irrgaussian}
\end{center}
\end{figure}

\begin{figure}
\begin{center}
\includegraphics[scale=0.3]{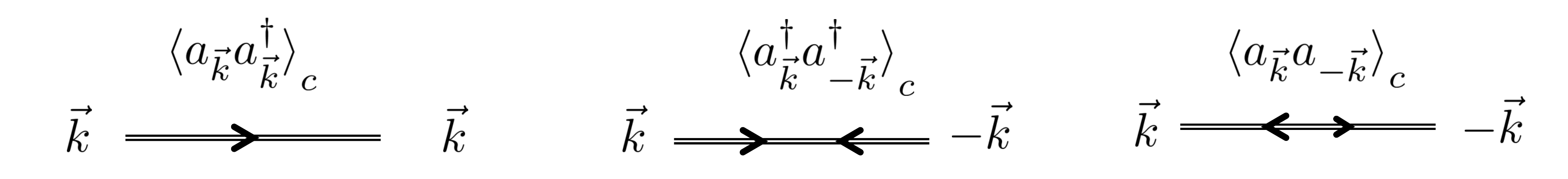}
\caption{Dressed propagators can be represented with two arrows and their value changes with their direction: with this notation we can recycle the old graphical rules and is still true that each vertex has all ingoing or all outgoing arrows.}\label{dressedprop}
\end{center}
\end{figure}

In the very simple case of a squeezed state ($\mathcal{K}_{n\ne2}=0$) we can exactly compute $\mathcal{G}_{\text{ir}}$, since we have only the two irreducible graphs of Figure \ref{irrgaussian}, therefore:

\begin{equation}
[\mathcal{G}_{\text{ir}}(\vec{k})]_{\text{S}}=\begin{pmatrix} 0 && \mathcal{K}_{2}(\vec{k},-\vec{k}) \\ \mathcal{K}^{*}_{2}(-\vec{k},\vec{k}) && 0\end{pmatrix}\label{squeezed}
\end{equation}

If we plug this expression in (\ref{sumirr}) we match exactly the result (\ref{gaussiancorrelators}) given in Appendix \ref{gaussian}, as it should.
In the general case there are many contributions to the irreducible graphs, but all the other contributions have one or more $\mathcal{K}_{n\ge 2}$ vertices, therefore the series of irreducible graphs must be approximated with a proper truncation.

Actually we can proceed even further: consider any graph for a generic correlator, then we can construct other graphs from it replacing the `bare' arrows with some irreducible graphs. This leads us to the concept of \emph{dressed propagator} \cite{Peskin} and to a definition of new graphical rules to compute propagators in which we have explicitly taken care of all the possible insertions of irreducible graphs.
In the usual Feynman graphs \cite{Peskin} there is only one kind of vertex, thus only one type of dressed propagator, here instead we have more. To recycle as far as we can the old rules, we represent the different propagators as in Figure \ref{dressedprop}: differently from the arrows of Section \ref{grtool} these dressed propagators have also a weight, determined by the connected two point correlators. 
Notice that the single arrow has the same momentum at the two extrema, instead the double arrows exchange their sign: as before the arrows departing from a vertex are all ingoing or outgoing. The presence of double arrows allows a vertex also to self interact; moreover a graph can contain single arrows: for example, it is clear from the definition that the two point correlators are graphs made of a single arrow.
Through these rules we do not compute correlators in the normal order, but rather in the opposite order: all the creation operator $a^{\dagger}$are on the right, the destruction operators $a$ on the left. This comes out from the fact that the first arrow of Figure \ref{dressedprop} brings a contribution $\braket{a_{\vec{k}}a^{\dagger}_{\vec{k}}}_{c}$.

\begin{figure}
\begin{center}
\includegraphics[scale=0.35]{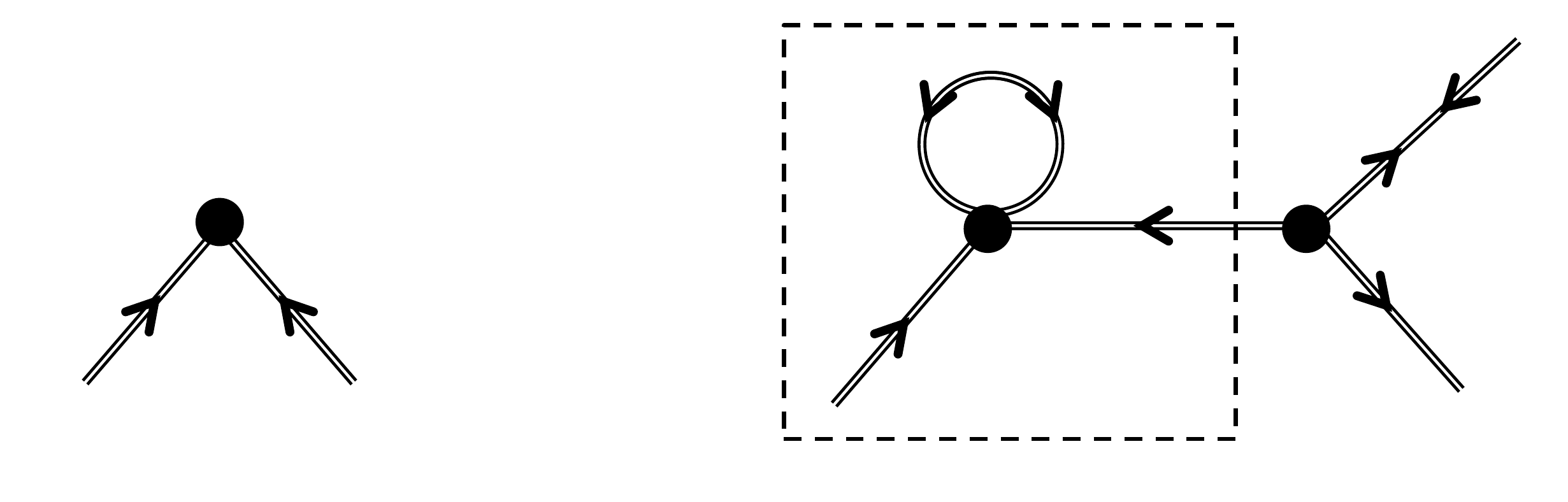}
\end{center}
\caption{When we draw graphs with the dressed propagator we must take care of having already summed on the irreducible graphs, this means that we cannot draw subgraphs with only two external legs. Therefore, neither of these graphs can be drawn: in the second, in the dotted lines, there is the subgraph with two external legs.}\label{toavoid}
\end{figure}

\begin{figure}
\begin{center}
\includegraphics[scale=0.3]{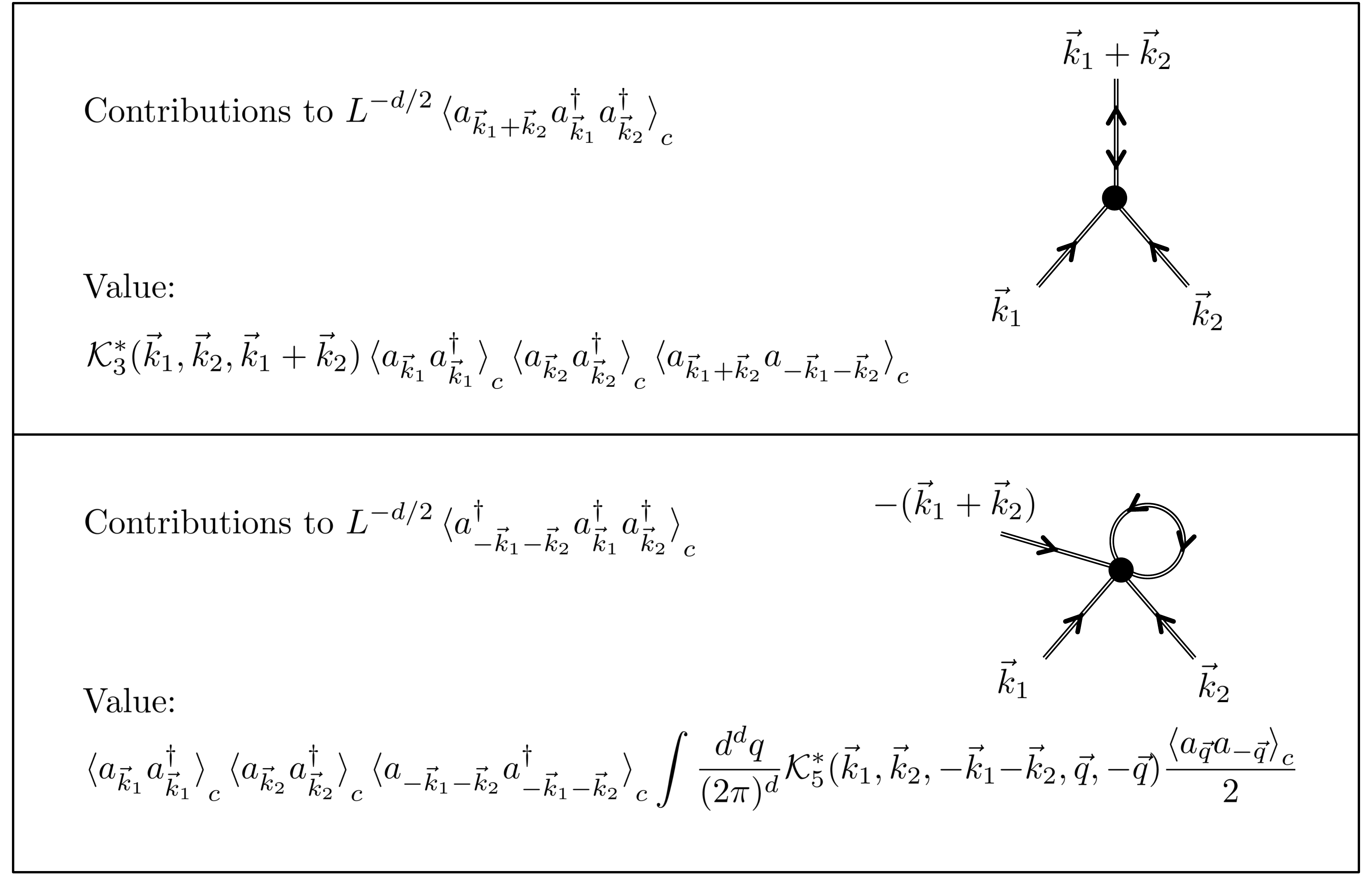}
\end{center}
\caption{Examples of graphs constructed with the dressed propagators and their value.}\label{exampledressed4}
\end{figure}

Since we have already summed on the one particle irreducible graphs, in the graphs with the dressed propagators we must avoid all the graphs in which we can find subgraphs with two external legs: in Figure \ref{toavoid} there are some examples of graphs to be avoided.
All the other rules (conservation of momenta, integration on internal lines, symmetry factors) are the same as before, in Figure \ref{exampledressed4} there are some allowed graphs with their values.

\section{The cluster expansion at criticality}
\label{criticality}

There is an important point that should be clarified and regards the presence of criticality. 
As we stated in Section \ref{sectech}, asking certain decaying properties for the cluster amplitudes (\ref{suffcond}) leads to the conclusion that any graph in the momentum space does not have any singularity. On the other hand, we know in advance that the two point correlators in critical systems typically exhibit divergences at zero momentum.
Of course, even in critical systems Properties 1, 2 and 3 are expected to be valid. There are two possible ways to acquire the singularity in the two point correlator: the amplitudes  $W_{n}^{c}$ do not decay fast enough (\ref{suffcond}) or the expansions we wrote in Section \ref{sectech} do not converge anymore. Note that it is very unlike that we can drop the decay condition about the cluster amplitudes: as a matter of fact (\ref{suffcond}) guarantees that all the graphs we compute do not have infrared divergences. 
If infrared divergences arise in some graphs in the expansion of correlators, in order to have the latter independent on the volume we would need cancellations of the infrared divergences among the graphs.
On the other hand, it is simpler to imagine that the expansions do not converge anymore: this is exactly what happens in the gaussian case of Appendix \ref{gaussian}.
 In particular consider the expression (\ref{gaussianity134}) 
$\braket{a_{\vec{k}}a^{\dagger}_{\vec{k}}}_{S}={(1-|\mathcal{K}_{2}(\vec{k},-\vec{k})|^{2})}^{-1}$. 
In that case it is clear that the singularities of the two point correlator are not associated with singularities of $|\mathcal{K}_{2}|$, but rather with the points where it equals 1: 
if we look at the geometric series, those are the points where it becomes divergent. Then, rephrasing the last sentence, we can say that a singularity in the two point correlator signals a problem in the convergence of a series. 
Actually, this interpretation is valid also in the most general case: consider the formal expression for the two point correlators (\ref{sumirr})  found in Appendix \ref{grtoolperturb}, in terms of the one particle irreducible graphs. The expression  (\ref{sumirr}) is the natural generalization of (\ref{gaussianity134}) and gives to the singularities in the two point correlators the same interpretation of the Gaussian case.
To have a singular correlator (\ref{sumirr}) we should ask $\det(\mathbb{1}-\mathcal{G}_{\text{ir}}(\vec{k}))=0$: this condition is associated to the lack of convergence of the geometric series (\ref{1pirr140}) and $\det(\mathbb{1}-\mathcal{G}_{\text{ir}}(\vec{k}))=0$ can be fulfilled even  if we ask for (\ref{suffcond}), therefore any graph that appears in $\mathcal{G}_{\text{ir}}(\vec{k})$ is not singular.
We can conclude that at criticality the state still has a cluster expansion whose amplitudes are expected to satisfy (\ref{suffcond}), but the graphical expansions of the correlators are not expected to be convergent any longer: in particular, we are at the boundary of the radius of convergence of (\ref{1pirr140}).

\section{Zero bare mass, failure of perturbation theory and non perturbative study}
\label{failure}

As we discussed in Section \ref{thprop} and Section (\ref{sectech}), the $\mathcal{K}_{n}$ cluster amplitudes of quantum local states should have no singularities in order to guarantee our locality properties and the perturbative calculation  of Section (\ref{scatmatsec}) confirms this behavior, at least in the massive case ($m\ne 0$).
The troubles arise when we let $m\rightarrow 0$: in this case the expressions (\ref{k4}) and (\ref{k2}) become singular because of poles at zero momenta. This is something not expected and we have a contradiction between the perturbative result and the supposed analytic structure of $\mathcal{K}_{n}$, on the other hand near the poles the perturbation theory is not reliable anymore. Perturbation theory makes sense if each contribution is small, but near a pole, regardless how small $\lambda$ is, the contributions (\ref{k4}) (\ref{k2}) are always diverging quantities.
This argument makes us suspect that the poles in (\ref{k4}) (\ref{k2}) are not really poles of $\mathcal{K}_{2}$ and $\mathcal{K}_{4}$, but rather signal a problem with the perturbation theory. To understand better this feature we should go beyond the perturbative regime and this is not a simple task. 
We will study directly the functional equations (\ref{excteq57}): we have seen in Section \ref{sectionexacteq} that (\ref{k4}) and (\ref{k2}) are simply the $\mathcal{O}(\lambda)$ solutions of (\ref{excteq57}). 
Before doing so, we should see what happens in the exactly solvable case of the gaussian state, therefore consider (\ref{k2mass}) of Appendix \ref{gaussian}. In particular we can extract the first order in $\lambda$:
\begin{equation}
\mathcal{K}_{2}(\vec{k},-\vec{k})=\frac{\sqrt{\vec{k}^{2}+m^{2}}-\sqrt{\vec{k}^{2}+m^{2}+\lambda}}{\sqrt{\vec{k}^{2}+m^{2}}+\sqrt{\vec{k}^{2}+m^{2}+\lambda}}=-\frac{\lambda}{2E(\vec{k})}+\mathcal{O}(\lambda^{2})\label{studgaussian}
\end{equation}

\begin{figure}
\begin{center}
\includegraphics[scale=0.25]{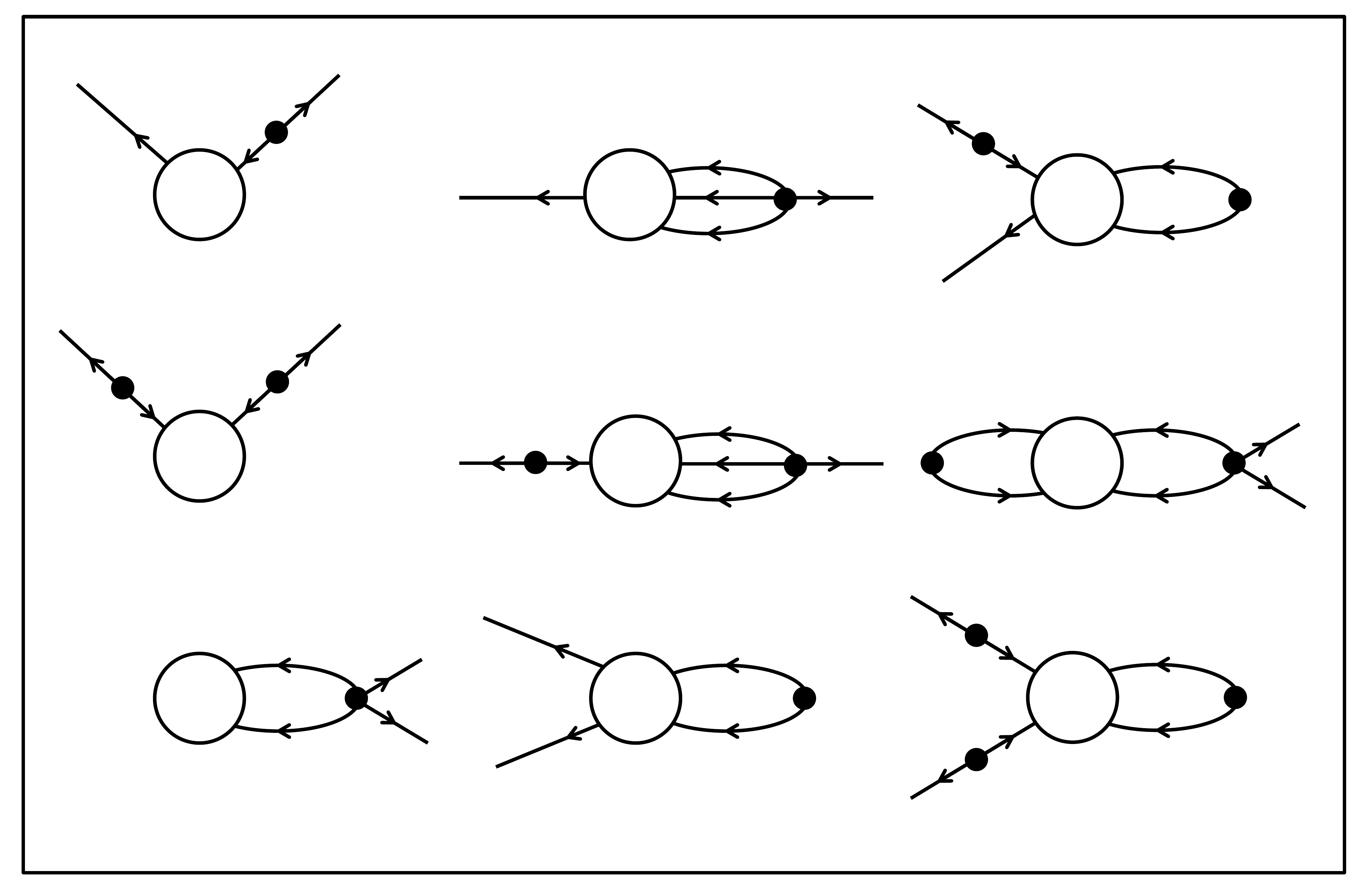}
\end{center}
\caption{Graphs for $\mathfrak{F}_{2}$ constructed using only $\mathcal{K}_{2}$ and $\mathcal{K}_{4}$.}\label{trunck2}
\end{figure}

We can see that when $m=0$, the first term of the expansion in $\lambda$ becomes singular at zero momentum, on the other hand the exact solution is finite in $\vec{k}=0$, in particular $\mathcal{K}_{2}(0,0)=-1$ regardless the value of $\lambda$.
Therefore, at least in the gaussian case, we see that the picture discussed above is correct and $\mathcal{K}_{2}$ has no singularities.
In an interacting theory like the $\phi^4$ model we cannot solve exactly the functional equations (\ref{excteq57}), nevertheless we can attempt a consistency check: we will suppose that the cluster amplitudes $\mathcal{K}_{n}$ are not singular, then check this is consistent with (\ref{excteq57}). In particular, we will check consistently that $\mathcal{K}_{2}$ does not have singularities at zero momentum, where the perturbation theory fails.
Such a consistency check is not exhaustive, since we will see only the absence of singularities in $\mathcal{K}_{2}$: it would be important to check also the other $\mathcal{K}_{n}$ and, even better, to study directly the cluster amplitudes in a non perturbative way. Nevertheless, because of the involved form of (\ref{excteq57}), even a consistency check for $\mathcal{K}_{2}$ is remarkable.
In order to proceed further, we need to use the equation (\ref{excteq57}) with $n=2$ and study the features of $\mathfrak{F}_{2}(\vec{k},-\vec{k})$: this quantity can be computed summing the contributions of the proper graphs, as explained in Section \ref{grexacteqsec}. As we are going to see, these graphs can be organized in terms of their singular behavior in the momentum $\vec{k}$. The rule is this: each outgoing arrow that departs directly from the `circles' vertices carries a singularity $\sim |\vec{k}|^{1/2}$, the same singularity is carried by the outgoing arrows associated with $\mathcal{K}_{2}$ vertices. Other outgoing arrows have no singularities associated with them. This counting of singularities holds in dimensions higher than one ($d>1$) and we will suppose that this is our case.
To understand this fact it is useful to see some explicit examples, then the general case will be obvious: in Figure \ref{trunck2} are drawn all the graphs that contribute to $\mathfrak{F}_{2}$ constructed using only $\mathcal{K}_{2}$ and $\mathcal{K}_{4}$.
Consider for example the first graph in the upper left corner of Figure \ref{trunck2}, its contribution can be readily written as:
\begin{equation}
\frac{\tilde{\beta}_{2}}{E(\vec{k})}\mathcal{K}_{2}(\vec{k},-\vec{k})
\end{equation}

As promised, it has a divergence as $|\vec{k}|^{-1}$, that is to say $\sim|\vec{k}|^{-1/2}$ for each of its external legs. Instead, the graph immediately below of it contribute as:
\begin{equation}
\frac{\tilde{\beta}_{2}}{2E(\vec{k})}(\mathcal{K}_{2}(\vec{k},-\vec{k}))^2
\end{equation}

Again, we have a singularity $\sim|\vec{k}|^{-1}$, that is to say $\sim|\vec{k}|^{-1/2}$ for each of its external legs, as we said.
Now, look instead to the contribution of the first graph from above, in the middle column of Figure \ref{trunck2}: our counting argument asserts the whole graphs should have a singularity $\sim|\vec{k}|^{-1/2}$. Its contribution is:
\begin{equation}
\frac{\tilde{\beta}_{4}}{3!}\int \frac{d^{d}q_{1}d^{d}q_{2}}{(2\pi)^{2}}\frac{\mathcal{K}_{4}(\vec{q}_{1},\vec{q}_{2},\vec{k}-\vec{q}_{1}-\vec{q}_{2},-\vec{k})}{\sqrt{2E(\vec{k})}\sqrt{2E(\vec{k}+\vec{q}_{1}+\vec{q}_{2})}\sqrt{2E(\vec{q}_{1})}\sqrt{2E(\vec{q}_{2})}}+(\vec{k}\leftrightarrow-\vec{k})
\end{equation}

This value diverges as $|\vec{k}|^{-1/2}$. As a matter of fact, if $d>1$ and the term $E(\vec{k})$ is omitted, the remain integration does not diverge when $\vec{k}\rightarrow 0$. It is not difficult to repeat the check also for the remaining graphs of Figure \ref{trunck2} and other graphs constructed with general cluster amplitudes $\mathcal{K}_{n}$. The reason why the counting of divergences is so simple is that integrations over closed loops remove singularities.
Now we should write down the equations for $\mathcal{K}_{2}$: their exact form does not matter, what is important is taking care explicitly of the pole structure and the presence of $\mathcal{K}_{2}$ attached to external legs.
We will see in a while that the equations  (\ref{excteq57}) with $n=2$  can be organized as:
\begin{equation}
\frac{[1+\mathcal{K}_{2}(\vec{k},-\vec{k})]^{2}}{|\vec{k}|}\xi_{2}(\vec{k})+\frac{1+\mathcal{K}_{2}(\vec{k},-\vec{k})}{|\vec{k}|^{1/2}}\xi_{1}(\vec{k})+\xi_{0}(\vec{k})=0\label{nonsing}
\end{equation}

Where $\xi_{0}$, $\xi_{1}$ and $\xi_{2}$ are non singular functions of $\vec{k}$.
This statement can be easily seen if we organize the graphs in a proper way. For example, consider the contribution of the first and second graphs from the above, in the second column of Figure \ref{trunck2}: the value of the second graph is simply $\mathcal{K}_{2}(\vec{k},-\vec{k})$ times the value of the first. Therefore, once we have summed these contributions we get $1+\mathcal{K}_{2}(\vec{k},-\vec{k})$ times the value of the first graph: this value has exactly the $|\vec{k}|^{-1/2}$ factor we need. In a similar way and being careful with the symmetry factors, we can organize also the other graphs in the same way and arrive to (\ref{nonsing}): we invite the reader to check some additional explicit examples.
Without a full knowledge of $\mathcal{K}_{2}$ and $\mathcal{K}_{4}$ it cannot be excluded that both $\xi_{1}$ and $\xi_{2}$ become zero at $\vec{k}=0$, but if at least one among $\xi_{1}$ and $\xi_{2}$ is not zero at zero momentum we conclude:
\begin{equation}
\mathcal{K}_{2}(\vec{k},-\vec{k}) =-1+\mathcal{O}(|\vec{k}|^{1/2})\label{k2at0}
\end{equation}

Therefore, $\mathcal{K}_{2}(0,0)=-1$. Notice that $\mathcal{K}_{2}$ is not singular at zero momentum, moreover we have obtained also a simple non perturbative result. If the bare mass is zero ($m=0$) and the dimension is larger than one ($d>1$), then we must have $\mathcal{K}_{2}(0,0)=-1$. Notice that the value of $\mathcal{K}_{2}(0,0)$ is not only non perturbative, but even independent from $\lambda$, exactly as happened in the gaussian case (\ref{studgaussian}).

\section{Consistency check for correlators in $\phi^4$ theory}
\label{check}

In Section \ref{grtool} we developed a method to obtain correlation functions once the cluster expansion is known, while in Section \ref{perturb} we saw how it is possible to compute the cluster expansion for ground states, at least perturbatively, and we gave the first order computation for the $\phi^{4}$ theory. A simple self-consistency test is to check if the correlators of the fields computed from this method match with the standard perturbative computation, i.e. the usual Feynman graphs \cite{Ma}. This is what we are going to do here, therefore in this section we deal with a $\phi^{4}$ Lagrangian (\ref{phi4action}).
The field correlators in the interacting ground state can be computed extracting the latter from the free ground state, i.e. the vacuum, using (\ref{thproj}). We choose to evaluate the normal ordered correlations to make the combinatorics easier, but the calculations can be done also without this requirement.
Using the same notations of Section \ref{scatmatsec} and evaluating the correlators in momentum space, we have:
\begin{equation}
\braket{:\prod_{i}\phi(\vec{k}_{i}):}{}_{G}=\lim_{\beta\rightarrow\infty}\frac{e^{2\beta E_{G}}}{\mathcal{N}_{G}}\bra{0}e^{-\beta H}:\prod_{i}\phi(\vec{k}_{i}):e^{-\beta H}\ket{0}
\end{equation}

This expression resembles very much (\ref{vaccumextr}), but notice that in (\ref{vaccumextr}) the exponential of the Hamiltonian was only on the left and not also on the right of the product of fields. Therefore, proceeding exactly as we did in Section \ref{scatmatsec} we arrive at the expression:
\begin{equation}
\braket{:\prod_{i}\phi(\vec{k}_{i}):}{}_{G}=\lim_{\beta\rightarrow\infty}\frac{\bra{0}\mathcal{T} \left(:\prod_{i}\phi_{I}(\vec{k}_{i},0): e^{-\lambda\int^{\beta}_{-\beta}d\tau V_{I}(\tau)}\right)\ket{0}}{\bra{0}\mathcal{T} e^{-\lambda\int^{\beta}_{-\beta}d\tau V_{I}(\tau)}\ket{0}}
\end{equation}

where we used that, by definition, $\phi_{I}(\vec{k},0)=\phi(\vec{k})$. Notice that the only difference with (\ref{scatmat40}) is that the domain of integration of the exponential is $(-\beta,\beta)$ instead of $(0,\beta)$.
From the expression above, it is not difficult to develop the formalism of Feynman graphs \cite{Ma} (that is different from Section \ref{scatmatsec} because of the different domain of integration of the euclidean time), but since we are interested only in the first order in $\lambda$, it is easier to expand directly the exponentials. The correlators can be easily computed using (\ref{tprop}). For example, for the four-point correlator we get:
\begin{equation}
\braket{:\prod_{i=1}^{4}\phi(\vec{k}_{i}):}{}_{G}=-\lambda \delta_{\sum_{i=1}^{4}\vec{k}_{i},0}\lim_{\beta\rightarrow\infty}\int_{-\beta}^{\beta}d\tau\prod_{i=1}^{4}\left(\int\frac{d k_{i}^{0}}{2\pi}\frac{ e^{ik_{i}^{0}\tau}}{[k_{i}^{0}]^{2}+[E(\vec{k}_{i})]^{2}}\right)
\end{equation}

The integrations can be easily performed giving:
\begin{equation}
\braket{:\prod_{i=1}^{4}\phi(\vec{k}_{i}):}{}_{G}=-\lambda \delta_{\sum_{i=1}^{4}\vec{k}_{i},0}\frac{2 L^{-d}}{\prod_{i=1}^{4}2E(\vec{k}_{i})}\frac{1}{\sum_{i=1}^{4}E(\vec{k}_{i})}+\mathcal{O}(\lambda^{2})\label{105}
\end{equation}

In a similar way we can also compute the two-point correlator. All other correlators vanish at this perturbation order. 

We will now compute the same object, but using the cluster expansion (\ref{expk}) of $\ket{G}$ with the expressions (\ref{k4}) and (\ref{k2}) for the functions $\mathcal{K}_n$. The graphical rules of Section \ref{grtool} allow us to find the expectation values of the fields $a_{\vec{k}}$, $a^{\dagger}_{\vec{k}}$ that are related to $\phi$ through the expression (\ref{phiI}), therefore the four point correlator on the ground state is:
\begin{equation}
\braket{:\prod_{i=1}^{4}\phi(\vec{k}_{i}):}{}_{G}=\prod_{i=1}^{4}\frac{1}{\sqrt{2E(\vec{k}_{i})}}\braket{:\prod_{i=1}^{4}\left(a^{\dagger}_{\vec{k}_{i}}+a_{-\vec{k}_{i}}\right):}{}_{G}
\end{equation}

\begin{figure}
\begin{center}
\includegraphics[scale=0.3]{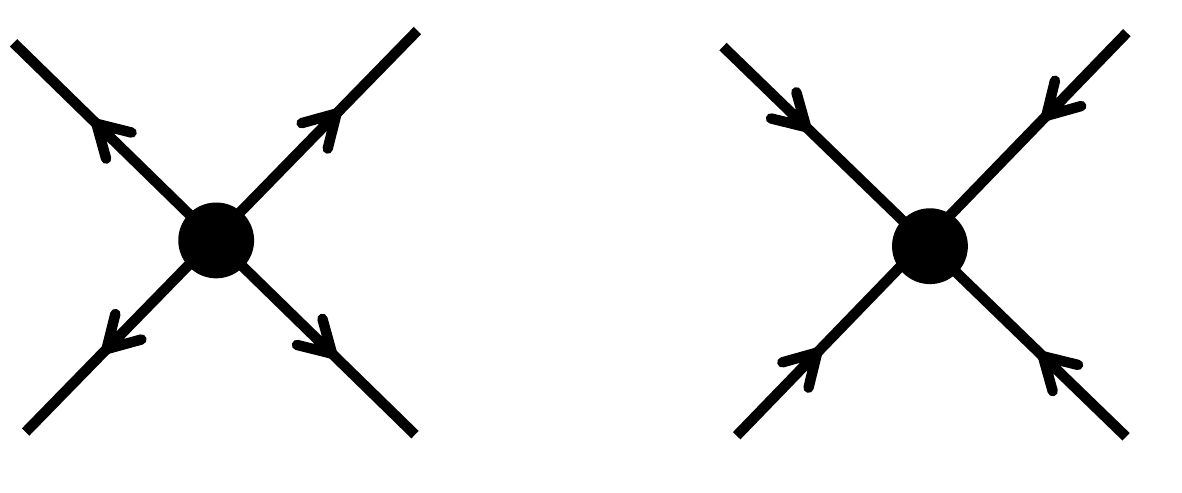}
\end{center}
\caption{First order contribution to the ground state four point connected correlation function in $\phi^{4}$}\label{phi41ord}
\end{figure}

At first order, only the graphs in Figure \ref{phi41ord} contribute, therefore among the four point correlators of the fields $a$ and $a^{\dagger}$ only $\braket{aaaa}_{G}$ and $\braket{a^{\dagger}a^{\dagger}a^{\dagger}a^{\dagger}}_{G}$ are of order $\lambda$, while all others are of higher order.
From Section \ref{grtool} we can read the value of the graphs in Figure \ref{phi41ord}:
\begin{equation}
L^{d}\braket{a_{\vec{k}_{1}}a_{\vec{k}_{2}}a_{\vec{k}_{3}}a_{\vec{k}_{4}}}_{G}=\delta_{\sum_{i=1}^{4}\vec{k}_{i},0}\hspace{3pt}\mathcal{K}_{4}(\vec{k}_{1},\vec{k}_{2},\vec{k}_{3},\vec{k}_{4})+\mathcal{O}(\lambda^{2})
\end{equation}

Thus, at first order in $\lambda$ we get:
\begin{equation}
\braket{:\prod_{i=1}^{4}\phi(\vec{k}_{i}):}{}_{G}= \delta_{\sum_{j=1}^{4}\vec{k}_{j},0} \frac{L^{-d}}{\prod_{i=1}^{4}\sqrt{2E(\vec{k}_{i})}}\left(\mathcal{K}^{*}_{4}(\vec{k}_{1},..,\vec{k}_{4})+\mathcal{K}_{4}(-\vec{k}_{1},..,-\vec{k}_{4})\right)+\mathcal{O}(\lambda^{2})
\end{equation}

Plugging the result for $\mathcal{K}_{4}$ of equation (\ref{k4}) into the expression above we get exactly the same result as in (\ref{105}), as it should be. 
With similar computation also the two point correlator can be computed and the two first order computations match, all the other correlators vanish at this order, as they should.

\end{document}